\documentclass[12pt]{article}
\usepackage{amsmath}
\usepackage{graphicx,psfrag,epsf}
\usepackage{enumerate}
\usepackage{natbib}
\usepackage{url} 
\usepackage{multirow}

\newtheorem{lemma}{Lemma}[section]
\newtheorem{theorem}[lemma]{Theorem}

\newtheorem{corollary}[lemma]{Corollary}
\newtheorem{proposition}[lemma]{Proposition}

\newcommand{\blind}{1}

\begin{document}

\def\spacingset#1{\renewcommand{\baselinestretch}%
{#1}\small\normalsize} \spacingset{1}


\if1\blind
{
  \title{\bf A Unified Approach on the Local Power of Panel Unit Root Tests}
  \author{Zhongwen Liang\thanks{Department of Economics, University at
 Albany, SUNY, Albany, NY 12222, USA. \newline \indent\hspace{0.5em} E-mail: zliang3@albany.edu} \\
    University at Albany, SUNY
    }
  \date{}
  \maketitle

  \begin{center}\large This version: \today \end{center}
} \fi

\if0\blind
{
  \bigskip
  \bigskip
  \bigskip
  \begin{center}
    {\LARGE\bf A Unified Approach on the Local Power of Panel Unit Root Tests}
\end{center}
  \medskip
} \fi

\bigskip
\begin{abstract}
In this paper, a unified approach is proposed to derive the exact local asymptotic power
for panel unit root tests, which is one of the most important issues in nonstationary panel data literature.
Two most widely used panel unit root tests known as Levin-Lin-Chu (LLC, Levin, Lin and Chu (2002))
and Im-Pesaran-Shin (IPS, Im, Pesaran and Shin (2003)) tests
are systematically studied for various situations to illustrate our method. Our approach is
characteristic function based, and can be used directly in deriving the moments of
the asymptotic distributions of these test statistics under the null and the local-to-unity alternatives. For the LLC test,
the approach provides an alternative way to obtain the results that can be derived by the existing method.
For the IPS test, the new results are obtained, which fills the gap in the literature where few results exist,
since the IPS test is non-admissible. Moreover, our approach has the advantage
in deriving Edgeworth expansions of these tests, which are also given in the paper.
The simulations are presented to illustrate our theoretical findings.
\end{abstract}

\noindent%

\medskip

\vfill

\newpage
\spacingset{1.45} 
 \section{Introduction}\label{sec:1}

 Since the circulation of working papers of Quah (1994), Breitung and Meyer (1994), and Levin and Lin (1992, unpublished manuscript),
 tremendous efforts have been made to construct and understand panel unit root tests. Levin et al. (2002)
 and Im et al. (2003) tests are among the ones that are the most widely used and influential, even though
 there are other tests proposed in the literature such as Breitung (2000),
 Ploberger and Phillips (2002), Moon and Phillips (2004), and Moon and Perron (2004). The LLC and IPS papers
 have received extremely high citations, which are among the most cited econometrics papers.
 There are excellent reviews in this area, such as Baltagi and Kao (2000),
 Phillips and Moon (2000), Choi (2006), Breitung and Pesaran (2008), and Westerlund and Breitung (2012).

 To evaluate the performance of these test statistics, the local power is the major concern.
 Moon, Perron and Phillips (2007) gave comprehensive discussions on deriving the Gaussian power envelopes
 for different scenarios, especially with incidental intercepts or incidental trends.
 They showed that under the homogeneous alternative, some tests such as a $t$-test constructed in the paper and the optimal invariant test
 in Ploberger and Phillips (2002) could achieve the power envelope in different situations. Furthermore, they proposed
 the corresponding point optimal invariant panel unit root test for each scenario. However, the local asymptotic power of
 IPS test was not discussed but compared with other tests using simulations, since it is shown in Bowman (2002) to be non-admissible.

 The majority of the literature on the local power of panel unit root tests rely on simulations,
 for example Maddala and Wu (1999) and Im et al. (2003),
 except some on the asymptotic limits such as Breitung (2000), Moon et al. (2006, 2007), Breitung and Pesaran (2008),
 Moon and Perron (2008), Harris et al. (2010), and Westerlund and Breitung (2012). In this paper, a new unified approach is proposed
 to explore the exact local asymptotic power of LLC and IPS tests, which utilizes the results of the Fredholm approach that targets directly on
 the characteristic functions of Dickey-Fuller tests, and that were extensively discussed in a series of papers
 by Nabeya and Tanaka (1988, 1990a, 1990b). Through this method,
 we are able to obtain the analytical forms of the local asymptotic power of
 LLC and IPS tests in different scenarios.

 In this paper, following similar setups in Levin et al. (2002), Im et al. (2003) and Moon et al. (2007), we discuss
 three scenarios in nonstationary panel data models, i.e., (i) without fixed effects; (ii) with incidental intercepts;
 (iii) with both incidental intercepts and incidental trends. We consider both homogeneous and heterogeneous
 alternatives. The analytical local asymptotic power of both LLC test and IPS test are derived for all scenarios.
 Moreover, since we directly target on moments, one advantage of our method is to obtain the Edgeworth expansion
 for the LLC and IPS tests under both null and local-to-unity alternatives.
 The one term Edgeworth expansions for the LLC and IPS tests are also derived in the paper.

 There is another strand of methods based on Fisher-type statistics such as Maddala and Wu (1999) and Choi (2001).
 As discussed in Bowman (2002), they are not admissible either. Until now, the local power of these tests are not very
 clear. Our method might also work for analyzing such tests. However, due to the extreme complexity of the problem,
 we leave it for further research. There is another literature on the second generation panel unit root tests for panel data models
 with cross section dependence, such as Bai and Ng (2004), Breitung and Das (2005, 2008) etc.,
 see Hurlin and Mignon (2007), Breitung and Pesaran (2008), Westerlund and Breitung (2012) and references therein. Our method could also be used
 in obtaining the exact local asymptotic power in these settings, which we leave for future research.

 The rest of paper is organized as follows. Section \ref{sec:2} introduces our unified approach in obtaining
 the asymptotic moments of test statistics involving unit root processes by summarizing the basic results of the Fredholm approach
 in Nabeya and Tanaka (1990a, 1990b) and extending these results to panel unit root data.
 Section \ref{sec:3} is devoted to derivations of the exact local asymptotic power of LLC and IPS tests under three different scenarios
 to illustrate our unified approach.
 In section \ref{sec:4}, we obtain the one term Edgeworth expansions
 for both LLC and IPS tests by utilizing our approach.
 Section \ref{sec:5} gives the simulation results. Section \ref{sec:6} concludes the paper. The main steps of proofs are gathered
 in the Appendix. A comprehensive supplemental material of proofs is also available.

 \section{A unified approach}\label{sec:2}

The traditional way to obtain the local asymptotic power is to derive the asymptotic limit of the test statistics under the local alternatives. However, this may lead to some expectations that are hard to compute if not possible at all, or it will keep the degenerate terms in the limiting expression which could be canceled out but cannot be seen directly. This is the case especially for the panel unit root tests. It will be clearer in the next two sections and the supplemental material. In contrast to the existing method, we propose a unified approach which is based on deriving the moments through the joint characteristic function. To fix the idea, we describe it here. The typical panel unit root tests take the form as
\[ T_1=\frac{N^{-1/2}\sum_{i=1}^N (A_i-E(A_i))}{\sqrt{N^{-1}\sum_{i=1}^N B_i}}, \quad \mbox{or} \quad T_2=N^{-1/2}\sum_{i=1}^N \left(\frac{A_i}{\sqrt{B_i}}-E\left(\frac{A_i}{\sqrt{B_i}}\right)\right),\]
if we take the sequential limit and let $T$ go to infinity. For instance, the LLC test takes the first expression,
and the IPS test takes the second expression.
Clearly, the limit of the first one under the local alternative is easier to evaluate, but not the second one.

We provide a unified approach here. Our idea is to compute the moments under the local alternatives. If we can derive the characteristic functions of $T_1$ and $T_2$, then the moments under both null and the local alternative can be obtained immediately. However, it's not easy to derive it especially in the unit root case. Fortunately, the joint characteristic function of $(A_i,B_i)$ can be obtained using the Fredholm approach below. Then, our unified approach consists of two steps. The first step is to obtain the joint characteristic function or the joint moment generating function (m.g.f.) of $(N^{-1/2}\sum_{i=1}^N A_i,N^{-1}\sum_{i=1}^N B_i)$ for $T_1$ and that of $(A_i,B_i)$ for $T_2$, respectively. The second step is the calculation of the asymptotic moments of the test statistic based on the m.g.f. obtained from the first step. These will
be better seen in the next section when the idea is illustrated using the LLC and IPS tests.

The first step of our approach can be accomplished using the Fredholm approach which is briefly summarized in the following.
For more detailed discussions, the readers are referred to the excellent monograph by Tanaka (1996). The Fredholm
approach with applications in deriving the characteristic functions of the Dickey-Fuller tests was systematically
studied in a series of papers by Nabeya and Tanaka (1988, 1990a, 1990b) and Tanaka (1990). However, it seems these results are largely overlooked in later analysis of the local power in the panel unit roots context.

In unit root time series literature, we typically consider the following three
setups:
\begin{eqnarray*}
\begin{array}{ll} \textrm{Model 2.1:}\ \ y_t = \rho y_{t-1}+\varepsilon_t, & (t=1,2,\dots,T) \\
\textrm{Model 2.2:}\ \ y_t = \alpha+\rho y_{t-1}+\varepsilon_t, & (t=1,2,\dots,T) \\
\textrm{Model 2.3:}\ \ y_t = \alpha+\beta t+\rho y_{t-1}+\varepsilon_t, & (t=1,2,\dots,T)\end{array}
\end{eqnarray*}
where the initial value $y_0$ is assumed to be a constant or a random variable whose distribution is independent of $T$, and
$\{\varepsilon_t\}$ is an i.i.d. sequence with $(0,\sigma^2)$.
It is well known that if the true value $\rho=1$, when we consider the OLS estimator $\hat \rho_i$ and
the corresponding $t$-statistic $t_{\hat \rho_i}$ for each model, we can obtain the following asymptotics:
\begin{eqnarray}
&&\textrm{Model 2.1:}\ \ T(\hat \rho_1-1)\Rightarrow \frac{U_1}{V_1}, \quad t_{\hat \rho_1}\Rightarrow \frac{U_1}{\sqrt{V_1}}, \label{eq2.1} \\
&&\textrm{Model 2.2:}\ \ T(\hat \rho_2-1)\Rightarrow \frac{U_2}{V_2}, \quad t_{\hat \rho_2}\Rightarrow \frac{U_2}{\sqrt{V_2}}, \label{eq2.2} \\
&&\textrm{Model 2.3:}\ \ T(\hat \rho_3-1)\Rightarrow \frac{U_3}{V_3}, \quad t_{\hat \rho_3}\Rightarrow \frac{U_3}{\sqrt{V_3}}, \label{eq2.3}
\end{eqnarray}
where $\Rightarrow$ stands for weak convergence,
\begin{eqnarray*}
U_1 &=& \int_0^1 W(r)dW(r)=\frac{1}{2}[W^2(1)-1], \quad V_1=\int_0^1 W^2(r)dr, \\
U_2 &=& \left|\begin{array}{ll} \int_0^1 W(r)dW(r) & \int_0^1 W(r)dr \\
\int_0^1 dW(r) & 1 \end{array}\right|, \quad V_2 = \left|\begin{array}{ll} \int_0^1 W^2(r)dr & \int_0^1 W(r)dr \\
\int_0^1 W(r)dr & 1 \end{array}\right|, \\
U_3 &=& 12\left|\begin{array}{lll} \int_0^1 W(r)dW(r) & \int_0^1 W(r)dr & \int_0^1 rW(r)dr \\
\int_0^1 dW(r) & 1 & \frac{1}{2} \\ \int_0^1 rdW(r) & \frac{1}{2} & \frac{1}{12} \end{array}\right|, \\
V_3 &=& 12\left|\begin{array}{lll} \int_0^1 W^2(r)dW(r) & \int_0^1 W(r)dr & \int_0^1 rW(r)dr \\
\int_0^1 W(r)dr & 1 & \frac{1}{2} \\ \int_0^1 rdW(r) & \frac{1}{2} & \frac{1}{12} \end{array}\right|,
\end{eqnarray*}
and $\{W(t): 0\leq t\leq 1\}$ is a standard Brownian motion. These results were obtained by various authors,
for example, the limiting expression for $\hat \rho_1$ in (\ref{eq2.1}) was obtained by Chan and Wei (1987) and Phillips (1987a),
the limiting expression for $t_{\hat \rho_1}$ in (\ref{eq2.1}) was obtained by Phillips (1987a). The
limiting expressions in (\ref{eq2.2}) and (\ref{eq2.3}) were obtained by Phillips and Perron (1988).

There was another strand of literature focusing directly on the limiting distributions of these statistics,
for example, White (1958) derived the joint m.g.f. for $(U_1,V_1)$ as
\[ \phi_1(u,v)=e^{-u/2}\left(\cos(\sqrt{2v})-u\frac{\sin(\sqrt{2v})}{\sqrt{2v}}\right)^{-1/2}.\]
Evans and Savin (1981) studied the moments of $U_1/V_1$ based on White's result. Dickey and Fuller (1979, 1981)
gave different expressions for the limit of
$t_{\hat \rho_1}$ and $t_{\hat \rho_2}$ in (\ref{eq2.1}) and (\ref{eq2.2}).

Nabeya and Tanaka (1988, 1990a, 1990b) extended the idea on
deriving the limiting distribution of $T(\hat \rho_i-1)$, for $i=1,2,3$. First,
they noticed that
\[ P\left(T(\hat \rho_i-1)<x\right)\to P(xV_i-U_i>0), \quad \mbox{as $T\to\infty$}.\]
Denote $Z_x=xV_i-U_i$. Then $Z_x$ could be
approximated by taking the limit of a quadratic form to reach the expression as
\[ Z_x=a^2\int_0^1\!\!\!\int_0^1 K_x(s,t)dW(s)dW(t)+b,\]
where $K_x(s,t)$ is the kernel associated with the eigenvalue integral equation
\[ f(t)=\lambda \int_0^1 K_x(s,t)f(s)ds\]
which is of the Fredholm type. Second, given the expression of $Z_x$, the characteristic function
of $Z_x/a^2$ could be expressed following Anderson and Darling (1952) as
\[ \phi_x(\theta)=e^{ir\theta}[D_x(2i\theta)]^{-1/2},\]
where $r=b/a^2$ and $D_x(\cdot)$ is the Fredholm determinant of $K_x(s,t)$. In the end,
the limiting distribution of $T(\hat \rho_i-1)$ could be calculated by Imhof (1961)'s formula, i.e.,
\[ \lim_{T\to\infty}P\left(T(\hat \rho-1)<x\right)=P(Z_x/a^2>0)=\frac{1}{2}+\frac{1}{\pi}\int_0^\infty \frac{1}{\theta}\mathrm{Im}(\phi_x(\theta))d\theta.\]
This is the so-called Fredholm approach. The key idea is to find the characteristic function of $Z_x$ using
the Fredholm determinant $D_x(\cdot)$ associated with the kernel $K_x(s,t)$. For models 1-3,
the expressions of $K_x(s,t)$, $a$ and $b$ can be derived. Nabeya and Tanaka (1988, 1990a, 1990b)
obtained the expressions of the Fredholm determinants for various cases.

The above-mentioned results can be used in detecting unit roots. Results in (\ref{eq2.1}),
(\ref{eq2.2}) and (\ref{eq2.3}) give the asymptotic expressions of the OLS estimators and $t$-statistics
under $H_0: \ \rho=1$. To consider the local power of these tests, the limiting distributions
under the local alternatives are also very important. Chan and Wei (1987) and Phillips (1987b) unified the asymptotics
through the local-to-unity alternatives $H_1:\ \rho=1-\frac{c}{T}$ or $H_1:\ \rho=e^c$, respectively. This unified
framework was also adopted in Nabeya and Tanaka (1990a). We summarize their results here for the convenience
of later reference. The following models were considered in Nabeya and Tanaka (1990a):
\begin{eqnarray*}
\begin{array}{ll} \textrm{Model }2.1':\ \ y_t = \eta_t, & (t=1,2,\dots,T) \\
\textrm{Model }2.2':\ \ y_t = \beta_0+\eta_t, & (t=1,2,\dots,T) \\
\textrm{Model }2.3':\ \ y_t = \beta_1 t+\eta_t, & (t=1,2,\dots,T) \\
\textrm{Model }2.4':\ \ y_t = \beta_0+\beta_1 t+\eta_t, & (t=1,2,\dots,T) \end{array}
\end{eqnarray*}
where $\eta_t=\rho\eta_{t-1}+u_t$, $\rho=1-c/T$, and $\{u_t\}$ is a linear process such that
\[ u_t=\sum_{j=0}^\infty \alpha_j \varepsilon_{t-j}, \quad \sum_{j=0}^\infty |\alpha_j|<\infty, \quad \sum_{j=0}^\infty \alpha_j\neq 0.\]
Here, $\{\varepsilon_t\}$ is a martingale difference process such that
\[ \frac{1}{T}\sum_{t=1}^T E(\varepsilon_t^2|\mathcal F_{t-1})\stackrel{p}{\to} \sigma^2,\]
where $\mathcal F_t=\sigma(\varepsilon_s,\ s\leq t)$. The initial value $\eta_0$ of $\{\eta_t\}$ was assumed to
be zero or a random variable whose distribution not depending on $T$. The main results are given in the following two lemmas,
which will be used in later discussion.

\begin{lemma}[Theorem 3 in Nabeya and Tanaka (1990a)]\label{lemma1}
Let $\hat \rho_j$ be the OLS estimator for Model $2.j'$ ($j=1,2,3,4$) correspondingly, with $\rho=1-c/T$.
Then
\[
\lim_{T\to\infty}P\left(T(\hat \rho_j-1)\right)=P\left(\left(\sum_{l=0}^\infty \alpha_l\right)^2W_j(c,x)+\sum_{l=0}^\infty \alpha_l^2>0\right)
=P\left(W_j(c,x)+r>0\right),
\]
where
\begin{eqnarray*}
r &=& \sum_{l=0}^\infty \alpha_l^2/\left(\sum_{l=0}^\infty \alpha_l\right)^2, \\
W_j(c,x) &=& \int_0^1\!\!\!\int_0^1 K_j(s,t;c,x)dW(s)dW(t),
\end{eqnarray*}
and $K_j(s,t;c,x)$ are defined as follows
{\small
\begin{eqnarray*}
K_1(s,t;c,x) &=& A_x(s,t)-e^{-c(2-s-t)}, \\
K_2(s,t;c,x) &=& A_x(s,t)-\frac{2x}{c^2}g(s)g(t)-e^{-c(2-s-t)}+\frac{1}{c}\left(e^{-c(1-s)}g(t)+e^{-c(1-r)}g(s)\right), \\
K_3(s,t;c,x) &=& A_x(s,t)-\frac{6x}{c^4}(g(s)+ch(s))(g(t)+ch(t))-\left(e^{-c(1-s)}-\frac{3}{c^2}(g(s)+ch(s))\right) \\
&&\times\left(e^{-c(1-r)}-\frac{3}{c^2}(g(t)+ch(t))\right), \\
K_4(s,t;c,x) &=& A_x(s,t)-x\left(\frac{8}{c^4}(c^2-3c+3)g(s)g(t)-\frac{12}{c^3}(c-2)(g(s)h(t)+g(t)h(s))+\frac{24}{c^2}h(s)h(t)\right) \\
&&-\left(e^{-c(1-s)}+\frac{2}{c}-\frac{6}{c^2}(g(s)+ch(s))\right)\left(e^{-c(1-r)}+\frac{2}{c}g(t)-\frac{6}{c^2}(g(t)+ch(t))\right) \\
&&+\frac{4}{c^4}(3c(1-s)-(c+3)g(s))(3c(1-t)-(c+3)g(t)),
\end{eqnarray*}}
where $A_x(s,t)=\frac{x}{c}(e^{-c|s-t|}-e^{-c(2-s-t)})$, $g(s)=1-e^{-c(1-s)}$, and $h(s)=s-e^{-c(1-s)}$.
\end{lemma}

The characteristic functions of corresponding $W_j(c,x)+r$ ($j=1,2,3,4)$ in Lemma \ref{lemma1} are given in the following lemma.
\begin{lemma}[Theorem 4 in Nabeya and Tanaka (1990a)]\label{lemma2}
The characteristic functions of $W_j(c,x)+r$ ($j=1,2,3,4)$ in Lemma \ref{lemma1} have the following
expression
\[ \varphi_j(\theta;c,r,x)=e^{ir\theta}\left[D_j(2i\theta;c,x)\right]^{-1/2},\]
where $D_j(\lambda;c,x)$ is the Fredholm determinant associated with $K_j(s,t;c,x)$, which is defined as follows
{\small
\begin{eqnarray*}
D_1(\lambda;c,x) &=& e^{-c}\left[\cos(\mu)+(c+\lambda)\frac{\sin(\mu)}{\mu}\right], \\
D_2(\lambda;c,x) &=& e^{-c}\left[\frac{\lambda^2+2\lambda x-c^2\lambda-c^3}{\mu^2}\frac{\sin(\mu)}{\mu}-c^2\frac{\cos(\mu)}{\mu}+
(2\lambda^2-4c\lambda x-2c^2\lambda)\frac{\cos(\mu)-1}{\mu^4}\right], \\
D_3(\lambda;c,x) &=& e^{-c}\bigg[-\frac{c^3+(c^2+3c+3)\lambda}{\mu^2}\frac{\sin(\mu)}{\mu}+\frac{3(c^2+3c+3+2(c+1)x)\lambda}{\mu^4}
\left(\frac{\sin(\mu)}{\mu}-\cos(\mu)\right) \\
&&-\frac{c^2}{\mu^2}\cos(\mu)\bigg], \\
D_4(\lambda;c,x) &=& e^{-c}\bigg[\frac{c^5+(c^4-4(c^2+3c+27)\lambda-8x(c^2-3c-3))\lambda}{\mu^4}\frac{\sin(\mu)}{\mu} \\
&&-\frac{24(c^4-8x\lambda^2+4(c+1)(x^2-3)\lambda)\lambda}{\mu^6}\left(\frac{\sin(\mu)}{\mu}+\frac{\cos(\mu)}{\mu^2}-\frac{1}{\mu^2}\right) \\
&&+\left(\frac{c^4}{\mu^4}+\frac{8(c^3(c+2x)-4(c^2+3c+6)\lambda)\lambda}{\mu^6}\right)\cos(\mu) \\
&&+\frac{4(c^4-4(c^2+3c-3)\lambda+2c^2x(c+3))\lambda}{\mu^6}\bigg],
\end{eqnarray*}}
where $\mu=\sqrt{2\lambda x-c^2}$.
\end{lemma}

The Fredholm approach could not only be applied to the Dickey-Fuller tests, but could also be used for understanding
other unit root tests, see Nabeya and Tanaka (1990b) for more discussions. The results in Lemma \ref{lemma2} serve as
the basis for the discussion in the following sections, since the related joint characteristic functions
of the panel unit root test statistics can be obtained with the adaption of the above-mentioned results.

 \section{Local powers}\label{sec:3}

To illustrate our approach, the exact local asymptotic powers of LLC and IPS tests are obtained in this section.
We consider the following general setting for the panel autoregressive model
\begin{eqnarray}\label{model}
z_{it} &=& d_{it}+y_{it}, \quad i=1,\dots,N; t=0,1,\dots,T, \nonumber \\
y_{it} &=& \rho_iy_{i,t-1}+u_{it}, \nonumber \\
d_{it} &=& \beta_{0i}+\beta_{1i}t, \nonumber \\
y_{i0} &=& \xi_i,
\end{eqnarray}
where $d_{it}$ is the deterministic component with possible trending,
$y_{it}$ is an autoregressive time series for each individual with
possibility to be unit root processes, and $y_{i0}$ gives the random initial conditions.
Specifically, the model (\ref{model}) includes three cases, i.e.,
\begin{eqnarray*}
\begin{array}{ll} \textrm{Model 3.1:}\ \ \beta_{0i}=\beta_{1i}=0, & \mbox{for all $i=1,\dots,N$} \\
\textrm{Model 3.2:}\ \ \beta_{1i}=0, & \mbox{for all $i=1,\dots,N$} \\
\textrm{Model 3.3:}\ \ \mbox{no restrictions on $\beta_{0i}$ and $\beta_{1i}$.} & \end{array}
\end{eqnarray*}
Our goal would be testing the presence of a common unit root against local alternatives.
The null and alternative hypotheses could be stated as follows.
\begin{eqnarray}
H_0: \rho_i=1, &&  \mbox{ for all $i$,} \label{H0} \\
H_1: \rho_i<1, &&  \mbox{ for $M$ number of $i$'s,} \label{H1}
\end{eqnarray}
where $M$ satisfies $\lim_{N\to \infty} M/N=p$, $0<p\leq 1$. As a special case for the alternative
hypothesis $H_1$, we can consider the homogeneous alternative
\begin{equation}\label{H1'}
H_1': \ \rho_1=\rho_2=\cdots=\rho_N=\rho<1.
\end{equation}
We discuss the LLC and IPS tests separately in the following subsections. We make the following assumptions before that.

\bigskip

 \noindent {\bf Assumption 1} The errors $u_{it}$ are i.i.d. with
 $(0,\sigma_{u ,i}^2)$ over $t=1,\dots,T$ and are also independent
 across $i=1,\dots,N$. $\sup_i E(u_{it}^8)<M_1<\infty$ and $\inf_i \sigma_{u,i}^2\geq M_2>0$ for
 some constants $M_1$ and $M_2$.

 \noindent {\bf Assumption 2} The initial points $y_{i0}$ are i.i.d. with $E(y_{i0}^8)<M_3<\infty$ for
 some constant $M_3$ and are independent of $\{u_{it}\}_{i=1}^T$ for all $i$.

 \noindent {\bf Assumption 3} For Model 3.1, let $\rho_i=1-c_i/(N^{1/2}T)$; for Model 3.2, let $\rho_i=1-c_i/(N^{1/2}T)$;
 for Model 3.3, let $\rho_i=1-c_i/(N^{1/4}T)$, where $c_i\geq 0$, $i=1,\dots,N$.

 \noindent {\bf Assumption 4} $1/T+1/N+N/T\to 0$.

\bigskip

\noindent {\bf Remark 3.1} Assumption 1 assumes i.i.d. errors, which is adopted for simplicity of
the derivation. From Lemma \ref{lemma2} in Section \ref{sec:2}, the i.i.d.
errors could be relaxed to be linear processes with no essential impact on the results. Assumption 2
assumes the nonexplosive initial conditions. We will relax this assumption and discuss the impact of initial conditions
in the next section. Assumption 3 gives local-to-unity alternatives with different rates
for different models and test statistics, which is well known in the literature. Assumption 4 is
the same as Assumption 3 in Moon et al. (2007), which is required for the convergence of test statistics.
The sequential convergence as $T\to \infty$ followed by $N\to \infty$ is adopted in this paper
for convenience. The joint convergence could also be obtained with strengthened conditions.
More comprehensive discussions could be found in Phillips and Moon (1999).

\subsection{LLC test}\label{sec:3.1}

Firstly, we discuss the LLC test.
In Levin et al. (2002), the following models were considered.
\begin{eqnarray*}
\textrm{Model }3.1':\ \ \Delta y_{it} &=& \delta y_{i,t-1}+\varepsilon_{i,t}, \\
\textrm{Model }3.2':\ \ \Delta y_{it} &=& \alpha_{0i}+\delta y_{i,t-1}+\varepsilon_{i,t}, \\
\textrm{Model }3.3':\ \ \Delta y_{it} &=& \alpha_{0i}+\alpha_{1i}t+\delta y_{i,t-1}+\varepsilon_{i,t}, \ \ \mbox{where $-2<\delta\leq 0$, for $i=1,\dots,N$}.
\end{eqnarray*}
Clearly, we can see that Model $3.1'$, Model $3.2'$ and Model $3.3'$ are equivalent to Model 3.1, Model 3.2
and Model 3.3, respectively, under the null hypothesis (\ref{H0}) and the homogeneous alternative (\ref{H1'}), i.e.,
\begin{eqnarray*}
\tilde H_0: \ \delta=0, \quad \mbox{v.s.} \quad \tilde H_1: \ \delta <0.
\end{eqnarray*}
LLC proposed a three-step pooled OLS estimation of $\delta$ and showed that the pooled $t$-statistic
converges to $N(0,1)$ under $H_0$ for Model 1 with some regularity conditions,
and converges to $N(0,1)$ under $H_0$ for Model 2 and Model 3 with corrections for the means and additional conditions.

Here, we focus on the local-to-unity alternatives such as those given in Assumption 3. Our goal is to achieve
analytical expressions for the local asymptotic power. For simplicity of exposition, we focus on the prototype
LLC test statistics. The construction of the LLC tests are discussed in the following. For Model $3.1'$, we can estimate $\delta$ by
\begin{equation}\label{3.1.1}
\hat \delta_1=\frac{\sum_{i=1}^N\sum_{t=2}^T (y_{i,t-1}\Delta y_{i,t}/\hat \sigma_{\varepsilon1,i}^2)}{\sum_{i=1}^N\sum_{t=2}^T (y_{i,t-1}^2/\hat \sigma_{\varepsilon1,i}^2)},
\end{equation}
where $\hat \sigma_{\varepsilon1,i}=\sqrt{\frac{1}{T-1}\sum_{t=2}^T (\Delta y_{i,t}-\tilde \delta_{1i} y_{i,t-1})^2}$
is a consistent estimator for $\sigma_{\varepsilon,i}$ to accommodate heteroscedasticity, and $\tilde \delta_{1i}$ is
the OLS estimator from the individual time series.
The $t$-statistic is given by
\begin{equation}\label{3.1.2}
\hat t_{\delta,1} = \frac{\hat \delta_1}{(\sum_{i=1}^N\sum_{t=2}^T (y_{i,t-1}^2/\hat \sigma_{\varepsilon1,i}^2))^{-1/2}}.
\end{equation}
Under $H_1$ that is specified in Assumption 3, we have
\begin{equation}\label{3.1.3}
\hat t_{\delta,1}=-\frac{1}{N^{1/2}T}\frac{\sum_{i=1}^N\sum_{t=2}^T c_i(y_{i,t-1}^2/\hat \sigma_{\varepsilon1,i}^2)}{\sqrt{\sum_{i=1}^N\sum_{t=2}^T (y_{i,t-1}^2/\hat \sigma_{\varepsilon1,i}^2)}}+\frac{N^{-1/2}T^{-1}\sum_{i=1}^N\sum_{t=2}^T (y_{i,t-1}\varepsilon_{i,t}/\hat \sigma_{\varepsilon1,i}^2)}{\sqrt{N^{-1}T^{-2}\sum_{i=1}^N\sum_{t=2}^T (y_{i,t-1}^2/\hat \sigma_{\varepsilon1,i}^2)}}.
\end{equation}

 From Phillips (1987b), we have that as $T\to \infty$
 \begin{equation}\label{3.1.4}
 T^{-1/2}(y_{i,[Tr]}-y_{i,0}) \Rightarrow \left\{\begin{array}{ll}
 \sigma_{\varepsilon,i}W_i(r) & \mbox{for $c_i=0$,} \\
 \sigma_{\varepsilon,i}\int_0^r e^{-c_iN^{-1/2}(r-s)}dW_i(s) & \mbox{else.}\end{array}\right.
 \end{equation}
Thus, as $T\to \infty$
\begin{equation}\label{3.1.5}
\hat t_{\delta,1}\Rightarrow -\frac{N^{-1}\sum_{i=1}^N c_i\int_0^1 K_{i,c_i}(r)^2dr}{\sqrt{N^{-1}\sum_{i=1}^N\int_0^1
 K_{i,c_i}(r)^2dr}}+\frac{N^{-1/2}\sum_{i=1}^N\int_0^1 K_{i,c_i}(r)dW_i(r)}{\sqrt{N^{-1}\sum_{i=1}^N\int_0^1
 K_{i,c_i}(r)^2dr}}\stackrel{def}{=}\frac{N^{-1/2}\sum_{i=1}^N \tilde{U}_{1i}}{\sqrt{N^{-1}\sum_{i=1}^N \tilde{V}_{1i}}},
\end{equation}
where $K_{i,c_i}(r) = \int_0^r e^{-c_iN^{-1/2}(r-s)}dW_i(s)$.

Now, we illustrate our approach to achieve the exact local asymptotic power. The idea is to calculate
the asymptotic moments of the statistic under the local-to-unity alternatives.
The asymptotic normality could be obtained by applying the standard Central Limit Theorem (CLT). To this end, we derive the
joint m.g.f. in the first step. By substituting $\theta=iu/2$, $x=-v/u$ and $c=c_iN^{-1/2}$ into
$\varphi_1(\theta;c,1,x)$ in Lemma \ref{lemma2},
we obtain the joint m.g.f. for $(\tilde{U}_{1i},\tilde{V}_{1i})$ as
\begin{equation}\label{3.1.6}
\psi_{1,i}(u,v) = e^{-\frac{u}{2}}\left[e^{-c_iN^{-\frac{1}{2}}}\left[\cos\sqrt{2v-c_i^2N^{-1}}
+(c_iN^{-1/2}-u)\frac{\sin\sqrt{2v-c_i^2N^{-1}}}{\sqrt{2v-c_i^2N^{-1}}}\right]\right]^{-1/2}.
\end{equation}
Hence, the joint m.g.f. for $(N^{-1/2}\sum_{i=1}^N \tilde{U}_{1i},N^{-1}\sum_{i=1}^N \tilde{V}_{1i})$ is
{\small
\[
\phi_1(u,v) =  e^{-\frac{\sqrt N u}{2}}\left[e^{-\sum_{i=1}^N c_iN^{-\frac{1}{2}}}\prod_{i=1}^N
\left[\cos\sqrt{\frac{2 v}{N}-c_i^2N^{-1}}+(c_iN^{-1/2}-\frac{u}{\sqrt N})\frac{\sin\sqrt{\frac{2 v}{N}-c_i^2N^{-1}}}{\sqrt{\frac{2v}{N}-c_i^2N^{-1}}}\right]\right]^{-1/2}.
\]}
Now, we take the second step by applying the relationship between moments and the m.g.f. of the random variable
in terms of a ratio, i.e., a formula given in Sawa (1972),
\begin{equation}\label{3.1.sawa}
E\left(\frac{U^p}{V^q}\right)=\frac{1}{\Gamma(q)}\int_0^\infty v^{q-1} \left.\frac{\partial^p}{\partial u^p}\phi(u,-v)\right|_{u=0}dv.
\end{equation}
The asymptotic moments $\hat t_{\delta,1}$ can be directly calculated using this formula.
From Taylor expansion (see the detailed derivations in the Appendix), we have
\begin{equation}\label{3.1.7}
\left.\frac{\partial }{\partial u}\phi_1(u,-v)\right|_{u=0} = -\frac{1}{2}e^{-v/2}N^{-1}\sum_{i=1}^N c_i+O(N^{-1/2}).
\end{equation}
Hence, combining (\ref{3.1.5}) and (\ref{3.1.sawa}) and plugging in (\ref{3.1.7}), we have
\begin{eqnarray*}
E(\hat t_{\delta,1}) &=& E\left(\frac{N^{-1/2}\sum_{i=1}^N \tilde{U}_{1i}}{\sqrt{N^{-1}\sum_{i=1}^N \tilde{V}_{1i}}}\right) = \frac{1}{\Gamma(\frac{1}{2})}\int_0^\infty \frac{1}{\sqrt v} \left.\frac{\partial }{\partial u}\phi_1(u,-v)\right|_{u=0}dv \nonumber \\
&=& -\left(N^{-1}\sum_{i=1}^N c_i\right)\frac{1}{2\sqrt \pi}\int_0^\infty \frac{e^{-v/2}}{\sqrt v}dv+O_p(N^{-1/2})=-\frac{\bar c}{\sqrt 2}+O_p(N^{-1/2}),
\end{eqnarray*}
where $\bar c=\lim_{N\to \infty}N^{-1}\sum_{i=1}^N c_i$. This moment gives the asymptotic bias that leads to the local power.
The result is summarized in Theorem \ref{LLC} below.

Next, we consider Model $3.2'$, $\delta$ can be estimated by the fixed-effects estimator
\begin{equation}\label{3.1.8}
\hat \delta_2=\frac{\sum_{i=1}^N\sum_{t=2}^T [(y_{i,t-1}-\bar y_{i,t-1})(\Delta y_{i,t}-\overline{\Delta y}_{i,t})/\hat \sigma_{\varepsilon2,i}^2]}{\sum_{i=1}^N\sum_{t=2}^T [(y_{i,t-1}-\bar y_{i,t-1})^2/\hat \sigma_{\varepsilon2,i}^2]},
\end{equation}
where $\hat \sigma_{\varepsilon2,i}=\sqrt{\frac{1}{T-1}\sum_{t=2}^T (\Delta y_{i,t}-\tilde \alpha_{0i}-\tilde \delta_{2i} y_{i,t-1})^2}$
is a consistent estimator for $\sigma_{\varepsilon,i}$ with that $\tilde \alpha_{0i}$ and $\tilde \delta_{2i}$ are OLS estimators
from the individual time series, $\bar y_{i,t-1}=(T-1)^{-1}\sum_{s=2}^T y_{i,s-1}$, and $\overline{\Delta y}_{i,t}=(T-1)^{-1}\sum_{s=2}^T \Delta y_{is}$.
However, $\hat \delta_2$ is biased, since under the null hypothesis, by (\ref{eq2.2}) as $T\to \infty$
\begin{eqnarray*}
T\hat \delta_2 &=& \frac{\sum_{i=1}^N\sum_{t=2}^T [(y_{i,t-1}-\bar y_{i,t-1})(\varepsilon_{i,t}-\bar \varepsilon_{i,t})/\hat \sigma_{\varepsilon2,i}^2]}{\sum_{i=1}^N\sum_{t=2}^T [(y_{i,t-1}-\bar y_{i,t-1})^2/\hat \sigma_{\varepsilon2,i}^2]} \\
&\Rightarrow&  \frac{N^{-1}\sum_{i=1}^N\int_0^1
 W_i^\mu(r)dW_i(r)}{N^{-1}\sum_{i=1}^N\int_0^1
 W_i^\mu(r)^2dr}\stackrel{def}{=} \frac{N^{-1}\sum_{i=1}^N U_{2i}}{N^{-1}\sum_{i=1}^N V_{2i}},
\end{eqnarray*}
where $W_i^\mu(r)=W_i(r)-\int_0^1 W_i(s)ds$, $\bar \varepsilon_{i,t}=(T-1)^{-1}\sum_{s=2}^T \varepsilon_{is}$
and $E(U_2)=-1/2$, $Var(U_2)=1/12$, $E(V_2)=1/6$, $Var(V_2)=1/45$ from Table 1 in Levin et al. (2002)
or the derivations in the supplementary material.
As discussed in Moon and Perron (2008), there are several ways for bias correction. The first way
is to correct the overall bias for the whole test statistic as what was proposed in Levin and Lin (1992).
With this type of bias correction, the $t$-statistic is given by
\begin{equation}\label{3.1.9}
\hat t_{\delta,21}
= \frac{\sqrt 5}{2}\frac{\hat \delta_2}{(\sum_{i=1}^N\sum_{t=2}^T [(y_{i,t-1}-\bar y_{i,t-1})^2/\hat \sigma_{\varepsilon2,i}^2])^{-1/2}}+\sqrt{\frac{15N}{8}}.
\end{equation}
Under the corresponding specification of $H_1$ in Assumption 3, we have
\begin{eqnarray*}
\hat t_{\delta,21}
&=& -\frac{\sqrt 5}{2N^{1/2}T}\frac{\sum_{i=1}^N\sum_{t=2}^T c_i[(y_{i,t-1}-\bar y_{i,t-1})^2/\hat \sigma_{\varepsilon2,i}^2]}{\sqrt{\sum_{i=1}^N\sum_{t=2}^T [(y_{i,t-1}-\bar y_{i,t-1})^2/\hat \sigma_{\varepsilon2,i}^2]}} \\
&&+\frac{\sqrt 5}{2}\frac{N^{-1/2}T^{-1}\sum_{i=1}^N\sum_{t=2}^T \left[\left((y_{i,t-1}-\bar y_{i,t-1})(\varepsilon_{i,t}-\bar\varepsilon_{i,t})/\hat \sigma_{\varepsilon2,i}^2\right)+\frac{1}{2}\right]}{\sqrt{N^{-1}T^{-2}\sum_{i=1}^N\sum_{t=2}^T [(y_{i,t-1}-\bar y_{i,t-1})^2/\hat \sigma_{\varepsilon2,i}^2]}} \\
&&-\frac{\sqrt 5}{2}\left(\frac{\sqrt N(T-1)}{2T\sqrt{N^{-1}T^{-2}\sum_{i=1}^N\sum_{t=2}^T [(y_{i,t-1}-\bar y_{i,t-1})^2/\hat \sigma_{\varepsilon2,i}^2]}}-\sqrt{\frac{3N}{2}}\right).
\end{eqnarray*}
Furthermore, as $T\to \infty$
\begin{eqnarray}\label{3.1.10}
\hat t_{\delta,21}
&\Rightarrow& -\frac{\sqrt 5}{2}\frac{N^{-1}\sum_{i=1}^N c_i\int_0^1
 K_{i,c_i}^\mu(r)^2dr}{\sqrt{N^{-1}\sum_{i=1}^N\int_0^1
 K_{i,c_i}^\mu(r)^2dr}}+\frac{\sqrt 5}{2}\frac{N^{-1/2}\sum_{i=1}^N \left(\int_0^1
 K_{i,c_i}^\mu(r)dW_i(r)+\frac{1}{2}\right)}{\sqrt{N^{-1}\sum_{i=1}^N\int_0^1
 K_{i,c_i}^\mu(r)^2dr}} \nonumber \\
 &&-\frac{\sqrt 5}{2}\left(\frac{\sqrt N}{2\sqrt{N^{-1}\sum_{i=1}^N \int_0^1
 K_{i,c_i}^\mu(r)^2dr}}-\sqrt{\frac{3N}{2}}\right)
 \stackrel{def}{=} \frac{\sqrt 5}{2}\frac{N^{-1/2}\sum_{i=1}^N \tilde{U}_{2i}}{\sqrt{N^{-1}\sum_{i=1}^N \tilde{V}_{2i}}}+\sqrt{\frac{15N}{8}},
 \nonumber \\
\end{eqnarray}
where $K_{i,c_i}^\mu(r)=K_{i,c_i}(r)-\int_0^1 K_{i,c_i}(s)ds$.

Our unified approach can be applied here. Substituting $\theta=iu/2$, $x=-v/u$ and $c=c_iN^{-1/2}$ into
$\varphi_2(\theta;c,1,x)$ in Lemma \ref{lemma2}, we have the joint m.g.f. for $(\tilde{U}_{2i},\tilde{V}_{2i})$ as
 {\small
 \begin{eqnarray}\label{3.1.11}
 \psi_{2,i}(u,v) &=& e^{-\frac{u}{2}}\Bigg[e^{-c_iN^{-\frac{1}{2}}}\Bigg[\frac{u^2+2v+c_i^2N^{-1}u-c_i^3N^{-3/2}}{2v-c_i^2N^{-1}}
 \frac{\sin\sqrt{2v-c_i^2N^{-1}}}{\sqrt{2v-c_i^2N^{-1}}} \nonumber \\
 &&-c_i^2N^{-1}\frac{\cos\sqrt{2v-c_i^2N^{-1}}}{2v-c_i^2N^{-1}}+(2u^2-4c_iN^{-1/2}v
 +2c_i^2N^{-1}u)\frac{\cos\sqrt{2v-c_i^2N^{-1}}-1}{(2v-c_i^2N^{-1})^2}\Bigg]\Bigg]^{-1/2}. \nonumber \\
 \end{eqnarray}}
 Hence, the joint m.g.f. for $(N^{-1/2}\sum_{i=1}^N \tilde{U}_{2i},N^{-1}\sum_{i=1}^N \tilde{V}_{2i})$ is
 {\footnotesize
 \begin{eqnarray*}
 &&\phi_2(u,v) \\
 &=& e^{-\frac{\sqrt N u}{2}}\Bigg[e^{-\sum_{i=1}^N c_iN^{-\frac{1}{2}}}\prod_{i=1}^N\Bigg[\frac{\frac{u^2}{N}
 +\frac{2v}{N}+c_i^2N^{-1}\frac{u}{\sqrt N}-c_i^3N^{-3/2}}{\frac{2v}{N}-c_i^2N^{-1}}
 \frac{\sin\sqrt{\frac{2v}{N}-c_i^2N^{-1}}}{\sqrt{\frac{2v}{N}-c_i^2N^{-1}}} \\
 &&-c_i^2N^{-1}\frac{\cos\sqrt{\frac{2v}{N}-c_i^2N^{-1}}}{\frac{2v}{N}-c_i^2N^{-1}}
 +(\frac{2u^2}{N}-4c_iN^{-1/2}\frac{v}{N}+2c_i^2N^{-1}\frac{u}{\sqrt N})\frac{\cos\sqrt{\frac{2v}{N}-c_i^2N^{-1}}-1}{(\frac{2v}{N}-c_i^2N^{-1})^2}\Bigg]\Bigg]^{-1/2}.
 \end{eqnarray*}}
 From Taylor expansion (see the Appendix), we have
 \begin{equation}\label{3.1.12}
 \left.\frac{\partial }{\partial u}\phi_2(u,-v)\right|_{u=0}
 =  -\frac{\sqrt N}{2}e^{-v/6}-\frac{1}{24}ve^{-v/6}N^{-1}\sum_{i=1}^N c_i+O_p(N^{-1/2}).
 \end{equation}
 Combining (\ref{3.1.10}), (\ref{3.1.sawa}) and (\ref{3.1.12}), it implies
 \begin{eqnarray*}
 &&E\left(\frac{\sqrt 5}{2}\frac{N^{-1/2}\sum_{i=1}^N \tilde{U}_{2i}}{\sqrt{N^{-1}\sum_{i=1}^N \tilde{V}_{2i}}}\right)
 =\frac{\sqrt 5}{2}\frac{1}{\Gamma(\frac{1}{2})}\int_0^\infty \frac{1}{\sqrt v} \left.\frac{\partial }{\partial u}\phi_2(u,-v)\right|_{u=0}dv \nonumber \\
 &=& -\frac{\sqrt 5}{2}\frac{\sqrt N}{2}\frac{1}{\sqrt \pi}\int_0^\infty \frac{e^{-v/6}}{\sqrt v}dv-\left(N^{-1}\sum_{i=1}^N c_i\right)\frac{\sqrt 5}{2}\frac{1}{24\sqrt \pi}\int_0^\infty \frac{ve^{-v/6}}{\sqrt v}dv+O_p(N^{-1/2}) \nonumber \\
 &=& -\sqrt{\frac{15N}{8}}-\frac{1}{8}\sqrt{\frac{15}{2}}\bar c+O_p(N^{-1/2}).
 \end{eqnarray*}
 Therefore, we have
 \[
  E\left(\hat t_{\delta,21}\right)=E\left(\frac{\sqrt 5}{2}\frac{N^{-1/2}\sum_{i=1}^N \tilde{U}_{2i}}{\sqrt{N^{-1}\sum_{i=1}^N \tilde{V}_{2i}}}\right)+\sqrt{\frac{15N}{8}}=-\frac{1}{8}\sqrt{\frac{15}{2}}\bar c+O_p(N^{-1/2}),
 \]
 which gives the local power.

Another way to correct the bias is to correct the overall bias for $\hat \delta_2$. With this type of bias correction, the $t$-statistic is given by
\begin{equation}\label{3.1.13}
\hat t_{\delta,22}
= \sqrt{\frac{10}{17}}\frac{\hat \delta_2+\frac{3}{T}}{(\sum_{i=1}^N\sum_{t=2}^T [(y_{i,t-1}-\bar y_{i,t-1})^2/\hat \sigma_{\varepsilon2,i}^2])^{-1/2}}.
\end{equation}
Under $H_1$ given in Assumption 3, we have
\begin{eqnarray*}
\hat t_{\delta,22}
&=& -\sqrt{\frac{10}{17}}\frac{1}{N^{1/2}T}\frac{\sum_{i=1}^N\sum_{t=2}^T c_i[(y_{i,t-1}-\bar y_{i,t-1})^2/\hat \sigma_{\varepsilon2,i}^2]}{\sqrt{\sum_{i=1}^N\sum_{t=2}^T [(y_{i,t-1}-\bar y_{i,t-1})^2/\hat \sigma_{\varepsilon2,i}^2]}} \\
&&+\sqrt{\frac{10}{17}}\frac{N^{-1/2}T^{-1}\sum_{i=1}^N\sum_{t=2}^T \left[\left((y_{i,t-1}-\bar y_{i,t-1})(\varepsilon_{i,t}-\bar\varepsilon_{i,t})/\hat \sigma_{\varepsilon2,i}^2\right)+\frac{1}{2}\right]}{\sqrt{N^{-1}T^{-2}\sum_{i=1}^N\sum_{t=2}^T [(y_{i,t-1}-\bar y_{i,t-1})^2/\hat \sigma_{\varepsilon2,i}^2]}} \\
&&-\sqrt{\frac{10}{17}}\left(\frac{\frac{\sqrt N(T-1)}{2T}-3\sqrt N \left(N^{-1}T^{-2}\sum_{i=1}^N\sum_{t=2}^T [(y_{i,t-1}-\bar y_{i,t-1})^2/\hat \sigma_{\varepsilon2,i}^2]\right)}{\sqrt{N^{-1}T^{-2}\sum_{i=1}^N\sum_{t=2}^T [(y_{i,t-1}-\bar y_{i,t-1})^2/\hat \sigma_{\varepsilon2,i}^2]}}\right).
\end{eqnarray*}
Moreover, as $T\to \infty$, we have
{\small
\begin{eqnarray}\label{3.1.14}
&&\hat t_{\delta,22}
\Rightarrow -\sqrt{\frac{10}{17}}\frac{N^{-1}\sum_{i=1}^N c_i\int_0^1
 K_{i,c_i}^\mu(r)^2dr}{\sqrt{N^{-1}\sum_{i=1}^N\int_0^1
 K_{i,c_i}^\mu(r)^2dr}}+\sqrt{\frac{10}{17}}\frac{N^{-1/2}\sum_{i=1}^N \left(\int_0^1
 K_{i,c_i}^\mu(r)dW_i(r)+\frac{1}{2}\right)}{\sqrt{N^{-1}\sum_{i=1}^N\int_0^1
 K_{i,c_i}^\mu(r)^2dr}} \nonumber \\
 &&+\sqrt{\frac{10}{17}}\frac{3\sqrt N\left[N^{-1}\sum_{i=1}^N \int_0^1
 K_{i,c_i}^\mu(r)^2dr-\frac{1}{6}\right]}{\sqrt{N^{-1}\sum_{i=1}^N \int_0^1
 K_{i,c_i}^\mu(r)^2dr}}\stackrel{def}{=} \sqrt{\frac{10}{17}}\left[\frac{N^{-1/2}\sum_{i=1}^N \tilde U_{2i}}{\sqrt{N^{-1}\sum_{i=1}^N \tilde V_{2i}}}+3\sqrt N\sqrt{N^{-1}\sum_{i=1}^N \tilde V_{2i}}\right]. \nonumber \\
\end{eqnarray}}

The local asymptotic power of $\hat t_{\delta,22}$ can be obtained in the same way
using our approach with an additional relationship on the moments.
It can shown that
 \begin{equation}\label{3.1.sawa1}
 E\left(V^{n-\alpha}\right)=\frac{(-1)^n}{\Gamma(\alpha)}\int_0^\infty v^{\alpha-1} \frac{\partial^n}{\partial v^n}\left[\phi(0,-v)\right]dv.
 \end{equation}
From Taylor expansion, we have
 \begin{equation}\label{3.1.15}
 \frac{\partial}{\partial v}\left[\phi_2(0,-v)\right]
 = -\frac{1}{6}e^{-v/6}+\frac{e^{-v/6}-\frac{1}{6}v e^{-v/6}}{12 N^{3/2}}\sum_{i=1}^N c_i+O(N^{-1}).
 \end{equation}
 Thus, combining (\ref{3.1.14}), (\ref{3.1.sawa}), (\ref{3.1.15}) and (\ref{3.1.sawa1}), we have
 \begin{eqnarray}\label{3.1.16}
 E\left(\hat t_{\delta,22}\right) &=& E\left(\sqrt{\frac{10}{17}}\left[\frac{N^{-1/2}\sum_{i=1}^N \tilde{U}_{2i}}{\sqrt{N^{-1}\sum_{i=1}^N \tilde{V}_{2i}}}+3\sqrt N\sqrt{N^{-1}\sum_{i=1}^N \tilde V_{2i}}\right]\right) \nonumber \\
 &=& \sqrt{\frac{10}{17}}\frac{1}{\Gamma(\frac{1}{2})}\int_0^\infty \frac{1}{\sqrt v} \left.\frac{\partial }{\partial u}\phi_2(u,-v)\right|_{u=0}dv-\sqrt{\frac{10}{17}}\frac{3\sqrt N}{\Gamma(\frac{1}{2})}\int_0^\infty \frac{1}{\sqrt v} \frac{\partial}{\partial v}\left[\phi_2(0,-v)\right]dv \nonumber \\
 &=& -\frac{1}{2}\sqrt{\frac{15}{17}}\bar c+O_p(N^{-1/2}).
 \end{eqnarray}
 The result in (\ref{3.1.16}) implies the result in Theorem \ref{LLC}
 with the standard CLT.

Finally, for Model $3.3'$, we can estimate $\delta$ by the following estimator
\begin{equation}\label{3.1.17}
\hat \delta_3 = \frac{\sum_{i=1}^N\sum_{t=2}^T [(y_{i,t-1}-\bar y_{i,t-1})-\hat \beta_{1i}(t-\bar t)][\Delta y_{i,t}-\overline{\Delta y}_{i,t}]/\hat \sigma_{\varepsilon3,i}^2}{\sum_{i=1}^N\sum_{t=2}^T [(y_{i,t-1}-\bar y_{i,t-1})-\hat \beta_{1i}(t-\bar t)]^2/\hat \sigma_{\varepsilon3,i}^2},
\end{equation}
where {\small $\hat \beta_{1i}=[\sum_{s=1}^T (s-\bar s)(\Delta y_{is}-\overline{\Delta y}_{is})]/[\sum_{s=1}^T (s-\bar s)^2]$}, {\scriptsize $\hat \sigma_{\varepsilon3,i}=\sqrt{(T-1)^{-1}\sum_{t=2}^T (\Delta y_{it}-\tilde
 \alpha_{0i}-\tilde \alpha_{1i} t-\tilde \delta_{3i} y_{i,t-1})^2}$} is a
 consistent estimator for $\sigma_{\varepsilon,3}$, and $\tilde \alpha_{0i}$, $\tilde \alpha_{1i}$ and $\tilde \delta_{3i}$
 are OLS estimators for each cross section.
It is well known that $\hat \delta_3$ is also biased, since under the null hypothesis as $T\to \infty$
{\footnotesize
\begin{eqnarray*}
&&T\hat \delta_3 \\
&=& \frac{\sum_{i=1}^N\left[\left(\sum_t (t-\bar t)^2\right)\left(\sum_t
 (y_{i,t-1}-\bar y_{i,t-1})(\varepsilon_{it}-\bar \varepsilon_{it})\right)-\left(\sum_t(t-\bar
 t)(y_{i,t-1}-\bar y_{i,t-1})\right)\left(\sum_t(t-\bar t)(\varepsilon_{it}-\bar \varepsilon_{it})\right)\right]/\hat \sigma_{\varepsilon3,i}^2}{\sum_{i=1}^N[\left(\sum_t (t-\bar t)^2\right)\left(\sum_t
 (y_{i,t-1}-\bar y_{i,t-1})^2\right)-\left(\sum_t(t-\bar
 t)(y_{i,t-1}-\bar y_{i,t-1})\right)^2]/\hat \sigma_{\varepsilon3,i}^2} \\
&\Rightarrow&  \frac{N^{-1}\sum_{i=1}^N \left[\int_0^1 W_i^\mu(r)dW_i(r)-12\int_0^1(r-\frac{1}{2})W_i(r)dr\int_0^1 (r-\frac{1}{2})dW_i(r)\right]}{N^{-1}\sum_{i=1}^N\left[\int_0^1 W_i^\mu(r)^2dr-12\left(\int_0^1(r-\frac{1}{2})W_i(r)dr\right)^2\right]}\stackrel{def}{=} \frac{N^{-1}\sum_{i=1}^N U_{3i}}{N^{-1}\sum_{i=1}^N V_{3i}},
\end{eqnarray*}}
where $E(U_3)=-1/2$, $Var(U_3)=1/60$, $E(V_3)=1/15$, $Var(V_3)=11/6300$ from Table 1 in Levin et al. (2002).

Similar as what we discussed for Model $3.2'$, there are different ways for bias correction.
The first way is to correct the overall bias for the whole test statistic. With this type of bias correction, the $t$-statistic is given by
\begin{equation}\label{3.1.18}
\hat t_{\delta,31}
= \sqrt{\frac{448}{277}}\frac{\hat \delta_3}{(\sum_{i=1}^N\sum_{t=2}^T [(y_{i,t-1}-\bar y_{i,t-1})-\frac{\sum_{s=1}^T (s-\bar s)(y_{is}-\bar y_{is})}{\sum_{s=1}^T (s-\bar s)^2}(t-\bar t)]^2/\hat \sigma_{\varepsilon3,i}^2)^{-1/2}}+\sqrt{\frac{1680N}{277}}.
\end{equation}
Following the corresponding specification of $H_1$ in Assumption 3, and taking $T\to \infty$, we have
as given in the Appendix
\begin{equation}\label{3.1.19}
\hat t_{\delta,31}
 \stackrel{def}{\Rightarrow} \sqrt{\frac{448}{277}}\frac{N^{-1/2}\sum_{i=1}^N \tilde{U}_{3i}}{\sqrt{N^{-1}\sum_{i=1}^N \tilde{V}_{3i}}}+\sqrt{\frac{448}{277}}\sqrt{\frac{15N}{4}}.
\end{equation}

To derive the local asymptotic power of $\hat t_{\delta,31}$, we apply the same approach.
The joint m.g.f. $\psi_{3,i}(u,v)$ of $(\tilde{U}_{3i},\tilde{V}_{3i})$ can be obtained
from Lemma \ref{lemma2} as given in the Appendix.
Hence, the joint m.g.f. $\phi_3(u,v)$ for $(N^{-1/2}\sum_{i=1}^N \tilde{U}_{3i},N^{-1}\sum_{i=1}^N \tilde{V}_{3i})$ can be
obtained by the relationship
\[ \phi_3(u,v)=\prod_{i=1}^N \psi_{3,i}(\frac{u}{\sqrt N},\frac{v}{N}).\]
 From Taylor expansion, we have
 \begin{equation}\label{3.1.20}
 \left.\frac{\partial }{\partial u}\phi_3(u,-v)\right|_{u=0}
 = -\frac{\sqrt N}{2}e^{-v/15}-\frac{1}{840}ve^{-v/15}N^{-1}\sum_{i=1}^N c_i^2+O_p(N^{-1/4}).
 \end{equation}
 Hence, by combining (\ref{3.1.19}), (\ref{3.1.sawa}) and (\ref{3.1.20}), we get
 \[
  E\left(\hat t_{\delta,31}\right)=E\left(\sqrt{\frac{448}{277}}\frac{N^{-1/2}\sum_{i=1}^N \tilde U_{3i}}{\sqrt{N^{-1}\sum_{i=1}^N \tilde V_{3i}}}\right)+\sqrt{\frac{448}{277}}\sqrt{\frac{15N}{4}}=-\frac{1}{14}\sqrt{\frac{105}{277}}\overline{c^2}
 +O_p(N^{-1/4}),
 \]
 where $\overline{c^2}=\lim_{N\to \infty} N^{-1}\sum_{i=1}^N c_i^2$. This implies the result in Theorem \ref{LLC} with the standard CLT.

 Another way to correct the bias is to correct the overall bias for $\hat \delta_3$. With this type of bias correction, the $t$-statistic is given by
\begin{equation}\label{3.1.21}
\hat t_{\delta,32}
= \sqrt{\frac{112}{193}}\frac{\hat \delta_3+\frac{15}{2T}}{(\sum_{i=1}^N\sum_{t=2}^T [(y_{i,t-1}-\bar y_{i,t-1})-\frac{\sum_{s=1}^T (s-\bar s)(y_{is}-\bar y_{is})}{\sum_{s=1}^T (s-\bar s)^2}(t-\bar t)]^2/\hat \sigma_{\varepsilon3,i}^2)^{-1/2}}.
\end{equation}
Based on the specification of $H_1$ given in Assumption 3, we have as $T\to \infty$,
\begin{equation}\label{3.1.22}
\hat t_{\delta,32}
 \stackrel{def}{\Rightarrow} \sqrt{\frac{112}{193}}\left[\frac{N^{-1/2}\sum_{i=1}^N \tilde U_{3i}}{\sqrt{N^{-1}\sum_{i=1}^N \tilde V_{3i}}}+\frac{15\sqrt N}{2}\sqrt{N^{-1}\sum_{i=1}^N \tilde V_{3i}}\right].
\end{equation}
 Further, we have
 \begin{equation}\label{3.1.23}
 \frac{\partial}{\partial v}\left[\phi_3(0,-v)\right]
 = -\frac{1}{15}e^{-v/15}+\frac{e^{-v/15}-\frac{1}{15}v e^{-v/15}}{420 N^{3/2}}\sum_{i=1}^N c_i^2+O(N^{-3/4}).
 \end{equation}
 Therefore, combining (\ref{3.1.22}), (\ref{3.1.sawa}), (\ref{3.1.sawa1}) and (\ref{3.1.23}), we have
 {\footnotesize
 \begin{eqnarray*}
 E\left(\hat t_{\delta,32}\right) &=& E\left(\sqrt{\frac{112}{193}}\left[\frac{N^{-1/2}\sum_{i=1}^N \tilde U_{3i}}{\sqrt{N^{-1}\sum_{i=1}^N \tilde V_{3i}}}+\frac{15\sqrt N}{2}\sqrt{N^{-1}\sum_{i=1}^N \tilde V_{3i}}\right]\right) \nonumber \\
 &=& \sqrt{\frac{112}{193}}\frac{1}{\Gamma(\frac{1}{2})}\int_0^\infty \frac{1}{\sqrt v} \left.\frac{\partial }{\partial u}\phi_3(u,-v)\right|_{u=0}dv
 -\sqrt{\frac{112}{193}}\frac{15\sqrt N}{2}\frac{1}{\Gamma(\frac{1}{2})}\int_0^\infty \frac{1}{\sqrt v} \frac{\partial}{\partial v}\left[\phi_3(0,-v)\right]dv \nonumber \\
 &=& -\frac{\sqrt{15}}{56}\sqrt{\frac{112}{193}}\overline{c^2}+O_p(N^{-1/4}),
 \end{eqnarray*}}
 which further implies the result in Theorem \ref{LLC}.

 The above-mentioned results are summarized in the following theorem.
\begin{theorem}\label{LLC}
Under Assumptions 1 to 4, when $T\to \infty$ followed by $N\to
 \infty$, we have the following asymptotic results.
\begin{itemize}
\item[(a)] For Model $3.1'$, $\hat t_{\delta,1}\Rightarrow N(0,1)-\frac{\bar c}{\sqrt 2}$;

\item[(b)] For Model $3.2'$, $\hat t_{\delta,21}\Rightarrow N(0,1)-\frac{1}{8}\sqrt{\frac{15}{2}}\bar c$, and $\hat t_{\delta,22} \Rightarrow N(0,1)-\frac{1}{2}\sqrt{\frac{15}{17}}\bar c$;

\item[(c)] For Model $3.3'$, $\hat t_{\delta,31}\Rightarrow N(0,1)-\frac{1}{14}\sqrt{\frac{105}{277}}\overline{c^2}$, and
$\hat t_{\delta,32} \Rightarrow N(0,1)-\frac{\sqrt{15}}{56}\sqrt{\frac{112}{193}}\overline{c^2}$.
\end{itemize}
\end{theorem}

\bigskip

\noindent {\bf Remark 3.2} Clearly, all of these results are obtained in a unified way. The same results were also obtained by different authors,
for example, Moon et al. (2007), Moon and Perron (2004), Moon and Perron (2008), and Westerlund and Breitung (2012),
using the computation of the expectations, which is different from our approach. The comparison with our approach
is given in the supplementary material. Also, from Moon et al. (2007), we know that none of these tests would
achieve the power envelope under the heterogeneous alternatives,
and $\hat t_{\delta,1}$ can achieve the optimal power for Model $3.1'$ under the homogeneous alternative $H_1'$, but not
$\hat t_{\delta,21}$, $\hat t_{\delta,22}$, $\hat t_{\delta,31}$ and $\hat t_{\delta,32}$ for Model $3.2'$ and Model $3.3'$
even under the homogeneous alternative. Also, clearly $\hat t_{\delta,22}$ and $\hat t_{\delta,32}$ have larger local powers
than $\hat t_{\delta,21}$ and $\hat t_{\delta,31}$, respectively. Our approach provides an alternative way.
The advantage of our approach can be better appreciated when we consider the IPS test.

\subsection{IPS test}\label{sec:3.2}

The IPS test is also one of the most widely used panel unit root tests. However, so far the literature on the local
power of the IPS test is rare. One of our major contributions in this paper is to derive the analytical
local asymptotic power of the IPS test for different scenarios. The advantage of our approach can be better seen in this section.
The model in (\ref{model}) was considered in Im et al. (2003). The idea is to form the standardized test statistics
from the OLS estimation of each individual time series.

For Model 3.1, The $t$ test statistic of $\rho_i$ is constructed by running the OLS estimation for each cross section.
Therefore, we have
\begin{equation}\label{3.2.1}
\hat \rho_i=\frac{\sum_{t=2}^T z_{i,t-1}z_{i,t}}{\sum_{t=2}^T z_{i,t-1}^2}, \quad \hat t_i=\frac{\hat \rho_i-1}{\hat \sigma_{u1,i}(\sum_{t=2}^T z_{i,t-1}^2)^{-1/2}},
\end{equation}
where $\hat \sigma_{u1,i}=\sqrt{(T-1)^{-1}\sum_{t=2}^T (z_{i,t}-\hat \rho_i z_{i,t-1})^2}$
is a consistent estimator for $\sigma_{u,i}$. Under $H_0$, the asymptotics of $\hat t_i$ is given in (\ref{eq2.1}).

The IPS test statistic is constructed as the standardized statistic
 of t-statistic, i.e.
 \[ Z=\frac{\sqrt N[N^{-1}\sum_{i=1}^N
 \hat t_i-E(t_0)]}{\sqrt{Var(t_0)}},\]
 where $E(t_0)$ and $Var(t_0)$ are the mean and the variance from the limiting distribution of the corresponding
 Dickey-Fuller statistic, respectively. We can find the approximated values of $E(t_0)=-0.42309565$ and
 $\sqrt{Var(t_0)}=0.98111424$ from Table 4 in Nabeya (1999).

 Under $H_1$, we have
 \[ \hat t_i=-\frac{c_i}{N^{1/2}\hat \sigma_{u1,i}}\sqrt{T^{-2}\sum_{t=2}^T y_{i,t-1}^2}+\frac{T^{-1}\sum_{t=2}^T y_{i,t-1}u_{i,t}}{\hat \sigma_{u1,i}\sqrt{T^{-2}\sum_{t=2}^T y_{i,t-1}^2}}.\]
 From (\ref{3.1.4}), we have that as $T\to \infty$
\begin{equation}\label{3.2.2}
\hat t_i\Rightarrow -\frac{c_i}{N^{1/2}}\sqrt{\int_0^1 K_{i,c_i}(r)^2dr}+\frac{\int_0^1 K_{i,c_i}(r)dW_i(r)}{\sqrt{\int_0^1
 K_{i,c_i}(r)^2dr}}\stackrel{def}{=}\frac{\tilde{U}_{1i}}{\sqrt{\tilde{V}_{1i}}}.
\end{equation}

 Using Taylor expansion, we could get the formal expression as
 {\small
 \begin{eqnarray*}
 Z &=& \frac{\sqrt N (N^{-1}\sum_{i=1}^N
 \hat t_i-E(t_0))}{\sqrt{Var(t_0)}} \Rightarrow N(0,1)-\bar c\Bigg[E\left(\sqrt{\int_0^1 W(r)^2dr}\right) \\
 &&+E\left(\frac{\int_0^1\!\!\int_0^r W(s)dsdW(r)}{\sqrt{\int_0^1 W(r)^2dr}}\right)
 -E\left(\frac{\int_0^1 W(r)dW(r)\int_0^1 W(r)\int_0^r W(s)dsdr}{\sqrt{(\int_0^1 W(r)^2dr)^3}}\right)
 \Bigg]/\sqrt{Var(t_0)}.
 \end{eqnarray*}}
 However, this is not very informative, since the expectations in this expression could not be calculated easily,
 which has to rely on simulations.

 Our approach can be readily applied here. In the first step, the joint m.g.f is directly given in (\ref{3.1.6}). Further, we get
 \begin{eqnarray*}
 \left.\frac{\partial }{\partial u}\psi_{1,i}(u,-v)\right|_{u=0}
 &=& -\frac{1}{2}\left[\cosh\sqrt{2 v}-c_iN^{-\frac{1}{2}}\left(\cosh\sqrt{2 v}-\frac{\sinh\sqrt{2 v}}{\sqrt{2 v}}\right)+O(N^{-1})\right]^{-1/2} \\
 &&+\frac{1}{2}\left[\cosh\sqrt{2 v}-c_iN^{-\frac{1}{2}}\left(\cosh\sqrt{2 v}-\frac{\sinh\sqrt{2 v}}{\sqrt{2 v}}\right)+O(N^{-1})\right]^{-3/2} \\
 &&\times \left(\frac{\sinh\sqrt{2 v}}{\sqrt{2 v}}-c_iN^{-1/2}\frac{\sinh\sqrt{2 v}}{\sqrt{2 v}}+O(N^{-1})\right).
 \end{eqnarray*}
 Therefore, by the change of variable as $x=\sqrt{2v}$ and Taylor expansion, with the formula in (\ref{3.1.sawa}) we have
 {\small
 \begin{eqnarray}\label{3.2.3}
 &&E(Z) \nonumber \\
 &=& (Var(t_0))^{-1/2} N^{1/2}N^{-1}\sum_{i=1}^N(E(\hat t_i)-E(t_0)) \nonumber \\
 &=& (Var(t_0))^{-1/2} N^{1/2}N^{-1}\sum_{i=1}^N \Bigg(\frac{1}{\Gamma(\frac{1}{2})}\int_0^\infty \frac{1}{\sqrt v} \left.\frac{\partial }{\partial u}\psi_{1,i}(u,-v)\right|_{u=0}dv-E(t_0)\Bigg) \nonumber \\
 &=& -(Var(t_0))^{-1/2} \frac{\bar c}{2\sqrt{2\pi}}\int_0^\infty (\cosh(x))^{-1/2}\left(1-\frac{2\sinh(x)}{x\cosh(x)}+\frac{3(\sinh(x))^2}{x^2(\cosh(x))^2}\right)dx+O_p(N^{-1/2}), \nonumber \\
 \end{eqnarray}}
 where
 \[ E(t_0)=-\frac{1}{\sqrt{2\pi}}\int_0^\infty (\cosh(x))^{-1/2}\left(1-\frac{\sinh(x)}{x\cosh(x)}\right) dx\]
 is also given in Nabeya (1999).

For Model 3.2, the $t$ test statistic of $\rho_i$ is constructed by running the OLS estimation for each cross section.
We have
\begin{eqnarray}\label{3.2.4}
\hat \rho_i^\mu &=& \frac{\sum_{t=2}^T (z_{i,t-1}-\bar z_{i,t-1})(z_{i,t}-\bar z_{i,t})}{\sum_{t=2}^T (z_{i,t-1}-\bar z_{i,t-1})^2}=\frac{\sum_{t=2}^T (y_{i,t-1}-\bar y_{i,t-1})(y_{i,t}-\bar y_{i,t})}{\sum_{t=2}^T (y_{i,t-1}-\bar y_{i,t-1})^2}, \nonumber \\
\hat t_i^\mu &=& \frac{\hat \rho_i^\mu-1}{\hat \sigma_{u2,i}(\sum_{t=2}^T (z_{i,t-1}-\bar z_{i,t-1})^2)^{-1/2}},
\end{eqnarray}
where $\bar z_{i,t-1}=(T-1)^{-1}\sum_{s=2}^T z_{i,s-1}$, $\bar z_{i,t}=(T-1)^{-1}\sum_{s=2}^T z_{i,s}$,
\[\hat \sigma_{u2,i}=\sqrt{(T-1)^{-1}\sum_{t=2}^T (z_{i,t}-\hat \alpha_i-\hat \rho_i^\mu z_{i,t-1})^2}\]
is a consistent estimator for $\sigma_{u,i}$, and $\hat \alpha_i$ is the OLS estimator from each cross section. Under $H_0$,
the asymptotics of $\hat t_i^\mu$ is given in (\ref{eq2.2}).

The IPS test statistic is constructed as the standardized statistic
 of t-statistic, i.e.
 \[ Z^\mu=\frac{\sqrt N[N^{-1}\sum_{i=1}^N
 \hat t_i^\mu-E(t_0^\mu)]}{\sqrt{Var(t_0^\mu)}},\]
 where $E(t_0^\mu)$ and $Var(t_0^\mu)$ are the mean and the variance from the limiting distribution of the corresponding
 Dickey-Fuller statistic for the model with an intercept, respectively. The approximated values of $E(t_0^\mu)=-1.53296244$ and
 $\sqrt{Var(t_0^\mu)}=0.84025086$ are given in Table 4 in Nabeya (1999).

 Under $H_1$, we have
 \[ \hat t_i^\mu = -\frac{c_i}{N^{1/2}\hat \sigma_{u2,i}}\sqrt{T^{-2}\sum_{t=1}^T (y_{i,t-1}-\bar y_{i,t-1})^2}+\frac{T^{-1}\sum_{t=2}^T (y_{i,t-1}-\bar y_{i,t-1})(u_{i,t}-\bar u_{i,t})}{\hat \sigma_{u2,i}\sqrt{T^{-2}\sum_{t=1}^T (y_{i,t-1}-\bar y_{i,t-1})^2}},\]
 where $\bar u_{i,t}=(T-1)^{-1}\sum_{s=2}^T u_{i,s}$. Moreover, as $T\to \infty$
\begin{equation}\label{3.2.5}
\hat t_i^\mu\Rightarrow -\frac{c_i}{N^{1/2}}\sqrt{\int_0^1
 K_{i,c_i}^\mu(r)^2dr}+\frac{\int_0^1 K_{i,c_i}^\mu(r)dW_i(r)}{\sqrt{\int_0^1
 K_{i,c_i}^\mu(r)^2dr}}\stackrel{def}{=}\frac{\tilde{U}_{2i}}{\sqrt{\tilde{V}_{2i}}}.
\end{equation}

 From the expression in (\ref{3.1.11}) and the derivations in the Appendix, we have
 \begin{eqnarray*}
 && \left.\frac{\partial }{\partial u}\psi_{2,i}(u,-v)\right|_{u=0} \\
 &=& -\frac{1}{2}\left[\frac{\sinh\sqrt{2 v}}{\sqrt{2 v}}-c_iN^{-\frac{1}{2}}\left(\frac{\sinh\sqrt{2 v}}{\sqrt{2 v}}-v^{-1}(\cosh\sqrt{2 v}-1)\right)+O(N^{-1})\right]^{-1/2}+O(N^{-1}).
 \end{eqnarray*}
 Applying our approach, we have
 \begin{eqnarray}\label{3.2.6}
 E(Z^\mu)  &=& (Var(t_0^\mu))^{-1/2} N^{1/2}N^{-1}\sum_{i=1}^N(E(\hat t_i^\mu)-E(t_0^\mu)) \nonumber \\
 &=& (Var(t_0^\mu))^{-1/2} N^{1/2}N^{-1}\sum_{i=1}^N \Bigg(\frac{1}{\Gamma(\frac{1}{2})}\int_0^\infty \frac{1}{\sqrt v} \left.\frac{\partial }{\partial u}\psi_{2,i}(u,-v)\right|_{u=0}dv-E(t_0^\mu)\Bigg) \nonumber \\
 &=& -(Var(t_0^\mu))^{-1/2} \frac{\bar c}{2\sqrt{2\pi}}\int_0^\infty \left(\frac{\sinh(x)}{x}\right)^{-1/2}\left(1-\frac{2(\cosh(x)-1)}{x\sinh(x)}\right)dx+O(N^{-1/2}), \nonumber \\
 \end{eqnarray}
 where
 \[ E(t_0^\mu)=-\frac{1}{\sqrt{2\pi}}\int_0^\infty \left(\frac{\sinh(x)}{x}\right)^{-1/2}dx,\]
 is given in (7) in Nabeya (1999).

 For Model 3.3, The $t$ test statistic of $\rho_i$ is constructed by running the OLS estimation for each cross section.
 We have
 \begin{equation}\label{3.2.7}
 \hat \rho_i^\tau =\frac{\left(\sum_t (t-\bar t)^2\right)\left(\sum_t
 (z_{i,t-1}-\bar z_{i,t-1})(z_{it}-\bar z_{it})\right)-\left(\sum_t(t-\bar
 t)(z_{i,t-1}-\bar z_{i,t-1})\right)\left(\sum_t(t-\bar
 t)(z_{it}-\bar z_{it})\right)}{\left(\sum_t (t-\bar t)^2\right)\left(\sum_t
 (z_{i,t-1}-\bar z_{i,t-1})^2\right)-\left(\sum_t(t-\bar
 t)(z_{i,t-1}-\bar z_{i,t-1})\right)^2},
 \end{equation}
 where all the summations are taken over 2 to $T$, and
 \begin{eqnarray*}
 \bar t &=& \frac{1}{T-1}\sum_{s=2}^T s=\frac{T+2}{2}, \ \
 \bar z_{i,t-1} = \frac{1}{T-1}\sum_{s=2}^T z_{i,s-1}, \ \
 \bar z_{it} = \frac{1}{T-1}\sum_{s=2}^T z_{i,s}.
 \end{eqnarray*}
 Then the t-statistic is given by
 \begin{equation}\label{3.2.8}
 \hat t_i^\tau=\frac{\hat \rho_i^\tau-1}{\hat \sigma_{u,i}\sqrt{\frac{\sum_t (t-\bar t)^2}{\left(\sum_t (t-\bar t)^2\right)\left(\sum_t
 (z_{i,t-1}-\bar z_{i,t-1})^2\right)-\left(\sum_t(t-\bar
 t)(z_{i,t-1}-\bar z_{i,t-1})\right)^2}}},
 \end{equation}
 where $\hat \sigma_{u,i}=\sqrt{(T-1)^{-1}\sum_{t=2}^T (z_{it}-\hat
 \alpha_i-\hat \gamma_i t-\hat \rho_i^\tau z_{i,t-1})^2}$ is a
 consistent estimator for $\sigma_{u,i}$, and $\hat \alpha_i$ and $\hat \gamma_i$ are
 OLS estimators from each cross section. Under $H_0$, the asymptotics of $\hat t_i^\tau$ is given in (\ref{eq2.3}).

 The IPS test statistic is constructed as the standardized statistic
 of t-statistic, i.e.
 \[ Z^\tau=\frac{\sqrt N[N^{-1}\sum_{i=1}^N
 \hat t_i^\tau-E(t_0^\tau)]}{\sqrt{Var(t_0^\tau)}},\]
 where $E(t_0^\tau)$ and $Var(t_0^\tau)$ are the mean and the variance from the limiting distribution of the
 Dickey-Fuller statistic for the model with both an intercept and an time trend, respectively. The approximated
 values of $E(t_0^\tau)=-2.18135582$ and $\sqrt{Var(t_0^\tau)}=0.74990847$ are given in Table 4 in Nabeya (1999).

 Under $H_1$, we have as $T\to \infty$
 \begin{eqnarray}\label{3.2.9}
 \hat t_i^\tau
 &\Rightarrow& -\frac{c_i}{N^{1/4}}\sqrt{\int_0^1
 K_{i,c_i}^\mu(r)^2dr-12\left(\int_0^1
 (r-\frac{1}{2})K_{i,c_i}^\mu(r)dr\right)^2} \nonumber \\
 &&+\frac{\int_0^1
 K_{i,c_i}^\mu(r)dW_i(r)-12\int_0^1
 (r-\frac{1}{2})K_{i,c_i}^\mu(r)dr\int_0^1 (r-\frac{1}{2})dW_i(r)}{\sqrt{\int_0^1
 K_{i,c_i}^\mu(r)^2dr-12\left(\int_0^1
 (r-\frac{1}{2})K_{i,c_i}^\mu(r)dr\right)^2}}\stackrel{def}{=}\frac{\tilde{U}_{3i}}{\sqrt{\tilde{V}_{3i}}}.
 \end{eqnarray}
 From the expression of $\psi_{3,i}(u,v)$ which is given in the Appendix, we obtain
 {\small
 \begin{eqnarray*}
 && \left.\frac{\partial }{\partial u}\psi_{3,i}(u,-v)\right|_{u=0} \nonumber \\
 &=& -\frac{1}{2}\Bigg[\Big[\frac{4v(5c_i^2N^{-1/2}-6)}{(-2v-c_i^2N^{-1/2})^2}
 \frac{\sin\sqrt{-2v-c_i^2N^{-1/2}}}{\sqrt{-2v-c_i^2N^{-1/2}}}+\frac{24(-4
 v^2+2c_i^2N^{-1/2}v^2)}{(-2v-c_i^2N^{-1/2})^3}\Bigg(\frac{\sin\sqrt{-2v-c_i^2N^{-1/2}}}{\sqrt{-2v-c_i^2N^{-1/2}}} \nonumber \\
 &&+\frac{\cos\sqrt{-2v-c_i^2N^{-1/2}}}{-2v-c_i^2N^{-1/2}}-\frac{1}{-2v-c_i^2N^{-1/2}}\Bigg)
 -\frac{4(6c_i^2N^{-1/2}v)}{(-2v-c_i^2N^{-1/2})^3}\Big]+O(N^{-3/4})\Bigg]^{-1/2}+O(N^{-1}). \nonumber \\
 \end{eqnarray*}}
 Applying our approach, we have
 {\footnotesize
 \begin{eqnarray}\label{3.2.10}
  E(Z^\tau)
 &=& (Var(t_0^\tau))^{-1/2} N^{1/2}N^{-1}\sum_{i=1}^N(E(\hat t_i^\tau)-E(t_0^\tau)) \nonumber \\
 &=& (Var(t_0^\tau))^{-1/2} N^{1/2}N^{-1}\sum_{i=1}^N \Bigg(\frac{1}{\Gamma(\frac{1}{2})}\int_0^\infty \frac{1}{\sqrt v} \left.\frac{\partial }{\partial u}\psi_{3,i}(u,-v)\right|_{u=0}dv-E(t_0^\tau)\Bigg) \nonumber \\
 &=& (Var(t_0^\tau))^{-1/2} \Bigg(\frac{\overline{c^2}}{\sqrt{2\pi}}\int_0^\infty [3f_{22}(x)]^{-3/2}\Big(-\frac{\sinh (x)}{x^3}+\frac{9\cosh (x)}{x^4}-\frac{33\sinh (x)}{x^5}+\frac{48(\cosh (x)-1)}{x^6}\Big)dx \nonumber \\
 &&+O(N^{-1/4})\Bigg),
 \end{eqnarray}}
 where
 \[f_{22}(x)=4\left(\frac{1}{x^3}\sinh (x)-\frac{2}{x^4}[\cosh (x)-1]\right), \quad \mbox{and} \quad E(t_0^\tau)=-\frac{1}{\sqrt{2\pi}}\int_0^\infty [3f_{22}(x)]^{-1/2} dx\]
 are given in equation (7) and page 147 in Nabeya (1999).

 We summarize these results in the following theorem.
\begin{theorem}\label{IPS}
Under the assumptions 1 to 4, when $T\to \infty$ followed by $N\to
 \infty$, we have the following asymptotic results.
\begin{itemize}
\item[(a)] For Model 3.1, \\
$Z\Rightarrow N(0,1)-(Var(t_0))^{-1/2} \frac{\bar c}{2\sqrt{2\pi}}\int_0^\infty (\cosh(x))^{-1/2}\left(1-\frac{2\sinh(x)}{x\cosh(x)}+\frac{3(\sinh(x))^2}{x^2(\cosh(x))^2}\right)dx$;

\item[(b)] For Model 3.2, $Z^\mu\Rightarrow N(0,1)-(Var(t_0^\mu))^{-1/2} \frac{\bar c}{2\sqrt{2\pi}}\int_0^\infty \left(\frac{\sinh(x)}{x}\right)^{-1/2}\left(1-\frac{2(\cosh(x)-1)}{x\sinh(x)}\right)dx$;

\item[(c)] For Model 3.3, $Z^\tau\Rightarrow N(0,1)-(Var(t_0^\tau))^{-1/2} \Bigg(\frac{\overline{c^2}}{\sqrt{2\pi}}\int_0^\infty [3f_{22}(x)]^{-3/2}\Big(\frac{\sinh (x)}{x^3}-\frac{9\cosh (x)}{x^4}+\frac{33\sinh (x)}{x^5}-\frac{48(\cosh (x)-1)}{x^6}\Big)dx\Bigg)$.
\end{itemize}
\end{theorem}
 The more detailed proofs are delegated in the Appendix.

\noindent {\bf Remark 3.3} The results in Theorem \ref{IPS} give the analytical forms of the asymptotic distributions of
 IPS tests under the local-to-unity alternatives, which imply the exact local asymptotic power of these tests. These results
 are new in the literature, which fills the gap for the IPS tests.
 Moreover, we can see that $Z$ and $Z^\mu$ have the local power in the neighborhood
 of unity with the order of $N^{-1/2}T^{-1}$, but $Z^\tau$ only has the local power
 in the neighborhood of unity with the order of $N^{-1/4}T^{-1}$. These are consistent
 with the general order results obtained in Moon et al. (2007).

 The integrals in Theorem \ref{IPS} can be evaluated numerically to further simplify the results.
 We adopt the numerical calculations stated in Nabeya (1999) to achieve the
 accuracy up to eight decimal places. From the numerical integrations using MATLAB, we have
 {\small
 \begin{eqnarray}\label{3.2.12}
 &&\frac{1}{2\sqrt{2\pi}}\int_0^\infty (\cosh(x))^{-1/2}\left(1-\frac{2\sinh(x)}{x\cosh(x)}+\frac{3(\sinh(x))^2}{x^2(\cosh(x))^2}\right)dx
 \approx 0.58198749, \\
 &&\frac{1}{2\sqrt{2\pi}}\int_0^\infty \left(\frac{\sinh(x)}{x}\right)^{-1/2}\left(1-\frac{2(\cosh(x)-1)}{x\sinh(x)}\right)dx
 \approx 0.23431142, \\
 &&\frac{1}{\sqrt{2\pi}}\int_0^\infty [3f_{22}(x)]^{-3/2}\Big(\frac{\sinh (x)}{x^3}-\frac{9\cosh (x)}{x^4}+\frac{33\sinh (x)}{x^5}-\frac{48(\cosh (x)-1)}{x^6}\Big)dx \approx 0.02854706, \nonumber \\
 \end{eqnarray}}

 Therefore, we obtain the following corollary.
 \begin{corollary}\label{coroll}
 Under Assumptions 1-4, when $T\to \infty$ followed by $N\to
 \infty$, we have that
 \begin{itemize}
\item[(a)] For model 3.1, $Z\Rightarrow N(0,1)-0.58198749 \bar c/\sqrt{Var(t_0)}$;

\item[(b)] For Model 3.2, $Z^\mu\Rightarrow N(0,1)-0.23431142 \bar c/\sqrt{Var(t_0^\mu)}$;

\item[(c)] For Model 3.3, $Z^\tau \Rightarrow N(0,1)-0.02854706 \overline{c^2}/\sqrt{Var(t_0^\tau)}$.
\end{itemize}
 \end{corollary}

 \noindent {\bf Remark 3.4} Compared with the results in Moon et al. (2007), we can see that the IPS tests
 would not achieve the power envelope for any case. Moreover, the IPS tests have
 lower power than the LLC tests in all scenarios. In Section \ref{sec:5}, we calculate the theoretical
 local asymptotic powers for different cases based on this corollary.

 \section{Edgeworth expansion}\label{sec:4}

 One advantage of our approach is to derive the Edgeworth expansion of panel unit root statistics, since our approach can be used
 to calculate the moments directly. Hall (1992) gave comprehensive discussions of conditions and results on the Edgeworth expansion.
 For LLC tests, Theorem 2.2 in Hall (1992) can be applied. The detailed derivations are collected in the Appendix.
 For $\hat t_{\delta,1}$, under $H_0$ we have the one term Edgeworth expansion as
 \begin{equation}\label{4.1}
 F_{1n}(x) = \Phi(x)+\frac{\sqrt 2}{3}N^{-1/2}\phi(x)+O(N^{-1}).
 \end{equation}
 Under $H_1$, we have
 \begin{eqnarray}\label{4.2}
 F_{1n,c}(x) &=& P\left(\hat t_{\delta,1}+\frac{\bar c}{\sqrt 2}-\frac{\sqrt 2}{6}\overline{c^2} N^{-1/2}+O(N^{-1})\leq x\right) \nonumber \\
 &=& F_{1n}(x)-\frac{\sqrt 2\bar c}{12}N^{-1}\phi(x)+O(N^{-1}).
 \end{eqnarray}

 Similarly, for $\hat t_{\delta,21}$, under $H_0$ we have the one term Edgeworth expansion as
 \begin{equation}\label{4.3}
 F_{21n}(x) = \Phi(x)+\frac{3\sqrt{30}}{40}N^{-1/2}\phi(x)-\frac{3\sqrt{30}}{560}N^{-1/2}(x^2-1)\phi(x)+O(N^{-1}).
 \end{equation}
 Under $H_1$, we have
 \begin{eqnarray}\label{4.4}
 F_{21n,c}(x) &=& P\left(\hat t_{\delta,21}+\frac{1}{8}\sqrt{\frac{15}{2}}\bar c-\frac{67\sqrt{30}}{1920}\overline{c^2}N^{-1/2}+O(N^{-1})\leq x\right) \nonumber \\
 &=& F_{21n}(x)-\frac{3\sqrt{30}\bar c}{160}N^{-1}\phi(x)-\frac{11\sqrt{30}\bar c}{17920}N^{-1}(x^2-1)\phi(x)+O(N^{-1}).
 \end{eqnarray}

 Moreover, for $\hat t_{\delta,22}$, under $H_0$ we have the one term Edgeworth expansion as
 \begin{equation}\label{4.5}
 F_{22n}(x) = \Phi(x)+\frac{\sqrt{1020}}{85}N^{-1/2}\phi(x)-\frac{27\sqrt{1020}}{20230}N^{-1/2}(x^2-1)\phi(x)+O(N^{-1}).
 \end{equation}
 Under $H_1$, we have
 \begin{eqnarray}\label{4.6}
 F_{22n,c}(x) &=& P\left(\hat t_{\delta,22}+\frac{1}{2}\sqrt{\frac{15}{17}}\bar c-\frac{31\sqrt{1020}}{4080}\overline{c^2}N^{-1/2}+O(N^{-1})\leq x\right) \nonumber \\
 &=& F_{22n}(x)-\frac{3\sqrt{1020}\bar c}{680}N^{-1}\phi(x)+\frac{13\sqrt{1020}\bar c}{20230}N^{-1}(x^2-1)\phi(x)+O(N^{-1}).
 \nonumber \\
 \end{eqnarray}

 Using the same method, for $\hat t_{\delta,31}$, under $H_0$ we have the one term Edgeworth expansion as
 \begin{equation}\label{4.7}
 F_{31n}(x) = \Phi(x)+\frac{33}{56}\sqrt{\frac{105}{277}}N^{-1/2}\phi(x)-\frac{491}{15512}\sqrt{\frac{105}{277}}N^{-1/2}(x^2-1)\phi(x)+O(N^{-1}).
 \end{equation}
 Under $H_1$, we have
 \begin{eqnarray}\label{4.8}
 F_{31n,c}(x) &=& P\left(\hat t_{\delta,21}+\frac{1}{14}\sqrt{\frac{105}{277}}\overline{c^2}+\frac{1}{12}\sqrt{\frac{105}{277}}\overline{c^3} N^{-1/4}+O(N^{-1/2})\leq x\right) \nonumber \\
 &=& F_{31n}(x)-\frac{59}{3136}\sqrt{\frac{105}{277}}\overline{c^2}N^{-1}\phi(x)
 +\frac{118445}{9555392}\sqrt{\frac{105}{277}}N^{-1}(x^2-1)\phi(x)+O(N^{-1}), \nonumber \\
 \end{eqnarray}
 where $\overline{c^3}=\lim_{N\to \infty}N^{-1}\sum_{i=1}^N c_i^3$.

 Moreover, for $\hat t_{\delta,32}$, under $H_0$ we have the one term Edgeworth expansion as
 \begin{equation}\label{4.9}
 F_{32n}(x) = \Phi(x)+\frac{11}{28}\sqrt{\frac{105}{193}}N^{-1/2}\phi(x)-\frac{397}{5404}\sqrt{\frac{105}{193}}N^{-1/2}(x^2-1)\phi(x)+O(N^{-1}).
 \end{equation}
 Under $H_1$, we have
 \begin{eqnarray}\label{4.10}
 F_{32n,c}(x) &=& P\left(\hat t_{\delta,22}+\frac{\sqrt{15}}{56}\sqrt{\frac{112}{193}}\overline{c^2}
 +\frac{1}{24}\sqrt{\frac{105}{193}}\overline{c^3}N^{-1/4}+O(N^{-1/2})\leq x\right) \nonumber \\
 &=& F_{32n}(x)-\frac{151}{9408}\sqrt{\frac{105}{193}}\overline{c^2}N^{-1}\phi(x)
 +\frac{84829}{6657728}\sqrt{\frac{105}{193}}\overline{c^2}N^{-1}(x^2-1)\phi(x)+O(N^{-1}). \nonumber \\
 \end{eqnarray}

 Next, we consider IPS tests. Theorem 2.1 in Hall (1992) can be applied directly. For $Z$, under $H_0$, we have
 the standard one term Edgeworth expansion as
 \begin{equation}\label{4.11}
 G_{1n}(x) = P\left(Z\leq x\right) = \Phi(x)-\frac{\lambda_1}{6}N^{-1/2}(x^2-1)\phi(x)+O(N^{-1}),
 \end{equation}
 where $\lambda_1=E(t_0-E(t_0))^3/(Var(t_0))^{3/2}=[E(t_0)^3-3E(t_0)^2E(t_0)+2(E(t_0))^3]/(Var(t_0))^{3/2}$
 is given in the Appendix.
 Under $H_1$, we have the one term Edgeworth expansion as
 \begin{eqnarray}\label{4.12}
 G_{1n,c}(x) &=& P\left(Z+N^{1/2}\sum_{i=1}^N (E(\hat t_i)-E(t_0))\leq x\right) \nonumber \\
 &=& \Phi(x)-\frac{\lambda_{1,c}}{6}N^{-1/2}(x^2-1)\phi(x)+O(N^{-1}),
 \end{eqnarray}
 where $\lambda_{1,c}=[E(\hat t_i)^3-3E(\hat t_i)^2E(\hat t_i)+2(E(\hat t_i))^3]/(Var(t_0))^{3/2}$
 is also given in the Appendix.

 Similarly, for $Z^\mu$, under $H_0$, we have
 the standard one term Edgeworth expansion as
 \begin{equation}\label{4.13}
 G_{2n}(x) = P\left(Z^\mu\leq x\right) = \Phi(x)-\frac{\lambda_2}{6}N^{-1/2}(x^2-1)\phi(x)+O(N^{-1}).
 \end{equation}
 where $\lambda_2=E(t_0^\mu-E(t_0^\mu))^3/(Var(t_0^\mu))^{3/2}$.
 Under $H_1$, we have
 the one term Edgeworth expansion as
 \begin{eqnarray}\label{4.14}
 G_{2n,c}(x) &=& P\left(Z^\mu+N^{1/2}\sum_{i=1}^N (E(\hat t_i^\mu)-E(t_0^\mu))\leq x\right) \nonumber \\
 &=& \Phi(x)-\frac{\lambda_{2,c}}{6}N^{-1/2}(x^2-1)\phi(x)+O(N^{-1}),
 \end{eqnarray}
 where
 \[ \lambda_{2,c}=[E(\hat t_i^\mu)^3-3E(\hat t_i^\mu)^2E(\hat t_i^\mu)+2(E(\hat t_i^\mu))^3]/(Var(t_0^\mu))^{3/2}.\]

 For $Z^\tau$, under $H_0$, we have
 the standard one term Edgeworth expansion as
 \begin{equation}\label{4.15}
 G_{3n}(x) = P\left(Z^\tau\leq x\right) = \Phi(x)-\frac{\lambda_3}{6}N^{-1/2}(x^2-1)\phi(x)+O(N^{-1}).
 \end{equation}
 where $\lambda_3=E(t_0^\tau-E(t_0^\tau))^3/(Var(t_0^\tau))^{3/2}$.
 Under $H_1$, we have
 the one term Edgeworth expansion as
 \begin{eqnarray}\label{4.16}
 G_{3n,c}(x) &=& P\left(Z^\tau+N^{1/2}\sum_{i=1}^N (E(\hat t_i^\tau)-E(t_0^\tau))\leq x\right) \nonumber \\
 &=& \Phi(x)-\frac{\lambda_{3,c}}{6}N^{-1/2}(x^2-1)\phi(x)+O(N^{-1}),
 \end{eqnarray}
 where
 \[ \lambda_{3,c}=[E(\hat t_i^\tau)^3-3E(\hat t_i^\tau)^2E(\hat t_i^\tau)+2(E(\hat t_i^\tau))^3]/(Var(t_0^\tau))^{3/2}.\]

Further, we can evaluate the numerical values of integrals in (\ref{4.11}) to (\ref{4.16}) following the way we discussed early.
The results are stated in the following corollary.
\begin{corollary}
 Under the Assumptions 1, 2 and 4, when $T\to \infty$ followed by $N\to
 \infty$, we have the one term Edgeworth expansion as follows.
\begin{itemize}
\item[(a)] For Model 3.1, for $Z$,
\begin{eqnarray*}
\mbox{under $H_0$}, && \quad G_{1n}(x) = P\left(Z\leq x\right) \approx \Phi(x)-0.0416N^{-1/2}(x^2-1)\phi(x)+O(N^{-1}), \\
\mbox{under $H_1$}, && \quad P\left(Z+N^{1/2}\sum_{i=1}^N (E(\hat t_i)-E(t_0))\leq x\right)
 = \Phi(x)-0.0416N^{-1/2}(x^2-1)\phi(x) \\
 &&+0.0672\bar cN^{-1}(x^2-1)\phi(x)+O(N^{-1});
\end{eqnarray*}

\item[(b)] For Model 3.2, for $Z^\mu$,
\begin{eqnarray*}
\mbox{under $H_0$}, && \quad G_{2n}(x) = P\left(Z^\mu\leq x\right) \approx \Phi(x)-0.0364N^{-1/2}(x^2-1)\phi(x)+O(N^{-1}), \\
\mbox{under $H_1$}, && \quad P\left(Z^\mu+N^{1/2}\sum_{i=1}^N (E(\hat t_i^\mu)-E(t_0^\mu))\leq x\right)
 = \Phi(x)-0.0364N^{-1/2}(x^2-1)\phi(x) \\
 &&+0.0354\bar cN^{-1}(x^2-1)\phi(x)+O(N^{-1});
\end{eqnarray*}

\item[(c)] For Model 3.3, for $Z^\tau$,
\begin{eqnarray*}
\mbox{under $H_0$}, && \quad G_{3n}(x) = P\left(Z^\tau\leq x\right) \approx \Phi(x)-0.0095N^{-1/2}(x^2-1)\phi(x)+O(N^{-1}), \\
\mbox{under $H_1$}, && \quad P\left(Z^\tau+N^{1/2}\sum_{i=1}^N (E(\hat t_i^\tau)-E(t_0^\tau))\leq x\right)
 = \Phi(x)-0.0095N^{-1/2}(x^2-1)\phi(x) \\
 &&+0.0058\overline{c^2}N^{-1}(x^2-1)\phi(x)+O(N^{-1}).
\end{eqnarray*}
\end{itemize}
\end{corollary}

 \section{Monte Carlo simulations}\label{sec:5}

In this section, we provide some simulations to illustrate our theoretical results.
The simulations on LLC tests were presented in Moon et al. (2007). Therefore, we only focus on the IPS tests.
The following data generating processes similar to that in Moon et al. (2007) are adopted.
 \begin{eqnarray*}
 z_{it} &=& b_{0i}+b_{1i}t+y_{it}, \\
 y_{it} &=& \left(1-\frac{c_i}{n^\alpha T}\right) y_{i,t-1}+\sigma_i e_{it}, \\
 y_{i,0} &=& 0, \ b_{0i},b_{1i},e_{it} \sim i.i.d. \ N(0,1),\
 \sigma_i^2 \sim U[0.5,1.5].
 \end{eqnarray*}
 Several different cases are considered where $c_i$ follows different
 distributions, i.e., (1) $c_i\sim iid \ U[0,1]$;
 (2) $c_i\sim iid \ U[0,8]$; (3) $c_i\sim iid \ \chi^2(1)$;
 (4) $c_i\sim iid \ \chi^2(6)$. Moreover, $N$ and $T$ are selected from $\{25,100,1000\}$
 and $\{50,100,250\}$, respectively. The results at 5\% significance level
 are reported with 2,000 replications. More simulations can be conducted with similar results.

\begin{figure}
\centering
\includegraphics[angle=0,width=\textwidth,height=2.3in]{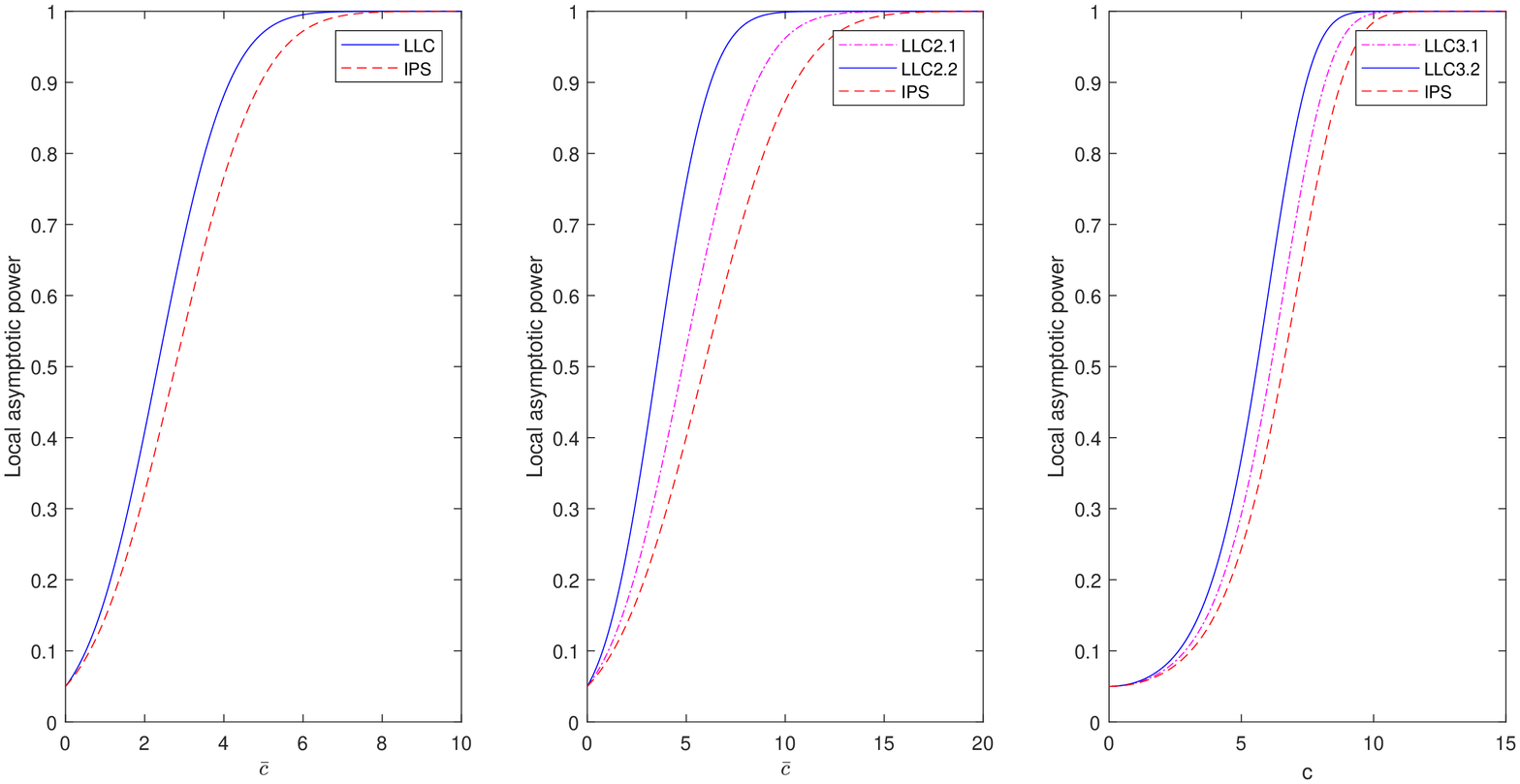}
\caption{Local asymptotic power of LLC and IPS}
\label{fig:1}
\end{figure}

 Firstly, the theoretical values of the local asymptotic powers are evaluated in Table \ref{tab1}
 based on Corollary \ref{coroll}. Clearly, we can see IPS tests have lower power than
 LLC tests. For $\hat t_{\delta,1}$, $\hat t_{\delta,21}$, $\hat t_{\delta,22}$, $Z$ and $Z^\mu$,
 the local asymptotic powers are increasing as $\bar c$ increases. For $\hat t_{\delta,31}$,
 $\hat t_{\delta,32}$ and $Z^\tau$, the local asymptotic powers are increasing as $\overline{c^2}$
 increases. This can also be seen in Figure \ref{fig:1}, where the theoretical local asymptotic powers of
 LLC and IPS tests are drawn given the values of $\bar c$ and $c_i\equiv c$ for all $i$.
 
 \begin{center}
\begin{table}[!thbp]
\caption{\label{tab1}Theoretical local asymptotic powers of LLC and IPS tests at 5\% level}\centering
\begin{tabular}{c|c|c|c|c|c|c|c|c}
\hline \hline
 $c_i\sim iid$ & \multicolumn{5}{c|}{LLC} & \multicolumn{3}{c}{IPS}   \\
\hline
 & $\hat t_{\delta,1}$ & $\hat t_{\delta,21}$ & $\hat t_{\delta,22}$ & $\hat t_{\delta,31}$ & $\hat t_{\delta,32}$ & $Z$ & $Z^\mu$ & $Z^\tau$ \\
 \hline
 $U[0,1]$ &  0.0983 &  0.0703 &  0.0792 &  0.0515 &  0.0518 &  0.0888 &  0.0661 &  0.0513  \\
 $U[0,8]$ &  0.8817 &  0.3914 &  0.5924 &  0.2398 &  0.3012 &  0.7666 &  0.2982 &  0.2025   \\
 $\chi^2(1)$ &  0.1741 &  0.0963 &  0.1199&  0.0651 &  0.0685 &  0.1464 &  0.0859 &  0.0629  \\
 $\chi^2(6)$ &  0.9953 &  0.6587 &  0.8796 &  0.6794 &  0.8116 &  0.9722 &  0.5112 &  0.5723  \\
\hline\hline
\end{tabular}
\end{table}\end{center}

Secondly, the simulated local asymptotic powers are reported in Table \ref{tab2} for different models.
The values are consistent with the theoretical values reported in Table \ref{tab1}. The difference
between the theoretical values and simulated values is due to the well known finite sample bias of
the unit root tests, see for instance Phillips (2012) and Hansen (2014).

Finally, in Table \ref{tab2}, the local power is increasing as $N$ increases.
Also, for $Z$ and $Z^\mu$, the local power is increasing as $\bar c$ increases.
For $Z^\tau$, the local power is increasing as $\overline{c^2}$ increases.

\begin{center}
\begin{table}[!thbp]
\caption{\label{tab2}Local power of IPS tests $Z$ and $Z^\mu$ in the neighborhood of unity with order $N^{-\alpha}T^{-1}$}
\footnotesize
\centering
\begin{tabular}{c|c|ccc|ccc|ccc}
\hline \hline \
 &  & \multicolumn{3}{c|}{$Z$ ($\alpha=1/2$)} & \multicolumn{3}{c|}{$Z^\mu$ ($\alpha=1/2$)} & \multicolumn{3}{c}{$Z^\tau$ ($\alpha=1/4$)} \\
\hline
 N & $c_i\sim$ & T=50 & T=100 & T=250 & T=50 & T=100 & T=250 & T=50 & T=100 & T=250 \\
\hline
\multirow{4}{*}{25} & $U[0,1]$ &  0.0800 &  0.0785 &  0.0820 &  0.0545 &  0.0685 &  0.0745  &  0.0510 &  0.0605 &  0.0540 \\
& $U[0,8]$ &  0.5985 &  0.5540 &  0.5160 &  0.1560 &  0.1535 &  0.1845  &  0.1190 &  0.0850 &  0.1105 \\
& $\chi^2(1)$ &  0.1520 &  0.0875 &  0.1430 &  0.0645 &  0.0665 &  0.0720 &  0.0510 &  0.0625 &  0.0615 \\
& $\chi^2(6)$ &  0.8345 &  0.8075 &  0.8115 &  0.2685 &  0.2535 &  0.2465 &  0.2750 &  0.1440 &  0.2155 \\
\hline
\multirow{4}{*}{100}  & $U[0,1]$ &  0.0745 &  0.0865 &  0.0795 &  0.0615 &  0.0615 &  0.0565 &  0.0565 &  0.0485 &  0.0500 \\
& $U[0,8]$ &  0.6150 &  0.6090 &  0.6555 &  0.1810 &  0.1995 &  0.2120  &  0.1635 &  0.1230 &  0.1345 \\
& $\chi^2(1)$ &  0.1070 &  0.1260 &  0.1380 &  0.0835 &  0.0655 &  0.0705 &  0.0615 &  0.0550 &  0.0605 \\
& $\chi^2(6)$ &  0.9465 &  0.9035 &  0.9135 &  0.2780 &  0.3205 &  0.3500 &  0.2430 &  0.3135 &  0.2645 \\
\hline
\multirow{4}{*}{1000} & $U[0,1]$  &  0.0640 &  0.0715 &  0.0825 &  0.0490 &  0.0575 &  0.0730 &  0.0520 &  0.0440 &  0.0530 \\
& $U[0,8]$  &  0.6635 &  0.7365 &  0.7035 &  0.2330 &  0.2520 &  0.2390 &  0.1535 &  0.1645 &  0.1575 \\
& $\chi^2(1)$ &  0.1035 &  0.1285 &  0.1550 &  0.0695 &  0.0780 &  0.0820 &  0.0610 &  0.0530 &  0.0585 \\
& $\chi^2(6)$ &  0.9175 &  0.9610 &  0.9725 &  0.3675 &  0.4065 &  0.4040 &  0.3410 &  0.3710 &  0.3550 \\
\hline\hline
\end{tabular}
\end{table}\end{center}

 \section{Conclusion}\label{sec:6}

In this paper, we propose a unified approach to study the local asymptotic power
of panel unit root tests. We use two most widely used panel unit root tests to
illustrate our method, i.e. LLC and IPS tests. We demonstrate how to apply our
approach to achieve the exact local asymptotic power of LLC and IPS tests for a variety of
scenarios. Moreover, the Edgeworth expansion of these test statistics can also be achieved with
our approach. Our approach can also be extended to other panel unit root tests.

\newpage

 \section*{Appendix}

 \renewcommand{\theequation}{A.\arabic{equation}}
 \renewcommand{\thelemma}{A.\arabic{lemma}}
 \setcounter{equation}{0}
 \setcounter{lemma}{0}

{\bf \large A.1 Proofs of Theorem \ref{LLC} and Theorem \ref{IPS}}

\bigskip

We give the detailed proofs of Theorem \ref{LLC} and Theorem \ref{IPS} in the following.

\bigskip

\noindent {\bf Proof of Theorem 3.1(a):}
Our goal is to calculate $E\left(\hat t_{\delta,1}\right)=E\left(\frac{N^{-1/2}\sum_{i=1}^N \tilde U_{1i}}{\sqrt{N^{-1}\sum_{i=1}^N \tilde V_{1i}}}\right)$ under $H_1$. Substituting $\theta=iu/2$, $x=-v/u$ and $c=c_iN^{-1/2}$ into $\varphi_1(\theta;c,1,x)$ in Lemma \ref{lemma2},
 we have the joint m.g.f. for $(\tilde U_{1i},\tilde V_{1i})$ as
 \begin{equation}\label{A.1}
 \psi_{1,i}(u,v) = e^{-\frac{u}{2}}\left[e^{-c_iN^{-\frac{1}{2}}}\left[\cos\sqrt{2v-c_i^2N^{-1}}
 +(c_iN^{-1/2}-u)\frac{\sin\sqrt{2v-c_i^2N^{-1}}}{\sqrt{2v-c_i^2N^{-1}}}\right]\right]^{-1/2}.
 \end{equation}
 Hence, the joint m.g.f. for $(N^{-1/2}\sum_{i=1}^N \tilde U_{1i},N^{-1}\sum_{i=1}^N \tilde V_{1i})$ is
 \begin{eqnarray*}
 &&\phi_1(u,v) = \prod_{i=1}^N\psi_{1,i}\left(\frac{u}{\sqrt N},\frac{v}{N}\right) \nonumber \\
 &=& e^{-\frac{N\frac{u}{\sqrt N}}{2}}\left[e^{-\sum_{i=1}^N c_iN^{-\frac{1}{2}}}\prod_{i=1}^N\left[\cos\sqrt{\frac{2 v}{N}-c_i^2N^{-1}}
 +(c_iN^{-1/2}-\frac{u}{\sqrt N})\frac{\sin\sqrt{\frac{2 v}{N}-c_i^2N^{-1}}}{\sqrt{\frac{2 v}{N}-c_i^2N^{-1}}}\right]\right]^{-1/2}.
 \end{eqnarray*}
 Then, we have
 \begin{eqnarray}\label{A.2}
 && \left.\frac{\partial }{\partial u}\phi_1(u,-v)\right|_{u=0} \nonumber\\
 &=& -\frac{\sqrt N}{2}\left[e^{-\sum_{i=1}^N c_iN^{-\frac{1}{2}}}\prod_{i=1}^N\left[\cos\sqrt{-\frac{2 v}{N}-c_i^2N^{-1}}
 +c_iN^{-1/2}\frac{\sin\sqrt{-\frac{2 v}{N}-c_i^2N^{-1}}}{\sqrt{-\frac{2 v}{N}-c_i^2N^{-1}}}\right]\right]^{-1/2} \nonumber\\
 &&-\frac{1}{2}\left[e^{-\sum_{i=1}^N c_iN^{-\frac{1}{2}}}\prod_{i=1}^N\left[\cos\sqrt{-\frac{2 v}{N}-c_i^2N^{-1}}
 +c_iN^{-1/2}\frac{\sin\sqrt{-\frac{2 v}{N}-c_i^2N^{-1}}}{\sqrt{-\frac{2 v}{N}-c_i^2N^{-1}}}\right]\right]^{-1/2} \nonumber\\
 &&\times\left(\sum_{i=1}^N\frac{\left(-\frac{1}{\sqrt N}\frac{\sin\sqrt{-\frac{2 v}{N}-c_i^2N^{-1}}}{\sqrt{-\frac{2 v}{N}-c_i^2N^{-1}}}\right)}{\cos\sqrt{-\frac{2 v}{N}-c_i^2N^{-1}} +c_iN^{-1/2}\frac{\sin\sqrt{-\frac{2 v}{N}-c_i^2N^{-1}}}{\sqrt{-\frac{2 v}{N}-c_i^2N^{-1}}}}\right).
 \end{eqnarray}
 From Taylor expansion, we have
 \begin{eqnarray*}
 && \cos\sqrt{-\frac{2 v}{N}-c_i^2N^{-1}}
 +c_iN^{-1/2}\frac{\sin\sqrt{-\frac{2 v}{N}-c_i^2N^{-1}}}{\sqrt{-\frac{2 v}{N}-c_i^2N^{-1}}} \nonumber\\
 &=& 1+c_iN^{-1/2}+\frac{1}{2N}(2v+c_i^2)+\frac{c_iN^{-3/2}}{6}(2v+c_i^2)+\frac{1}{24N^2}(2v+c_i^2)^2+O(N^{-5/2}).
 \end{eqnarray*}
 Thus,
 \begin{eqnarray}\label{A.3}
 &&\log\Bigg(1+c_iN^{-1/2}+\frac{1}{2N}(2v+c_i^2)+\frac{c_iN^{-3/2}}{6}(2v+c_i^2)+O(N^{-2})\Bigg) \nonumber\\
 &=& c_iN^{-1/2}+\frac{v}{N}+\frac{1}{2}c_i^2N^{-1}+\frac{1}{3}c_ivN^{-3/2}+\frac{1}{6}c_i^3N^{-3/2}
 -\frac{1}{2}\left[c_i^2N^{-1}+2c_ivN^{-3/2}+c_i^3N^{-3/2}\right] \nonumber\\
 &&+\frac{1}{3}c_i^3N^{-3/2}+O(N^{-2}) \nonumber\\
 &=& c_iN^{-1/2}+\frac{v}{N}-\frac{2}{3}c_ivN^{-3/2}+O(N^{-2}).
 \end{eqnarray}
 Further, combining (\ref{A.2}) and (\ref{A.3}), we have
 {\footnotesize
 \begin{eqnarray}\label{A.4}
 && \left.\frac{\partial }{\partial u}\phi_1(u,-v)\right|_{u=0} \nonumber\\
 &=& -\frac{\sqrt N}{2}\left[e^{-\sum_{i=1}^N c_iN^{-\frac{1}{2}}}e^{\sum_{i=1}^N\log\Big(1+c_iN^{-1/2}+\frac{1}{2N}(2v+c_i^2)+\frac{c_iN^{-3/2}}{6}(2v+c_i^2)+O(N^{-2})\Big)}\right]^{-1/2} \nonumber\\
 &&-\frac{1}{2}\Bigg[e^{-\sum_{i=1}^N c_iN^{-\frac{1}{2}}}e^{\sum_{i=1}^N\log\Big(1+c_iN^{-1/2}+\frac{1}{2N}(2v+c_i^2)+\frac{c_iN^{-3/2}}{6}(2v+c_i^2)+O(N^{-2})\Big)}\Bigg]^{-1/2} \nonumber\\
 &&\times\left(-\sqrt N+N^{-1}\sum_{i=1}^N c_i-N^{-1/2}\left(\frac{v}{3}+\frac{1}{6}N^{-1}\sum_{i=1}^N c_i^2\right)+\frac{v}{\sqrt N}+\frac{1}{2N^{3/2}}\sum_{i=1}^N c_i^2-\frac{1}{N^{3/2}}\sum_{i=1}^N c_i^2+O(N^{-1})\right) \nonumber\\
 &=&  -\frac{\sqrt N}{2}\Bigg[e^{-\sum_{i=1}^N c_iN^{-\frac{1}{2}}}e^{\sum_{i=1}^Nc_iN^{-1/2}+v-\frac{2v}{3}N^{-1/2}\sum_{i=1}^N c_i+O(N^{-1})}\Bigg]^{-1/2} \nonumber\\
 &&-\frac{1}{2}\Bigg[e^{-\sum_{i=1}^N c_iN^{-\frac{1}{2}}}e^{\sum_{i=1}^Nc_iN^{-1/2}+v-\frac{2v}{3}N^{-1/2}\sum_{i=1}^N c_i+O(N^{-1})}\Bigg]^{-1/2} \nonumber\\
 &&\times\left(-\sqrt N+N^{-1}\sum_{i=1}^N c_i-N^{-1/2}\left(\frac{v}{3}+\frac{1}{6}N^{-1}\sum_{i=1}^N c_i^2\right)+\frac{v}{\sqrt N}+\frac{1}{2N^{3/2}}\sum_{i=1}^N c_i^2-\frac{1}{N^{3/2}}\sum_{i=1}^N c_i^2+O(N^{-1})\right) \nonumber\\
 &=&  -\frac{1}{2}\left[e^{v+O(N^{-1/2})}\right]^{-1/2}
 \left(N^{-1}\sum_{i=1}^N c_i+\frac{2v}{3\sqrt N}-\frac{2}{3N^{3/2}}\sum_{i=1}^N c_i^2+O(N^{-1})\right) \nonumber\\
 &=& -\frac{1}{2}\left(e^{-v/2}+O(N^{-1/2})\right)
 \left(N^{-1}\sum_{i=1}^N c_i+\frac{2v}{3\sqrt N}-\frac{2}{3N^{3/2}}\sum_{i=1}^N c_i^2+O(N^{-1})\right) \nonumber\\
 &=& -\frac{1}{2}e^{-v/2}N^{-1}\sum_{i=1}^N c_i+O(N^{-1/2}).
 \end{eqnarray}}

 From Sawa (1972), we have
 \begin{equation}\label{A.5}
 E\left(\frac{U^p}{V^q}\right)=\frac{1}{\Gamma(q)}\int_0^\infty v^{q-1} \left.\frac{\partial^p}{\partial u^p}\phi(u,-v)\right|_{u=0}dv.
 \end{equation}
 Hence, from (\ref{A.5}) and plugging in (\ref{A.4}), we get
 \begin{eqnarray*}
 &&E\left(\hat t_{\delta,1}\right)=E\left(\frac{N^{-1/2}\sum_{i=1}^N \tilde U_{1i}}{\sqrt{N^{-1}\sum_{i=1}^N \tilde V_{1i}}}\right) = \frac{1}{\Gamma(\frac{1}{2})}\int_0^\infty \frac{1}{\sqrt v} \left.\frac{\partial }{\partial u}\phi_1(u,-v)\right|_{u=0}dv \\
 &=& -\left(N^{-1}\sum_{i=1}^N c_i\right)\frac{1}{2\sqrt \pi}\int_0^\infty \frac{e^{-v/2}}{\sqrt v}dv+O_p(N^{-1/2})=-\frac{\bar c}{\sqrt 2}+O_p(N^{-1/2}).
 \end{eqnarray*}

\noindent {\bf Proof of Theorem 3.1(b):} Substituting $\theta=iu/2$, $x=-v/u$ and $c=c_iN^{-1/2}$
into $\varphi_2(\theta;c,1,x)$ in Lemma \ref{lemma2}, we have the joint m.g.f. for $(\tilde U_{2i},\tilde V_{2i})$ as
 {\footnotesize
 \begin{eqnarray}\label{A.6}
 \psi_{2,i}(u,v) &=& e^{-\frac{u}{2}}\Bigg[e^{-c_iN^{-\frac{1}{2}}}\Bigg[\frac{u^2+2v+c_i^2N^{-1}u-c_i^3N^{-3/2}}{2v-c_i^2N^{-1}}
 \frac{\sin\sqrt{2v-c_i^2N^{-1}}}{\sqrt{2v-c_i^2N^{-1}}}-c_i^2N^{-1}\frac{\cos\sqrt{2v-c_i^2N^{-1}}}{2v-c_i^2N^{-1}} \nonumber \\
 &&+(2u^2-4c_iN^{-1/2}v+2c_i^2N^{-1}u)\frac{\cos\sqrt{2v-c_i^2N^{-1}}-1}{(2v-c_i^2N^{-1})^2}\Bigg]\Bigg]^{-1/2}.
 \end{eqnarray}}
 Hence, the joint m.g.f. for $(N^{-1/2}\sum_{i=1}^N \tilde U_{2i},N^{-1}\sum_{i=1}^N \tilde V_{2i})$ is
 {\footnotesize
 \begin{eqnarray*}
 &&\phi_2(u,v) = \prod_{i=1}^N\psi_{2,i}\left(\frac{u}{\sqrt N},\frac{v}{N}\right) \nonumber \\
 &=& e^{-\frac{N\frac{u}{\sqrt N}}{2}}\Bigg[e^{-\sum_{i=1}^N c_iN^{-\frac{1}{2}}}\prod_{i=1}^N\Bigg[\frac{\frac{u^2}{N}+\frac{2v}{N}+c_i^2N^{-1}\frac{u}{\sqrt N}-c_i^3N^{-3/2}}{\frac{2v}{N}-c_i^2N^{-1}}
 \frac{\sin\sqrt{\frac{2v}{N}-c_i^2N^{-1}}}{\sqrt{\frac{2v}{N}-c_i^2N^{-1}}} \nonumber \\
 &&-c_i^2N^{-1}\frac{\cos\sqrt{\frac{2v}{N}-c_i^2N^{-1}}}{\frac{2v}{N}-c_i^2N^{-1}}
 +(\frac{2u^2}{N}-4c_iN^{-1/2}\frac{v}{N}+2c_i^2N^{-1}\frac{u}{\sqrt N})\frac{\cos\sqrt{\frac{2v}{N}-c_i^2N^{-1}}-1}{(\frac{2v}{N}-c_i^2N^{-1})^2}\Bigg]\Bigg]^{-1/2}. \nonumber
 \end{eqnarray*}}
 Then, we have
 {\footnotesize
 \begin{eqnarray*}
 && \left.\frac{\partial }{\partial u}\phi_2(u,-v)\right|_{u=0} \\
 &=& -\frac{\sqrt N}{2}\Bigg[e^{-\sum_{i=1}^N c_iN^{-\frac{1}{2}}}\prod_{i=1}^N\Bigg[\frac{-\frac{2v}{N}-c_i^3N^{-3/2}}{-\frac{2v}{N}-c_i^2N^{-1}}
 \frac{\sin\sqrt{-\frac{2v}{N}-c_i^2N^{-1}}}{\sqrt{-\frac{2v}{N}-c_i^2N^{-1}}}-c_i^2N^{-1}
 \frac{\cos\sqrt{-\frac{2v}{N}-c_i^2N^{-1}}}{-\frac{2v}{N}-c_i^2N^{-1}} \\
 &&+(4c_iN^{-1/2}\frac{v}{N})\frac{\cos\sqrt{-\frac{2v}{N}-c_i^2N^{-1}}-1}{(-\frac{2v}{N}-c_i^2N^{-1})^2}\Bigg]\Bigg]^{-1/2} \\
 &&-\frac{1}{2}\Bigg[e^{-\sum_{i=1}^N c_iN^{-\frac{1}{2}}}\prod_{i=1}^N\Bigg[\frac{-\frac{2v}{N}-c_i^3N^{-3/2}}{-\frac{2v}{N}-c_i^2N^{-1}}
 \frac{\sin\sqrt{-\frac{2v}{N}-c_i^2N^{-1}}}{\sqrt{-\frac{2v}{N}-c_i^2N^{-1}}}-c_i^2N^{-1}
 \frac{\cos\sqrt{-\frac{2v}{N}-c_i^2N^{-1}}}{-\frac{2v}{N}-c_i^2N^{-1}}
 \end{eqnarray*}
 \begin{eqnarray}\label{A.7}
 &&+(4c_iN^{-1/2}\frac{v}{N})\frac{\cos\sqrt{-\frac{2v}{N}-c_i^2N^{-1}}-1}{(-\frac{2v}{N}-c_i^2N^{-1})^2}\Bigg]\Bigg]^{-1/2} \nonumber\\
 &&\times\left(\sum_{i=1}^N\frac{\frac{c_i^2N^{-3/2}}{-\frac{2v}{N}-c_i^2N^{-1}}
 \frac{\sin\sqrt{-\frac{2v}{N}-c_i^2N^{-1}}}{\sqrt{-\frac{2v}{N}-c_i^2N^{-1}}}
 +(2c_i^2N^{-3/2})\frac{\cos\sqrt{-\frac{2v}{N}-c_i^2N^{-1}}-1}{(-\frac{2v}{N}-c_i^2N^{-1})^2}}
 {\frac{-\frac{2v}{N}-c_i^3N^{-3/2}}{-\frac{2v}{N}-c_i^2N^{-1}}
 \frac{\sin\sqrt{-\frac{2v}{N}-c_i^2N^{-1}}}{\sqrt{-\frac{2v}{N}-c_i^2N^{-1}}}
 -c_i^2N^{-1}\frac{\cos\sqrt{-\frac{2v}{N}-c_i^2N^{-1}}}{-\frac{2v}{N}-c_i^2N^{-1}} +(4c_iN^{-1/2}\frac{v}{N})\frac{\cos\sqrt{-\frac{2v}{N}-c_i^2N^{-1}}-1}{(-\frac{2v}{N}-c_i^2N^{-1})^2}}\right). \nonumber\\
 \end{eqnarray}}
 From Taylor expansion, we have
 {\footnotesize
 \begin{eqnarray*}
 && \frac{-\frac{2v}{N}-c_i^3N^{-3/2}}{-\frac{2v}{N}-c_i^2N^{-1}}
 \frac{\sin\sqrt{-\frac{2v}{N}-c_i^2N^{-1}}}{\sqrt{-\frac{2v}{N}-c_i^2N^{-1}}}-c_i^2N^{-1}
 \frac{\cos\sqrt{-\frac{2v}{N}-c_i^2N^{-1}}}{-\frac{2v}{N}-c_i^2N^{-1}} +(4c_iN^{-1/2}\frac{v}{N})\frac{\cos\sqrt{-\frac{2v}{N}-c_i^2N^{-1}}-1}{(-\frac{2v}{N}-c_i^2N^{-1})^2} \\
 &=& 1+c_iN^{-1/2}+\frac{1}{6N}(2v+3c_i^2)+\frac{c_iN^{-3/2}}{6}(v+c_i^2)+O(N^{-2}).
 \end{eqnarray*}}
 Thus,
 \begin{eqnarray}\label{A.8}
 &&\log\Bigg(1+c_iN^{-1/2}+\frac{1}{6N}(2v+3c_i^2)+\frac{c_iN^{-3/2}}{6}(v+c_i^2)+O(N^{-2})\Bigg) \nonumber\\
 &=& c_iN^{-1/2}+\frac{v}{3N}+\frac{1}{2}c_i^2N^{-1}+\frac{1}{6}c_ivN^{-3/2}+\frac{1}{6}c_i^3N^{-3/2}
 -\frac{1}{2}\left[c_i^2N^{-1}+\frac{2c_iv}{3}N^{-3/2}+c_i^3N^{-3/2}\right] \nonumber\\
 &&+\frac{1}{3}c_i^3N^{-3/2}+O(N^{-2}) \nonumber\\
 &=& c_iN^{-1/2}+\frac{v}{3N}-\frac{1}{6}c_ivN^{-3/2}+O(N^{-2}).
 \end{eqnarray}
 Further, combining (\ref{A.7}) and (\ref{A.8}), we have
 \begin{eqnarray*}
 \left.\frac{\partial }{\partial u}\phi_2(u,-v)\right|_{u=0}
 &=& -\frac{\sqrt N}{2}\left[e^{-\sum_{i=1}^N c_iN^{-\frac{1}{2}}}e^{\sum_{i=1}^N\log\Big(1+c_iN^{-1/2}+\frac{1}{6N}(2v+3c_i^2)+\frac{c_iN^{-3/2}}{6}(v+c_i^2)+O(N^{-2})\Big)}\right]^{-1/2} \\
 &&-\frac{1}{2}\Bigg[e^{-\sum_{i=1}^N c_iN^{-\frac{1}{2}}}e^{\sum_{i=1}^N\log\Big(1+c_iN^{-1/2}+\frac{1}{6N}(2v+3c_i^2)+\frac{c_iN^{-3/2}}{6}(v+c_i^2)+O(N^{-2})\Big)}\Bigg]^{-1/2} \\
 &&\times\left(-\frac{1}{12N^{3/2}}\sum_{i=1}^N c_i^2+\frac{1}{12N^{2}}\sum_{i=1}^N c_i^3+O(N^{-5/2})\right) \\
 &=&  -\frac{\sqrt N}{2}\Bigg[e^{-\sum_{i=1}^N c_iN^{-\frac{1}{2}}}e^{\sum_{i=1}^N c_iN^{-1/2}+\frac{v}{3}-\frac{v}{6}N^{-3/2}\sum_{i=1}^N c_i+O(N^{-1})}\Bigg]^{-1/2} \\
 &&-\frac{1}{2}\Bigg[e^{-\sum_{i=1}^N c_iN^{-\frac{1}{2}}}e^{\sum_{i=1}^N c_iN^{-1/2}+\frac{v}{3}-\frac{v}{6}N^{-3/2}\sum_{i=1}^N c_i+O(N^{-1})}\Bigg]^{-1/2} \\
  &&\times\left(-\frac{1}{12N^{3/2}}\sum_{i=1}^N c_i^2+\frac{1}{12N^{2}}\sum_{i=1}^N c_i^3+O(N^{-5/2})\right)
 \end{eqnarray*}
 \begin{eqnarray}\label{A.9}
 &=&   -\frac{\sqrt N}{2}\left[e^{v/3-\frac{v}{6 N^{3/2}}\sum_{i=1}^N c_i+O(N^{-1})}\right]^{-1/2}-\frac{1}{2}\left[e^{v/3-\frac{v}{6 N^{3/2}}\sum_{i=1}^N c_i+O(N^{-1})}\right]^{-1/2} \nonumber\\
 &&\times\left(-\frac{1}{12N^{3/2}}\sum_{i=1}^N c_i^2+\frac{1}{12N^{2}}\sum_{i=1}^N c_i^3+O(N^{-5/2})\right) \nonumber\\
 &=& -\frac{\sqrt N}{2}\left(e^{-v/6}+\frac{v e^{-v/6}}{12 N^{3/2}}\sum_{i=1}^N c_i+O(N^{-1})\right)-\frac{1}{2}\left(e^{-v/6}+O(N^{-1/2})\right)O_p(N^{-1/2}) \nonumber\\
 &=& -\frac{\sqrt N}{2}e^{-v/6}-\frac{1}{24}ve^{-v/6}N^{-1}\sum_{i=1}^N c_i+O_p(N^{-1/2}).
 \end{eqnarray}

 Hence, from (\ref{A.5}) and plugging in (\ref{A.9}), we have
 \begin{eqnarray*}
 &&E\left(\frac{\sqrt 5}{2}\frac{N^{-1/2}\sum_{i=1}^N \tilde U_{2i}}{\sqrt{N^{-1}\sum_{i=1}^N \tilde V_{2i}}}\right)
 =\frac{\sqrt 5}{2}\frac{1}{\Gamma(\frac{1}{2})}\int_0^\infty \frac{1}{\sqrt v} \left.\frac{\partial }{\partial u}\phi_2(u,-v)\right|_{u=0}dv \\
 &=& -\frac{\sqrt 5}{2}\frac{\sqrt N}{2}\frac{1}{\sqrt \pi}\int_0^\infty \frac{e^{-v/6}}{\sqrt v}dv-\left(N^{-1}\sum_{i=1}^N c_i\right)\frac{\sqrt 5}{2}\frac{1}{24\sqrt \pi}\int_0^\infty \frac{ve^{-v/6}}{\sqrt v}dv+O_p(N^{-1/2}) \\
 &=& -\sqrt{\frac{15N}{8}}-\frac{1}{8}\sqrt{\frac{15}{2}}\bar c+O_p(N^{-1/2}).
 \end{eqnarray*}
 Therefore,
 \[ E\left(\hat t_{\delta,21}\right)=E\left(\frac{\sqrt 5}{2}\frac{N^{-1/2}\sum_{i=1}^N \tilde U_{2i}}{\sqrt{N^{-1}\sum_{i=1}^N \tilde V_{2i}}}\right)+\sqrt{\frac{15N}{8}}=-\frac{1}{8}\sqrt{\frac{15}{2}}\bar c+O_p(N^{-1/2}).\]

 Next, we consider $\hat t_{\delta,22}$.
 We have
 \begin{eqnarray}\label{A.10}
 && \frac{\partial}{\partial v}\left[\phi_2(0,-v)\right] \nonumber\\
 &=& \frac{\partial}{\partial v}\left[e^{-\sum_{i=1}^N c_iN^{-\frac{1}{2}}}e^{\sum_{i=1}^N\log\Big(1+c_iN^{-1/2}+\frac{1}{6N}(2v+3c_i^2)+\frac{c_iN^{-3/2}}{6}(v+c_i^2)+O(N^{-2})\Big)}\right]^{-1/2} \nonumber\\
 &=& \frac{\partial}{\partial v}\Bigg[e^{-\sum_{i=1}^N c_iN^{-\frac{1}{2}}}e^{\sum_{i=1}^N\left(c_iN^{-1/2}+\frac{1}{6N}(2v+3c_i^2)-\frac{1}{2N} c_i^2+\frac{1}{6 N^{3/2}} c_i(v+c_i^2)-\frac{1}{6 N^{3/2}} c_i(2v+3c_i^2)+\frac{1}{3N^{3/2}}c_i^3\right)+O(N^{-1})}\Bigg]^{-1/2} \nonumber\\
 &=&   \frac{\partial}{\partial v}\left[e^{v/3-\frac{v}{6 N^{3/2}}\sum_{i=1}^N c_i+O(N^{-1})}\right]^{-1/2}
 = -\frac{1}{6}e^{-v/6}+\frac{e^{-v/6}-\frac{1}{6}v e^{-v/6}}{12 N^{3/2}}\sum_{i=1}^N c_i+O(N^{-1}).
 \end{eqnarray}
 It can be shown
 \begin{equation}\label{A.11}
 E\left(V^{n-\alpha}\right)=\frac{(-1)^n}{\Gamma(\alpha)}\int_0^\infty v^{\alpha-1} \frac{\partial^n}{\partial v^n}\left[\phi(0,-v)\right]dv.
 \end{equation}
 Hence, from (\ref{A.5}) and (\ref{A.11}), and plugging in (\ref{A.9}) and (\ref{A.10}), we receive
 {\small
 \begin{eqnarray*}
 &&E\left(\sqrt{\frac{10}{17}}\left[\frac{N^{-1/2}\sum_{i=1}^N \tilde U_{2i}}{\sqrt{N^{-1}\sum_{i=1}^N \tilde V_{2i}}}+3\sqrt N\sqrt{N^{-1}\sum_{i=1}^N \tilde V_{2i}}\right]\right) \\
 &=& \sqrt{\frac{10}{17}}\frac{1}{\Gamma(\frac{1}{2})}\int_0^\infty \frac{1}{\sqrt v} \left.\frac{\partial }{\partial u}\phi_2(u,-v)\right|_{u=0}dv-\sqrt{\frac{10}{17}}\frac{3\sqrt N}{\Gamma(\frac{1}{2})}\int_0^\infty \frac{1}{\sqrt v} \frac{\partial}{\partial v}\left[\phi_2(0,-v)\right]dv \\
 &=& -\sqrt{\frac{10}{17}}\frac{\sqrt N}{2}\frac{1}{\sqrt \pi}\int_0^\infty \frac{e^{-v/6}}{\sqrt v}dv-\left(N^{-1}\sum_{i=1}^N c_i\right)\sqrt{\frac{10}{17}}\frac{1}{24\sqrt \pi}\int_0^\infty \frac{ve^{-v/6}}{\sqrt v}dv \\
 &&+\sqrt{\frac{10}{17}}\frac{\sqrt N}{2\sqrt \pi}\int_0^\infty \frac{e^{-v/6}}{\sqrt v}dv-\left(N^{-1}\sum_{i=1}^N c_i\right)\sqrt{\frac{10}{17}}\frac{1}{4\sqrt \pi}\left(\int_0^\infty \frac{e^{-v/6}}{\sqrt v}dv-\frac{1}{6}\int_0^\infty \frac{ve^{-v/6}}{\sqrt v}dv\right)+O_p(N^{-1/2}) \\
 &=& -\sqrt{\frac{30N}{34}}-\frac{1}{4}\sqrt{\frac{15}{17}}\bar c+\sqrt{\frac{30N}{34}}-\frac{1}{4}\sqrt{\frac{15}{17}}\bar c+O_p(N^{-1/2})
 =-\frac{1}{2}\sqrt{\frac{15}{17}}\bar c+O_p(N^{-1/2}).
 \end{eqnarray*}}
 Therefore,
 \[ E\left(\hat t_{\delta,22}\right)=E\left(\sqrt{\frac{10}{17}}\left[\frac{N^{-1/2}\sum_{i=1}^N \tilde U_{2i}}{\sqrt{N^{-1}\sum_{i=1}^N \tilde V_{2i}}}+3\sqrt N\sqrt{N^{-1}\sum_{i=1}^N \tilde V_{2i}}\right]\right)=-\frac{1}{2}\sqrt{\frac{15}{17}}\bar c+O_p(N^{-1/2}).\]

\noindent {\bf Proof of Theorem 3.1(c):}
Following the discussion in Section \ref{sec:3.1}, we have
\begin{equation}\label{A.12}
\hat t_{\delta,31}
= \sqrt{\frac{448}{277}}\frac{\hat \delta_3}{(\sum_{i=1}^N\sum_{t=2}^T [(y_{i,t-1}-\bar y_{i,t-1})-\frac{\sum_{s=1}^T (s-\bar s)(y_{is}-\bar y_{is})}{\sum_{s=1}^T (s-\bar s)^2}(t-\bar t)]^2/\hat \sigma_{\varepsilon3,i}^2)^{-1/2}}+\sqrt{\frac{1680N}{277}}.
\end{equation}
Following the corresponding specification of $H_1$ in Assumption 3, we have
{\footnotesize
\begin{eqnarray*}
\hat t_{\delta,31}
&=& -\sqrt{\frac{448}{277}}\frac{1}{N^{1/4}T}\frac{\sum_{i=1}^N c_i\left[\left(\sum_t
 (y_{i,t-1}-\bar y_{i,t-1})^2\right)-\frac{\left(\sum_t(t-\bar
 t)(y_{i,t-1}-\bar y_{i,t-1})\right)^2}{\sum_t (t-\bar t)^2}\right]/\hat \sigma_{\varepsilon3,i}^2}{\sqrt{\sum_{i=1}^N\left[\left(\sum_t
 (y_{i,t-1}-\bar y_{i,t-1})^2\right)-\frac{\left(\sum_t(t-\bar
 t)(y_{i,t-1}-\bar y_{i,t-1})\right)^2}{\sum_t (t-\bar t)^2}\right]/\hat \sigma_{\varepsilon3,i}^2}} \\
&&+\sqrt{\frac{448}{277}}\left(N^{-1}\sum_{i=1}^N T^{-2}\left[\left(\sum_t(y_{i,t-1}-\bar y_{i,t-1})^2\right)-\frac{\left(\sum_t(t-\bar
 t)(y_{i,t-1}-\bar y_{i,t-1})\right)^2}{\sum_t (t-\bar t)^2}\right]/\hat \sigma_{\varepsilon3,i}^2\right)^{-1/2} \\
 &&\times\Bigg(N^{-1/2}\sum_{i=1}^N \Bigg[T^{-1}\Bigg(\left(\sum_t
 (y_{i,t-1}-\bar y_{i,t-1})(\varepsilon_{it}-\bar \varepsilon_{it})\right)-\left(\sum_t (t-\bar t)^2\right)^{-1}\left(\sum_t(t-\bar
 t)(y_{i,t-1}-\bar y_{i,t-1})\right) \\
 &&\times\left(\sum_t(t-\bar t)(\varepsilon_{it}-\bar \varepsilon_{it})\right)\Bigg)/\hat \sigma_{\varepsilon3,i}^2+\frac{1}{2}\Bigg]\Bigg) \\
&&-\sqrt{\frac{448}{277}}\left(\frac{\sqrt N}{2\sqrt{N^{-1}T^{-2}\sum_{i=1}^N\left[\left(\sum_t
 (y_{i,t-1}-\bar y_{i,t-1})^2\right)-\frac{\left(\sum_t(t-\bar
 t)(y_{i,t-1}-\bar y_{i,t-1})\right)^2}{\sum_t (t-\bar t)^2}\right]/\hat \sigma_{\varepsilon3,i}^2}}-\sqrt{\frac{15N}{4}}\right).
\end{eqnarray*}}
Moreover, as $T\to \infty$, we have
\begin{eqnarray}\label{A.13}
\hat t_{\delta,31}
&\Rightarrow& -\sqrt{\frac{448}{277}}\frac{N^{-3/4}\sum_{i=1}^N c_i\left[\int_0^1
 K_{i,c_i}^\mu(r)^2dr-12\left(\int_0^1
 (r-\frac{1}{2})K_{i,c_i}^\mu(r)dr\right)^2\right]}{\sqrt{N^{-1}\sum_{i=1}^N\left[\int_0^1
 K_{i,c_i}^\mu(r)^2dr-12\left(\int_0^1
 (r-\frac{1}{2})K_{i,c_i}^\mu(r)dr\right)^2\right]}} \nonumber \\
 &&+\sqrt{\frac{448}{277}}\frac{N^{-1/2}\sum_{i=1}^N \left(\int_0^1
 K_{i,c_i}^\mu(r)dW_i(r)-12\int_0^1
 (r-\frac{1}{2})K_{i,c_i}^\mu(r)dr\int_0^1 (r-\frac{1}{2})dW_i(r)+\frac{1}{2}\right)}{\sqrt{N^{-1}\sum_{i=1}^N\left[\int_0^1
 K_{i,c_i}^\mu(r)^2dr-12\left(\int_0^1
 (r-\frac{1}{2})K_{i,c_i}^\mu(r)dr\right)^2\right]}} \nonumber \\
 &&-\sqrt{\frac{448}{277}}\left(\frac{\sqrt N}{2\sqrt{N^{-1}\sum_{i=1}^N \left[\int_0^1
 K_{i,c_i}^\mu(r)^2dr-12\left(\int_0^1
 (r-\frac{1}{2})K_{i,c_i}^\mu(r)dr\right)^2\right]}}-\sqrt{\frac{15N}{4}}\right) \nonumber \\
 &\stackrel{def}{=}& \sqrt{\frac{448}{277}}\frac{N^{-1/2}\sum_{i=1}^N \tilde U_{3i}}{\sqrt{N^{-1}\sum_{i=1}^N \tilde V_{3i}}}+\sqrt{\frac{448}{277}}\sqrt{\frac{15N}{4}}.
\end{eqnarray}

Substituting $\theta=iu/2$, $x=-v/u$ and $c=c_iN^{-1/4}$ into $\varphi_4(\theta;c,1,x)$ in Lemma \ref{lemma2},
 we have the joint m.g.f. for $(\tilde U_{3i},\tilde V_{3i})$ as
 {\scriptsize
 \begin{eqnarray}\label{A.14}
 && \psi_{3,i}(u,v) \nonumber \\
 &=& e^{-\frac{u}{2}}\Bigg[e^{-c_iN^{-\frac{1}{4}}}\Big[\frac{(c_iN^{-1/4})^5-(c_iN^{-1/4})^4u-4((c_iN^{-1/4})^2+3c_iN^{-1/4}+27)u^2
 -8v((c_iN^{-1/4})^2-3c_iN^{-1/4}-3)}{(2v-c_i^2N^{-1/2})^2} \nonumber \\
 &&\times\frac{\sin\sqrt{2v-c_i^2N^{-1/2}}}{\sqrt{2v-c_i^2N^{-1/2}}}+\frac{24((c_iN^{-1/4})^4u+8vu^2-4(c_iN^{-1/4}+1)
 (v^2-3 u^2))}{(2v-c_i^2N^{-1/2})^3}\Bigg(\frac{\sin\sqrt{2v-c_i^2N^{-1/2}}}{\sqrt{2v-c_i^2N^{-1/2}}} \nonumber \\
 &&
 +\frac{\cos\sqrt{2v-c_i^2N^{-1/2}}}{2v-c_i^2N^{-1/2}}-\frac{1}{2v-c_i^2N^{-1/2}}\Bigg)+\Bigg(\frac{c_i^4N^{-1}}{(2v-c_i^2N^{-1/2})^2}-
 \frac{8(c_i^4N^{-1}u-c_i^3N^{-3/4}2v+4(c_i^2N^{-1/2}+3c_iN^{-1/4}+6)u^2)}{(2v-c_i^2N^{-1/2})^3}\Bigg) \nonumber \\
 &&\times\cos\sqrt{2v-c_i^2N^{-1/2}}-\frac{4(c_i^4N^{-1}u+4(c_i^2N^{-1/2}+3c_iN^{-1/4}-3)u^2-2c_i^2N^{-1/2}v(c_iN^{-1/4}+3))}
 {(2v-c_i^2N^{-1/2})^3}\Big]\Bigg]^{-1/2}.
 \end{eqnarray}}
 Hence, the joint m.g.f. for $(N^{-1/2}\sum_{i=1}^N \tilde U_{3i},N^{-1}\sum_{i=1}^N \tilde V_{3i})$ is
 {\tiny
 \begin{eqnarray*}
 &&\phi_3(u,v) = \prod_{i=1}^N\psi_{3,i}\left(\frac{u}{\sqrt N},\frac{v}{N}\right) \nonumber \\
 &=& e^{-\frac{N\frac{u}{\sqrt N}}{2}}\Bigg[e^{-\sum_{i=1}^N c_iN^{-\frac{1}{4}}}\prod_{i=1}^N\Bigg[\frac{(c_iN^{-1/4})^5-(c_iN^{-1/4})^4\frac{u}{\sqrt N}-4((c_iN^{-1/4})^2+3c_iN^{-1/4}+27)\frac{u^2}{N}
 -\frac{8v}{N}((c_iN^{-1/4})^2-3c_iN^{-1/4}-3)}{(\frac{2v}{N}-c_i^2N^{-1/2})^2} \nonumber \\
 &&\times\frac{\sin\sqrt{\frac{2v}{N}-c_i^2N^{-1/2}}}{\sqrt{\frac{2v}{N}-c_i^2N^{-1/2}}}+\frac{24((c_iN^{-1/4})^4\frac{u}{\sqrt N}+8\frac{vu^2}{N^2}-4(c_iN^{-1/4}+1)
 (\frac{v^2}{N^2}-3 \frac{u^2}{N}))}{(\frac{2v}{N}-c_i^2N^{-1/2})^3}\Bigg(\frac{\sin\sqrt{\frac{2v}{N}-c_i^2N^{-1/2}}}{\sqrt{\frac{2v}{N}-c_i^2N^{-1/2}}} \nonumber \\
 &&
 +\frac{\cos\sqrt{\frac{2v}{N}-c_i^2N^{-1/2}}}{\frac{2v}{N}-c_i^2N^{-1/2}}-\frac{1}{\frac{2v}{N}-c_i^2N^{-1/2}}\Bigg)
 +\Bigg(\frac{c_i^4N^{-1}}{(\frac{2v}{N}-c_i^2N^{-1/2})^2}-
 \frac{8(c_i^4N^{-1}\frac{u}{\sqrt N}-c_i^3N^{-3/4}\frac{2v}{N}+4(c_i^2N^{-1/2}+3c_iN^{-1/4}+6)\frac{u^2}{N})}{(\frac{2v}{N}-c_i^2N^{-1/2})^3}\Bigg) \nonumber \\
 &&\times\cos\sqrt{\frac{2v}{N}-c_i^2N^{-1/2}}-\frac{4(c_i^4N^{-1}\frac{u}{\sqrt N}+4(c_i^2N^{-1/2}+3c_iN^{-1/4}-3)\frac{u^2}{N}-2c_i^2N^{-1/2}\frac{v}{N}(c_iN^{-1/4}+3))}
 {(\frac{2v}{N}-c_i^2N^{-1/2})^3}\Bigg]\Bigg]^{-1/2}.
 \end{eqnarray*}}
 Then, we have
 {\footnotesize
 \begin{eqnarray}\label{A.15}
 && \left.\frac{\partial }{\partial u}\phi_3(u,-v)\right|_{u=0} \nonumber\\
 &=& -\frac{\sqrt N}{2}\Bigg[e^{-\sum_{i=1}^N c_iN^{-\frac{1}{4}}}\prod_{i=1}^N\Bigg[\frac{(c_iN^{-1/4})^5
 +\frac{8v}{N}((c_iN^{-1/4})^2-3c_iN^{-1/4}-3)}{(-\frac{2v}{N}-c_i^2N^{-1/2})^2}
 \frac{\sin\sqrt{-\frac{2v}{N}-c_i^2N^{-1/2}}}{\sqrt{-\frac{2v}{N}-c_i^2N^{-1/2}}} \nonumber\\
 &&+\frac{24(-4(c_iN^{-1/4}+1)\frac{v^2}{N^2})}{(-\frac{2v}{N}-c_i^2N^{-1/2})^3}
 \Bigg(\frac{\sin\sqrt{-\frac{2v}{N}-c_i^2N^{-1/2}}}{\sqrt{-\frac{2v}{N}-c_i^2N^{-1/2}}}
 +\frac{\cos\sqrt{-\frac{2v}{N}-c_i^2N^{-1/2}}}{-\frac{2v}{N}-c_i^2N^{-1/2}}-\frac{1}{-\frac{2v}{N}-c_i^2N^{-1/2}}\Bigg) \nonumber\\
 && +\Bigg(\frac{c_i^4N^{-1}}{(-\frac{2v}{N}-c_i^2N^{-1/2})^2}-
 \frac{8(c_i^3N^{-3/4}\frac{2v}{N})}{(-\frac{2v}{N}-c_i^2N^{-1/2})^3}\Bigg)\cos\sqrt{-\frac{2v}{N}-c_i^2N^{-1/2}}
 -\frac{4(2c_i^2N^{-1/2}\frac{v}{N}(c_iN^{-1/4}+3))}
 {(-\frac{2v}{N}-c_i^2N^{-1/2})^3}\Bigg]\Bigg]^{-1/2} \nonumber\\
 &&-\frac{1}{2}\Bigg[e^{-\sum_{i=1}^N c_iN^{-\frac{1}{4}}}\prod_{i=1}^N\Bigg[\frac{(c_iN^{-1/4})^5
 +\frac{8v}{N}((c_iN^{-1/4})^2-3c_iN^{-1/4}-3)}{(-\frac{2v}{N}-c_i^2N^{-1/2})^2}
 \frac{\sin\sqrt{-\frac{2v}{N}-c_i^2N^{-1/2}}}{\sqrt{-\frac{2v}{N}-c_i^2N^{-1/2}}} \nonumber\\
 &&+\frac{24(-4(c_iN^{-1/4}+1)\frac{v^2}{N^2})}{(-\frac{2v}{N}-c_i^2N^{-1/2})^3}
 \Bigg(\frac{\sin\sqrt{-\frac{2v}{N}-c_i^2N^{-1/2}}}{\sqrt{-\frac{2v}{N}-c_i^2N^{-1/2}}}
 +\frac{\cos\sqrt{-\frac{2v}{N}-c_i^2N^{-1/2}}}{-\frac{2v}{N}-c_i^2N^{-1/2}}-\frac{1}{-\frac{2v}{N}-c_i^2N^{-1/2}}\Bigg) \nonumber\\
 &&+\Bigg(\frac{c_i^4N^{-1}}{(-\frac{2v}{N}-c_i^2N^{-1/2})^2}-
 \frac{8(c_i^3N^{-3/4}\frac{2v}{N})}{(-\frac{2v}{N}-c_i^2N^{-1/2})^3}\Bigg)\cos\sqrt{-\frac{2v}{N}-c_i^2N^{-1/2}}
 -\frac{4(2c_i^2N^{-1/2}\frac{v}{N}(c_iN^{-1/4}+3))}
 {(-\frac{2v}{N}-c_i^2N^{-1/2})^3}\Bigg]\Bigg]^{-1/2} \nonumber\\
 &&\times\Bigg(\sum_{i=1}^N\Bigg[\frac{(c_iN^{-1/4})^5
 +\frac{8v}{N}((c_iN^{-1/4})^2-3c_iN^{-1/4}-3)}{(-\frac{2v}{N}-c_i^2N^{-1/2})^2}
 \frac{\sin\sqrt{-\frac{2v}{N}-c_i^2N^{-1/2}}}{\sqrt{-\frac{2v}{N}-c_i^2N^{-1/2}}} \nonumber\\
 &&+\frac{24(-4(c_iN^{-1/4}+1)\frac{v^2}{N^2})}{(-\frac{2v}{N}-c_i^2N^{-1/2})^3}
 \Bigg(\frac{\sin\sqrt{-\frac{2v}{N}-c_i^2N^{-1/2}}}{\sqrt{-\frac{2v}{N}-c_i^2N^{-1/2}}}
 +\frac{\cos\sqrt{-\frac{2v}{N}-c_i^2N^{-1/2}}}{-\frac{2v}{N}-c_i^2N^{-1/2}}-\frac{1}{-\frac{2v}{N}-c_i^2N^{-1/2}}\Bigg) \nonumber\\
 && +\Bigg(\frac{c_i^4N^{-1}}{(-\frac{2v}{N}-c_i^2N^{-1/2})^2}-
 \frac{8(c_i^3N^{-3/4}\frac{2v}{N})}{(-\frac{2v}{N}-c_i^2N^{-1/2})^3}\Bigg)\cos\sqrt{-\frac{2v}{N}-c_i^2N^{-1/2}}
 -\frac{4(2c_i^2N^{-1/2}\frac{v}{N}(c_iN^{-1/4}+3))}
 {(-\frac{2v}{N}-c_i^2N^{-1/2})^3}\Bigg]^{-1} \nonumber\\
 &&\times\Bigg(\frac{-(c_iN^{-1/4})^4\frac{1}{\sqrt N}}{(-\frac{2v}{N}-c_i^2N^{-1/2})^2}
 \frac{\sin\sqrt{-\frac{2v}{N}-c_i^2N^{-1/2}}}{\sqrt{-\frac{2v}{N}-c_i^2N^{-1/2}}}
 +\frac{24(c_iN^{-1/4})^4\frac{1}{\sqrt N}}{(-\frac{2v}{N}-c_i^2N^{-1/2})^3}\Bigg(\frac{\sin\sqrt{-\frac{2v}{N}-c_i^2N^{-1/2}}}{\sqrt{-\frac{2v}{N}-c_i^2N^{-1/2}}} \nonumber\\
 &&+\frac{\cos\sqrt{-\frac{2v}{N}-c_i^2N^{-1/2}}}{-\frac{2v}{N}-c_i^2N^{-1/2}}-\frac{1}{-\frac{2v}{N}-c_i^2N^{-1/2}}\Bigg)
 -\frac{8c_i^4N^{-1}\frac{1}{\sqrt N}}{(-\frac{2v}{N}-c_i^2N^{-1/2})^3}\cos\sqrt{-\frac{2v}{N}-c_i^2N^{-1/2}} \nonumber\\
 &&-\frac{4c_i^4N^{-1}\frac{1}{\sqrt N}}{(-\frac{2v}{N}-c_i^2N^{-1/2})^3}\Bigg)\Bigg).
 \end{eqnarray}}
 From Taylor expansion, we have
 {\small
 \begin{eqnarray*}
 &&\frac{(c_iN^{-1/4})^5
 +\frac{8v}{N}((c_iN^{-1/4})^2-3c_iN^{-1/4}-3)}{(-\frac{2v}{N}-c_i^2N^{-1/2})^2}
 \frac{\sin\sqrt{-\frac{2v}{N}-c_i^2N^{-1/2}}}{\sqrt{-\frac{2v}{N}-c_i^2N^{-1/2}}}
 \end{eqnarray*}
 \begin{eqnarray*}
 &&+\frac{24(-4(c_iN^{-1/4}+1)\frac{v^2}{N^2})}{(-\frac{2v}{N}-c_i^2N^{-1/2})^3}
 \Bigg(\frac{\sin\sqrt{-\frac{2v}{N}-c_i^2N^{-1/2}}}{\sqrt{-\frac{2v}{N}-c_i^2N^{-1/2}}}
 +\frac{\cos\sqrt{-\frac{2v}{N}-c_i^2N^{-1/2}}}{-\frac{2v}{N}-c_i^2N^{-1/2}}-\frac{1}{-\frac{2v}{N}-c_i^2N^{-1/2}}\Bigg) \\
 && +\Bigg(\frac{c_i^4N^{-1}}{(-\frac{2v}{N}-c_i^2N^{-1/2})^2}-
 \frac{8(c_i^3N^{-3/4}\frac{2v}{N})}{(-\frac{2v}{N}-c_i^2N^{-1/2})^3}\Bigg)\cos\sqrt{-\frac{2v}{N}-c_i^2N^{-1/2}}
 -\frac{4(2c_i^2N^{-1/2}\frac{v}{N}(c_iN^{-1/4}+3))}
 {(-\frac{2v}{N}-c_i^2N^{-1/2})^3} \\
 &=& 1+c_iN^{-1/4}+\frac{1}{2}c_i^2N^{-1/2}+\frac{1}{6}c_i^3N^{-3/4}+N^{-1}\left(\frac{2v}{15}+\frac{c_i^4}{24}\right)
 +N^{-5/4}\left(\frac{2c_iv}{15}+\frac{c_i^5}{120}\right) \\
 &&+N^{-3/2}\left(\frac{13c_i^2v}{210}+\frac{c_i^6}{720}\right)+O(N^{-7/4}).
 \end{eqnarray*}}
 Further, we have
 {\small
 \begin{eqnarray}\label{A.16}
 &&\log\Bigg(1+c_iN^{-1/4}+\frac{1}{2}c_i^2N^{-1/2}+\frac{1}{6}c_i^3N^{-3/4}+\frac{2v}{15N}+\frac{1}{24}c_i^4N^{-1}
 +\frac{2}{15}c_ivN^{-5/4}+\frac{1}{120}c_i^5N^{-5/4} \nonumber\\
 &&+\frac{13c_i^2 v }{210}N^{-3/2}+\frac{c_i^6}{720}N^{-3/2}+O(N^{-7/4})\Bigg) \nonumber\\
 &=& c_iN^{-1/4}+\frac{1}{2}c_i^2N^{-1/2}+\frac{1}{6}c_i^3N^{-3/4}+\frac{2v}{15N}+\frac{1}{24}c_i^4N^{-1}
 +\frac{2}{15}c_ivN^{-5/4}+\frac{1}{120}c_i^5N^{-5/4}+\frac{13c_i^2 v }{210}N^{-3/2} \nonumber\\
 &&+\frac{c_i^6}{720}N^{-3/2}-\frac{1}{2}\Bigg[c_i^2N^{-1/2}+c_i^3N^{-3/4}+\frac{1}{4}c_i^4N^{-1}+\frac{1}{3}c_i^4N^{-1}
 +\frac{1}{6}c_i^5N^{-5/4}+\frac{1}{36}c_i^6N^{-3/2}+\frac{4 c_iv}{15}N^{-5/4} \nonumber\\
 &&+\frac{1}{12}c_i^5N^{-5/4}+\frac{2c_i^2v}{5}N^{-3/2}+\frac{7c_i^6}{120}N^{-3/2}\Bigg]+\frac{1}{3}\Bigg[c_i^3N^{-3/4}
 +\frac{3}{2}c_i^4N^{-1}+\frac{5}{4}c_i^5N^{-5/4}+\frac{3}{4}c_i^6N^{-3/2} \nonumber\\
 &&+\frac{2}{5}c_i^2vN^{-3/2}\Bigg]-\frac{1}{4}\Bigg[c_i^4N^{-1}+2c_i^5N^{-5/4}+\frac{13}{6}c_i^6N^{-3/2}\Bigg]+\frac{1}{5}\Bigg[c_i^5N^{-5/4}
 +\frac{5}{2}c_i^6N^{-3/2}\Bigg]-\frac{1}{6}c_i^6N^{-3/2} \nonumber\\
 &&+O(N^{-7/4}) \nonumber\\
 &=& c_iN^{-1/4}+\frac{2v}{15N}-\frac{c_i^2v}{210}N^{-3/2}+O(N^{-7/4}),
 \end{eqnarray}}
 and combining (\ref{A.15}) and (\ref{A.16}), we have
 {\scriptsize
 \begin{eqnarray*}
 && \left.\frac{\partial }{\partial u}\phi_3(u,-v)\right|_{u=0} \nonumber\\
 &=& -\frac{\sqrt N}{2}\Bigg[e^{-\sum_{i=1}^N c_iN^{-\frac{1}{4}}}\exp\Big(\sum_{i=1}^N\log\Big(1+c_iN^{-1/4}+\frac{1}{2}c_i^2N^{-1/2}+\frac{1}{6}c_i^3N^{-3/4}
 +N^{-1}\left(\frac{2v}{15}+\frac{c_i^4}{24}\right)
 +N^{-5/4}\left(\frac{2c_iv}{15}+\frac{c_i^5}{120}\right) \nonumber\\
 &&+N^{-3/2}\left(\frac{13c_i^2v}{210}+\frac{c_i^6}{720}\right)+O(N^{-7/4})\Big)\Big)\Bigg]^{-1/2}-\frac{1}{2}\Bigg[e^{-\sum_{i=1}^N c_iN^{-\frac{1}{4}}}\exp\Big(\sum_{i=1}^N\log\Big(1+c_iN^{-1/4}+\frac{1}{2}c_i^2N^{-1/2}+\frac{1}{6}c_i^3N^{-3/4}
 \nonumber\\
 &&+N^{-1}\left(\frac{2v}{15}+\frac{c_i^4}{24}\right)
 +N^{-5/4}\left(\frac{2c_iv}{15}+\frac{c_i^5}{120}\right)+N^{-3/2}\left(\frac{13c_i^2v}{210}+\frac{c_i^6}{720}\right)+O(N^{-7/4})\Big)\Big)\Bigg]^{-1/2}
 \left(-\frac{1}{720N^{3/2}}\sum_{i=1}^N c_i^4+O(N^{-3/4})\right) \nonumber\\
 &=&  -\frac{\sqrt N}{2}\Bigg[e^{-\sum_{i=1}^N c_iN^{-\frac{1}{4}}}e^{\sum_{i=1}^Nc_iN^{-1/4}+\frac{2v}{15}
 -\sum_{i=1}^N\frac{c_i^2v}{210}N^{-3/2}+O(N^{-3/4})}\Bigg]^{-1/2} \nonumber\\
 &&-\frac{1}{2}\Bigg[e^{-\sum_{i=1}^N c_iN^{-\frac{1}{4}}}e^{\sum_{i=1}^Nc_iN^{-1/4}+\frac{2v}{15}
 -\sum_{i=1}^N\frac{c_i^2v}{210}N^{-3/2}+O(N^{-3/4})}\Bigg]^{-1/2} \nonumber
 \end{eqnarray*}}
 \begin{eqnarray}\label{A.17}
 &&\times\left(-\frac{1}{720N^{3/2}}\sum_{i=1}^N c_i^4+O(N^{-3/4})\right) \nonumber\\
 &=&   -\frac{\sqrt N}{2}\left[e^{\frac{2v}{15}-N^{-3/2}\sum_{i=1}^N \frac{c_i^2v}{210}+O(N^{-3/4})}\right]^{-1/2}-\frac{1}{2}\left[e^{\frac{2v}{15}-N^{-3/2}\sum_{i=1}^N\frac{c_i^2v}{210}+O(N^{-3/4})}\right]^{-1/2} \nonumber\\
 &&\times\left(-\frac{1}{720N^{3/2}}\sum_{i=1}^N c_i^4+O(N^{-3/4})\right) \nonumber\\
 &=& -\frac{\sqrt N}{2}\left(e^{-v/15}+\frac{v e^{-v/15}}{420 N^{3/2}}\sum_{i=1}^N c_i^2+O(N^{-3/4})\right)-\frac{1}{2}\left(e^{-v/15}+O(N^{-1/2})\right)O_p(N^{-1/2}) \nonumber\\
 &=& -\frac{\sqrt N}{2}e^{-v/15}-\frac{1}{840}ve^{-v/15}N^{-1}\sum_{i=1}^N c_i^2+O_p(N^{-1/4}).
 \end{eqnarray}

 Hence, applying (\ref{A.5}) and plugging in (\ref{A.17}), we have
 \begin{eqnarray*}
 &&E\left(\sqrt{\frac{448}{277}}\frac{N^{-1/2}\sum_{i=1}^N \tilde U_{3i}}{\sqrt{N^{-1}\sum_{i=1}^N \tilde V_{3i}}}\right)
 =\sqrt{\frac{448}{277}}\frac{1}{\Gamma(\frac{1}{2})}\int_0^\infty \frac{1}{\sqrt v} \left.\frac{\partial }{\partial u}\phi_3(u,-v)\right|_{u=0}dv \\
 &=& -\sqrt{\frac{448}{277}}\frac{\sqrt N}{2}\frac{1}{\sqrt \pi}\int_0^\infty \frac{e^{-v/15}}{\sqrt v}dv-\left(N^{-1}\sum_{i=1}^N c_i^2\right)\sqrt{\frac{448}{277}}\frac{1}{840\sqrt \pi}\int_0^\infty \frac{ve^{-v/15}}{\sqrt v}dv +O_p(N^{-1/4}) \\
 &=& -\sqrt{\frac{448}{277}}\sqrt{\frac{15N}{4}}-\frac{1}{14}\sqrt{\frac{105}{277}}c^2+O_p(N^{-1/4}).
 \end{eqnarray*}
 Therefore,
 \[ E\left(\hat t_{\delta,31}\right)=E\left(\sqrt{\frac{448}{277}}\frac{N^{-1/2}\sum_{i=1}^N \tilde U_{3i}}{\sqrt{N^{-1}\sum_{i=1}^N \tilde V_{3i}}}\right)+\sqrt{\frac{448}{277}}\sqrt{\frac{15N}{4}}=-\frac{1}{14}\sqrt{\frac{105}{277}}c^2
 +O_p(N^{-1/4}).\]

 Next, we consider $\hat t_{\delta,32}$. Recall that
 \begin{equation}\label{A.18}
\hat t_{\delta,32}
= \sqrt{\frac{112}{193}}\frac{\hat \delta_3+\frac{15}{2T}}{(\sum_{i=1}^N\sum_{t=2}^T [(y_{i,t-1}-\bar y_{i,t-1})-\frac{\sum_{s=1}^T (s-\bar s)(y_{is}-\bar y_{is})}{\sum_{s=1}^T (s-\bar s)^2}(t-\bar t)]^2/\hat \sigma_{\varepsilon3,i}^2)^{-1/2}}.
\end{equation}
Based on the specification of $H_1$ given in Assumption 3, we have
{\footnotesize
\begin{eqnarray*}
\hat t_{\delta,32}
&=& -\sqrt{\frac{112}{193}}\frac{1}{N^{1/4}T}\frac{\sum_{i=1}^N c_i\left[\left(\sum_t
 (y_{i,t-1}-\bar y_{i,t-1})^2\right)-\frac{\left(\sum_t(t-\bar
 t)(y_{i,t-1}-\bar y_{i,t-1})\right)^2}{\sum_t (t-\bar t)^2}\right]/\hat \sigma_{\varepsilon3,i}^2}{\sqrt{\sum_{i=1}^N\left[\left(\sum_t
 (y_{i,t-1}-\bar y_{i,t-1})^2\right)-\frac{\left(\sum_t(t-\bar
 t)(y_{i,t-1}-\bar y_{i,t-1})\right)^2}{\sum_t (t-\bar t)^2}\right]/\hat \sigma_{\varepsilon3,i}^2}}
 \end{eqnarray*}
 \begin{eqnarray*}
 &&+\sqrt{\frac{112}{193}}\left(N^{-1}\sum_{i=1}^N T^{-2}\left[\left(\sum_t(y_{i,t-1}-\bar y_{i,t-1})^2\right)-\frac{\left(\sum_t(t-\bar
 t)(y_{i,t-1}-\bar y_{i,t-1})\right)^2}{\sum_t (t-\bar t)^2}\right]/\hat \sigma_{\varepsilon3,i}^2\right)^{-1/2} \\
 &&\times\Bigg(N^{-1/2}\sum_{i=1}^N \Bigg[T^{-1}\Bigg(\left(\sum_t
 (y_{i,t-1}-\bar y_{i,t-1})(\varepsilon_{it}-\bar \varepsilon_{it})\right)-\left(\sum_t (t-\bar t)^2\right)^{-1}\left(\sum_t(t-\bar
 t)(y_{i,t-1}-\bar y_{i,t-1})\right) \\
 &&\times\left(\sum_t(t-\bar t)(\varepsilon_{it}-\bar \varepsilon_{it})\right)\Bigg)/\hat \sigma_{\varepsilon3,i}^2+\frac{1}{2}\Bigg]\Bigg) \\
 &&+\sqrt{\frac{112}{193}}\left(\frac{\frac{\sqrt N}{2}-\frac{15\sqrt N}{2} \left(N^{-1}T^{-2}\sum_{i=1}^N\left[\left(\sum_t
 (y_{i,t-1}-\bar y_{i,t-1})^2\right)-\frac{\left(\sum_t(t-\bar
 t)(y_{i,t-1}-\bar y_{i,t-1})\right)^2}{\sum_t (t-\bar t)^2}\right]/\hat \sigma_{\varepsilon3,i}^2\right)}{\sqrt{N^{-1}T^{-2}\sum_{i=1}^N\left[\left(\sum_t
 (y_{i,t-1}-\bar y_{i,t-1})^2\right)-\frac{\left(\sum_t(t-\bar
 t)(y_{i,t-1}-\bar y_{i,t-1})\right)^2}{\sum_t (t-\bar t)^2}\right]/\hat \sigma_{\varepsilon3,i}^2}}\right).
\end{eqnarray*}}
Hence, we have that as $T\to \infty$
{\small
\begin{eqnarray}\label{A.19}
\hat t_{\delta,32}
&\Rightarrow& -\sqrt{\frac{112}{193}}\frac{N^{-3/4}\sum_{i=1}^N c_i\left[\int_0^1
 K_{i,c_i}^\mu(r)^2dr-12\left(\int_0^1
 (r-\frac{1}{2})K_{i,c_i}^\mu(r)dr\right)^2\right]}{\sqrt{N^{-1}\sum_{i=1}^N\left[\int_0^1
 K_{i,c_i}^\mu(r)^2dr-12\left(\int_0^1
 (r-\frac{1}{2})K_{i,c_i}^\mu(r)dr\right)^2\right]}} \nonumber \\
 &&+\sqrt{\frac{112}{193}}\frac{N^{-1/2}\sum_{i=1}^N \left(\int_0^1
 K_{i,c_i}^\mu(r)dW_i(r)-12\int_0^1
 (r-\frac{1}{2})K_{i,c_i}^\mu(r)dr\int_0^1 (r-\frac{1}{2})dW_i(r)+\frac{1}{2}\right)}{\sqrt{N^{-1}\sum_{i=1}^N\left[\int_0^1
 K_{i,c_i}^\mu(r)^2dr-12\left(\int_0^1
 (r-\frac{1}{2})K_{i,c_i}^\mu(r)dr\right)^2\right]}} \nonumber \\
 &&-\sqrt{\frac{112}{193}}\left(\frac{\frac{15\sqrt N}{2} \left(N^{-1}\sum_{i=1}^N \left[\int_0^1
 K_{i,c_i}^\mu(r)^2dr-12\left(\int_0^1
 (r-\frac{1}{2})K_{i,c_i}^\mu(r)dr\right)^2\right]-\frac{1}{15}\right)}{\sqrt{N^{-1}\sum_{i=1}^N \left[\int_0^1
 K_{i,c_i}^\mu(r)^2dr-12\left(\int_0^1
 (r-\frac{1}{2})K_{i,c_i}^\mu(r)dr\right)^2\right]}}\right) \nonumber \\
 &\stackrel{def}{=}& \sqrt{\frac{112}{193}}\left[\frac{N^{-1/2}\sum_{i=1}^N \tilde U_{3i}}{\sqrt{N^{-1}\sum_{i=1}^N \tilde V_{3i}}}+\frac{15\sqrt N}{2}\sqrt{N^{-1}\sum_{i=1}^N \tilde V_{3i}}\right]. \nonumber \\
\end{eqnarray}}

 We have
 \begin{eqnarray*}
 && \frac{\partial}{\partial v}\left[\phi_3(0,-v)\right] \\
 &=& \frac{\partial}{\partial v}\Bigg[e^{-\sum_{i=1}^N c_iN^{-\frac{1}{4}}}\exp\Big(\sum_{i=1}^N\log\Big(1+c_iN^{-1/4}+\frac{1}{2}c_i^2N^{-1/2}+\frac{1}{6}c_i^3N^{-3/4}
 +N^{-1}\left(\frac{2v}{15}+\frac{c_i^4}{24}\right)
 \\
 &&+N^{-5/4}\left(\frac{2c_iv}{15}+\frac{c_i^5}{120}\right) +N^{-3/2}\left(\frac{13c_i^2v}{210}+\frac{c_i^6}{720}\right)+O(N^{-7/4})\Big)\Big)\Bigg]^{-1/2}
 \end{eqnarray*}
 \begin{eqnarray}\label{A.20}
 &=& \frac{\partial}{\partial v}\Bigg[e^{-\sum_{i=1}^N c_iN^{-\frac{1}{4}}}e^{\sum_{i=1}^Nc_iN^{-1/4}+\frac{2v}{15}
 -\sum_{i=1}^N\frac{c_i^2v}{210}N^{-3/2}+O(N^{-3/4})}\Bigg]^{-1/2} \nonumber\\
 &=&  \frac{\partial}{\partial v}\left(\left[e^{\frac{2v}{15}-N^{-3/2}\sum_{i=1}^N \frac{c_i^2v}{210}+O(N^{-3/4})}\right]^{-1/2}-\frac{1}{2}\left[e^{\frac{2v}{15}-N^{-3/2}\sum_{i=1}^N\frac{c_i^2v}{210}+O(N^{-3/4})}\right]^{-1/2}\right) \nonumber\\
 &=& -\frac{1}{15}e^{-v/15}+\frac{e^{-v/15}-\frac{1}{15}v e^{-v/15}}{420 N^{3/2}}\sum_{i=1}^N c_i^2+O(N^{-3/4}).
 \end{eqnarray}

 Hence, from (\ref{A.5}) and (\ref{A.11}), and plugging in (\ref{A.17}) and (\ref{A.20}), we have
 {\small
 \begin{eqnarray*}
 &&E\left(\sqrt{\frac{112}{193}}\left[\frac{N^{-1/2}\sum_{i=1}^N \tilde U_{3i}}{\sqrt{N^{-1}\sum_{i=1}^N \tilde V_{3i}}}+\frac{15\sqrt N}{2}\sqrt{N^{-1}\sum_{i=1}^N \tilde V_{3i}}\right]\right) \\
 &=& \sqrt{\frac{112}{193}}\frac{1}{\Gamma(\frac{1}{2})}\int_0^\infty \frac{1}{\sqrt v} \left.\frac{\partial }{\partial u}\phi_3(u,-v)\right|_{u=0}dv
 -\sqrt{\frac{112}{193}}\frac{15\sqrt N}{2}\frac{1}{\Gamma(\frac{1}{2})}\int_0^\infty \frac{1}{\sqrt v} \frac{\partial}{\partial v}\left[\phi_3(0,-v)\right]dv \\
 &=& -\sqrt{\frac{112}{193}}\frac{\sqrt N}{2}\frac{1}{\sqrt \pi}\int_0^\infty \frac{e^{-v/15}}{\sqrt v}dv-\left(N^{-1}\sum_{i=1}^N c_i^2\right)\sqrt{\frac{112}{193}}\frac{1}{840\sqrt \pi}\int_0^\infty \frac{ve^{-v/15}}{\sqrt v}dv \\
 &&+\sqrt{\frac{112}{193}}\frac{\sqrt N}{2}\frac{1}{\sqrt \pi}\int_0^\infty \frac{e^{-v/15}}{\sqrt v}dv-\left(N^{-1}\sum_{i=1}^N c_i^2\right)\sqrt{\frac{112}{193}}\frac{1}{56\sqrt \pi}\left(\int_0^\infty \frac{e^{-v/15}}{\sqrt v}dv-\frac{1}{15}\int_0^\infty \frac{ve^{-v/15}}{\sqrt v}dv\right) \\
 &&+O_p(N^{-1/4}) \\
 &=& -\sqrt{\frac{112}{193}}\sqrt{\frac{15N}{4}}-\frac{\sqrt{15}}{112}\sqrt{\frac{112}{193}}\overline{c^2}
 +\sqrt{\frac{112}{193}}\sqrt{\frac{15N}{4}}-\frac{\sqrt{15}}{112}\sqrt{\frac{112}{193}}\overline{c^2}+O_p(N^{-1/4}).
 \end{eqnarray*}}
 Therefore,
 \begin{eqnarray*}
 E\left(\hat t_{\delta,32}\right) &=& E\left(\sqrt{\frac{112}{193}}\left[\frac{N^{-1/2}\sum_{i=1}^N \tilde U_{3i}}{\sqrt{N^{-1}\sum_{i=1}^N \tilde V_{3i}}}+\frac{15\sqrt N}{2}\sqrt{N^{-1}\sum_{i=1}^N \tilde V_{3i}}\right]\right) \\
 &=& -\frac{\sqrt{15}}{56}\sqrt{\frac{112}{193}}\overline{c^2}
 +O_p(N^{-1/4}).
 \end{eqnarray*}

\noindent {\bf Proof of Theorem 3.2(a):} Let
 \begin{eqnarray*}
 F_i &=& \int_0^1 W_i(r)^2dr, \\
 G_i &=& -2c_iN^{-1/2}\int_0^1W_i(r)\int_0^r W_i(s)dsdr+O_p(N^{-1}).
 \end{eqnarray*}
 We have that
 \begin{eqnarray*}
 && (F_i+G_i)^{-1/2} \\
 &=& \frac{1}{\sqrt F_i}-\frac{G_i}{2\sqrt{F_i^3}}+O_p(N^{-1}) \\
 &=& \frac{1}{\sqrt{\int_0^1 W_i(r)^2dr}}+\frac{2c_iN^{-1/2}\int_0^1 W_i(r)\int_0^r W_i(s)dsdr}{2\sqrt{(\int_0^1 W_i(r)^2dr)^3}}+O_p(N^{-1}).
 \end{eqnarray*}
Hence,
\begin{eqnarray*}
\hat t_i&\Rightarrow&\frac{\tilde U_{1i}}{\sqrt{\tilde V_{1i}}} =-\frac{c_i}{N^{1/2}}\sqrt{\int_0^1 W_i(r)^2dr}+\frac{\int_0^1 W_i(r)dW_i(r)-c_iN^{-1/2}\int_0^1\!\!\int_0^r W_i(s)dsdW_i(r)}{\sqrt{\int_0^1 W_i(r)^2dr}} \\
&&+\frac{c_iN^{-1/2}\int_0^1 W_i(r)dW_i(r)\int_0^1 W_i(r)\int_0^r W_i(s)dsdr}{\sqrt{(\int_0^1 W_i(r)^2dr)^3}}+O_p(N^{-1}).
\end{eqnarray*}
 From the standard CLT and LLN, we have
 \begin{eqnarray*}
 Z &=& \frac{\sqrt N (N^{-1}\sum_{i=1}^N
 \hat t_i-E(t_0))}{\sqrt{Var(t_0)}} \Rightarrow
 N(0,1)-\bar c\Bigg[E\left(\sqrt{\int_0^1 W(r)^2dr}\right) \\
 &&+E\left(\frac{\int_0^1\!\!\int_0^r W(s)dsdW(r)}{\sqrt{\int_0^1 W(r)^2dr}}\right)
 -E\left(\frac{\int_0^1 W(r)dW(r)\int_0^1 W(r)\int_0^r W(s)dsdr}{\sqrt{(\int_0^1 W(r)^2dr)^3}}\right)
 \Bigg]/\sqrt{Var(t_0)}.
 \end{eqnarray*}
 However, this is not very informative, since the expectations in this expression could not be calculated easily,
 which has to rely on simulations.

 Now, we apply our approach. From (\ref{A.1}), we have
 {\footnotesize
 \begin{eqnarray*}
 && \left.\frac{\partial }{\partial u}\psi_{1,i}(u,-v)\right|_{u=0} \\
 &=& -\frac{1}{2}\left[e^{-c_iN^{-\frac{1}{2}}}\left[\cos\sqrt{-2 v-c_i^2N^{-1}}
 +c_iN^{-1/2}\frac{\sin\sqrt{-2 v-c_i^2N^{-1}}}{\sqrt{-2 v-c_i^2N^{-1}}}\right]\right]^{-1/2} \\
 &&-\frac{1}{2}\left[e^{-c_iN^{-\frac{1}{2}}}\left[\cos\sqrt{-2 v-c_i^2N^{-1}}
 +c_iN^{-1/2}\frac{\sin\sqrt{-2 v-c_i^2N^{-1}}}{\sqrt{-2 v-c_i^2N^{-1}}}\right]\right]^{-3/2}
 \left(-\frac{\sin\sqrt{-2 v-c_i^2N^{-1}}}{\sqrt{-2 v-c_i^2N^{-1}}}\right)e^{-c_iN^{-\frac{1}{2}}} \\
 &=& -\frac{1}{2}\left[\left(1-c_iN^{-\frac{1}{2}}+O(N^{-1})\right)\left(\cosh\sqrt{2 v}+c_iN^{-1/2}\frac{\sinh\sqrt{2 v}}{\sqrt{2 v}}
 +O(N^{-1})\right)\right]^{-1/2} \\
 &&+\frac{1}{2}\left[\left(1-c_iN^{-\frac{1}{2}}+O(N^{-1})\right)\left(\cosh\sqrt{2 v}+c_iN^{-1/2}\frac{\sinh\sqrt{2 v}}{\sqrt{2 v}}+O(N^{-1})\right)\right]^{-3/2} \\
 &&\times\left(\frac{\sinh\sqrt{2 v}}{\sqrt{2 v}}-c_iN^{-1/2}\frac{\sinh\sqrt{2 v}}{\sqrt{2 v}}+O(N^{-1})\right)
 \end{eqnarray*}
 \begin{eqnarray*}
 &=& -\frac{1}{2}\left[\cosh\sqrt{2 v}-c_iN^{-\frac{1}{2}}\left(\cosh\sqrt{2 v}-\frac{\sinh\sqrt{2 v}}{\sqrt{2 v}}\right)+O(N^{-1})\right]^{-1/2} \\
 &&+\frac{1}{2}\left[\cosh\sqrt{2 v}-c_iN^{-\frac{1}{2}}\left(\cosh\sqrt{2 v}-\frac{\sinh\sqrt{2 v}}{\sqrt{2 v}}\right)+O(N^{-1})\right]^{-3/2}
 \left(\frac{\sinh\sqrt{2 v}}{\sqrt{2 v}}-c_iN^{-1/2}\frac{\sinh\sqrt{2 v}}{\sqrt{2 v}}+O(N^{-1})\right).
 \end{eqnarray*}}

 Recall that $\hat t_i=\frac{\tilde U_{1i}}{\sqrt{\tilde V_{1i}}}$.
 Therefore, applying (\ref{A.5}) and by the change of variable as $x=\sqrt{2v}$ and Taylor expansion, we have
 {\small
 \begin{eqnarray*}
 && E(Z) \nonumber \\
 &=& (Var(t_0))^{-1/2} N^{1/2}N^{-1}\sum_{i=1}^N(E(\hat t_i)-E(t_0)) \nonumber \\
 &=& (Var(t_0))^{-1/2} N^{1/2}N^{-1}\sum_{i=1}^N \Bigg(\frac{1}{\Gamma(\frac{1}{2})}\int_0^\infty \frac{1}{\sqrt v} \left.\frac{\partial }{\partial u}\psi_{1,i}(u,-v)\right|_{u=0}dv-E(t_0)\Bigg) \nonumber \\
 &=& (Var(t_0))^{-1/2} N^{1/2}\Bigg(-\frac{1}{\sqrt{2\pi}}\int_0^\infty \Bigg[(\cosh(x))^{-1/2}\left(1-\frac{\sinh(x)}{x\cosh(x)}\right) \nonumber \\
 &&+\frac{1}{2}\left(N^{-1}\sum_{i=1}^N c_i\right)N^{-1/2}(\cosh(x))^{-1/2}\left(1-\frac{2\sinh(x)}{x\cosh(x)}+\frac{3(\sinh(x))^2}{x^2(\cosh(x))^2}\right)\Bigg] dx+O(N^{-1})-E(t_0)\Bigg) \nonumber \\
 &=& -(Var(t_0))^{-1/2} \frac{\bar c}{2\sqrt{2\pi}}\int_0^\infty (\cosh(x))^{-1/2}\left(1-\frac{2\sinh(x)}{x\cosh(x)}+\frac{3(\sinh(x))^2}{x^2(\cosh(x))^2}\right)dx+O_p(N^{-1/2}),
 \end{eqnarray*}}
 where
 \[ E(t_0)=-\frac{1}{\sqrt{2\pi}}\int_0^\infty (\cosh(x))^{-1/2}\left(1-\frac{\sinh(x)}{x\cosh(x)}\right) dx.\]

 \noindent {\bf Proof of Theorem 3.2(b):} Let
 \begin{eqnarray*}
 F_i &=& \int_0^1 W_i^\mu(r)^2dr, \\
 G_i &=& -2c_iN^{-1/2}\int_0^1 W_i^\mu(r)\left(\int_0^r W_i(s)ds-\int_0^1\!\!\!\int_0^t W_i(s)dsdt\right)dr+O_p(N^{-1}).
 \end{eqnarray*}
 We have that
 \begin{eqnarray*}
 && (F_i+G_i)^{-1/2} \\
 &=& \frac{1}{\sqrt F_i}-\frac{G_i}{2\sqrt{F_i^3}}+O_p(N^{-1}) \\
 &=& \frac{1}{\sqrt{\int_0^1 W_i^\mu(r)^2dr}}+\frac{2c_iN^{-1/2}\int_0^1 W_i^\mu(r)\left(\int_0^r W_i(s)ds-\int_0^1\!\!\int_0^t W_i(s)dsdt\right)dr}{2\sqrt{(\int_0^1 W_i^\mu(r)^2dr)^3}}+O_p(N^{-1}).
 \end{eqnarray*}
Hence,
{\footnotesize
\begin{eqnarray*}
\hat t_i^\mu &\Rightarrow&\frac{U_{2i}}{\sqrt{V_{2i}}} =-\frac{c_i}{N^{1/2}}\sqrt{\int_0^1 W_i^\mu(r)^2dr}+\frac{\int_0^1 W_i^\mu(r)dW_i(r)-c_iN^{-1/2}\int_0^1\left( \int_0^r W_i(s)ds-\int_0^1\!\!\int_0^t W_i(s)dsdt\right)dW_i(r)}{\sqrt{\int_0^1 W_i^\mu(r)^2dr}} \\
&&+\frac{c_iN^{-1/2}\int_0^1 W_i^\mu(r)dW_i(r)\int_0^1W_i^\mu(r)\left(\int_0^r W_i(s)ds-\int_0^1\!\!\int_0^t W_i(s)dsdt\right)dr}{\sqrt{(\int_0^1 W_i^\mu(r)^2dr)^3}}+O_p(N^{-1}).
\end{eqnarray*}}
 From the standard CLT and LLN, we have
 \begin{eqnarray*}
 &&Z^\mu=\frac{\sqrt N (N^{-1}\sum_{i=1}^N
 \hat t_i^\mu-E(t_0^\mu))}{\sqrt{Var(t_0^\mu)}} \\
 &\Rightarrow&
 N(0,1)-\bar c\Bigg[E\left(\sqrt{\int_0^1 W^\mu(r)^2dr}\right)+E\left(\frac{\int_0^1 \left( \int_0^r W(s)ds-\int_0^1\!\!\int_0^t W(s)dsdt\right)dW(r)}{\sqrt{\int_0^1 W^\mu(r)^2dr}}\right) \\
 &&-E\left(\frac{\int_0^1 W^\mu(r)dW(r)\int_0^1 W^\mu(r)\left(\int_0^r W(s)ds-\int_0^1\!\!\int_0^t W(s)dsdt\right)dr}{\sqrt{(\int_0^1 W^\mu(r)^2dr)^3}}\right)
 \Bigg]/\sqrt{Var(t_0^\mu)},
 \end{eqnarray*}
 which is almost impossible to evaluate.

 Now, we apply our approach. From (\ref{A.6}), we have
 {\small
 \begin{eqnarray*}
 && \left.\frac{\partial }{\partial u}\psi_{2,i}(u,-v)\right|_{u=0} \\
 &=& -\frac{1}{2}\Bigg[e^{-c_iN^{-\frac{1}{2}}}\Bigg[\frac{-2v-c_i^3N^{-3/2}}{-2v-c_i^2N^{-1}}
 \frac{\sin\sqrt{-2v-c_i^2N^{-1}}}{\sqrt{-2v-c_i^2N^{-1}}}-c_i^2N^{-1}\frac{\cos\sqrt{-2v-c_i^2N^{-1}}}{-2v-c_i^2N^{-1}} \\
 &&+(4c_iN^{-1/2}v)\frac{\cos\sqrt{-2v-c_i^2N^{-1}}-1}{(-2v-c_i^2N^{-1})^2}\Bigg]\Bigg]^{-1/2} \\
 &&-\frac{1}{2}\Bigg[e^{-c_iN^{-\frac{1}{2}}}\Bigg[\frac{-2v-c_i^3N^{-3/2}}{-2v-c_i^2N^{-1}}
 \frac{\sin\sqrt{-2v-c_i^2N^{-1}}}{\sqrt{-2v-c_i^2N^{-1}}}-c_i^2N^{-1}\frac{\cos\sqrt{-2v-c_i^2N^{-1}}}{-2v-c_i^2N^{-1}} \\
 &&+(4c_iN^{-1/2}v)\frac{\cos\sqrt{-2v-c_i^2N^{-1}}-1}{(-2v-c_i^2N^{-1})^2}\Bigg]\Bigg]^{-3/2}
 \Bigg(\frac{c_i^2N^{-1}}{-2v-c_i^2N^{-1}}\frac{\sin\sqrt{-2 v-c_i^2N^{-1}}}{\sqrt{-2 v-c_i^2N^{-1}}} \\
 &&+2c_i^2N^{-1}\frac{\cos\sqrt{-2v-c_i^2N^{-1}}-1}{(-2v-c_i^2N^{-1})^2}\Bigg)e^{-c_iN^{-\frac{1}{2}}} \\
 &=& -\frac{1}{2}\left[\left(1-c_iN^{-\frac{1}{2}}+O(N^{-1})\right)\left(\frac{\sinh\sqrt{2 v}}{\sqrt{2 v}}+c_iN^{-1/2}v^{-1}[\cosh\sqrt{2 v}-1]
 +O(N^{-1})\right)\right]^{-1/2}
 \end{eqnarray*}
 \begin{eqnarray*}
 &&-\frac{1}{2}\left[\left(1-c_iN^{-\frac{1}{2}}+O(N^{-1})\right)\left(\frac{\sinh\sqrt{2 v}}{\sqrt{2 v}}+c_iN^{-1/2}v^{-1}[\cosh\sqrt{2 v}-1]+O(N^{-1})\right)\right]^{-3/2}O(N^{-1}) \\
 &=& -\frac{1}{2}\left[\frac{\sinh\sqrt{2 v}}{\sqrt{2 v}}-c_iN^{-\frac{1}{2}}\left(\frac{\sinh\sqrt{2 v}}{\sqrt{2 v}}-v^{-1}(\cosh\sqrt{2 v}-1)\right)+O(N^{-1})\right]^{-1/2}+O(N^{-1}).
 \end{eqnarray*}}

 Recall that $\hat t_i^\mu=\frac{\tilde U_{2i}}{\sqrt{\tilde V_{2i}}}$, applying (\ref{A.5}) and by the change of variable as $x=\sqrt{2v}$ and Taylor expansion, we have
 \begin{eqnarray*}
 && E(Z^\mu) \nonumber \\
 &=& (Var(t_0^\mu))^{-1/2} N^{1/2}N^{-1}\sum_{i=1}^N(E(\hat t_i^\mu)-E(t_0^\mu)) \nonumber \\
 &=& (Var(t_0^\mu))^{-1/2} N^{1/2}N^{-1}\sum_{i=1}^N \Bigg(\frac{1}{\Gamma(\frac{1}{2})}\int_0^\infty \frac{1}{\sqrt v} \left.\frac{\partial }{\partial u}\psi_{2,i}(u,-v)\right|_{u=0}dv-E(t_0^\mu)\Bigg) \nonumber \\
 &=& (Var(t_0^\mu))^{-1/2} N^{1/2}\Bigg(-\frac{1}{\sqrt{2\pi}}\int_0^\infty \Bigg[\left(\frac{\sinh(x)}{x}\right)^{-1/2}+\frac{1}{2}\left(N^{-1}\sum_{i=1}^N c_i\right)N^{-1/2} \nonumber \\
 &&\times\left(\frac{\sinh(x)}{x}\right)^{-1/2}\left(1-\frac{2(\cosh(x)-1)}{x\sinh(x)}\right)\Bigg] dx+O(N^{-1})-E(t_0^\mu)\Bigg) \nonumber \\
 &=& -(Var(t_0^\mu))^{-1/2} \frac{\bar c}{2\sqrt{2\pi}}\int_0^\infty \left(\frac{\sinh(x)}{x}\right)^{-1/2}\left(1-\frac{2(\cosh(x)-1)}{x\sinh(x)}\right)dx+O(N^{-1/2}), \nonumber \\
 \end{eqnarray*}
 where
 \[ E(t_0^\mu)=-\frac{1}{\sqrt{2\pi}}\int_0^\infty \left(\frac{\sinh(x)}{x}\right)^{-1/2}dx.\]

 \noindent {\bf Proof of Theorem 3.2(c):} Plugging (\ref{3.2.8}) into (\ref{3.2.7}) and applying (12), we have that as $T\to \infty$
 {\small
 \begin{eqnarray*}
 \hat t_i^\tau &=& \frac{\hat \delta_i^\tau-1}{\hat \sigma_{\varepsilon,i}\sqrt{\frac{\sum_t (t-\bar t)^2}{\left(\sum_t (t-\bar t)^2\right)\left(\sum_t
 (z_{i,t-1}-\bar z_{i,t-1})^2\right)-\left(\sum_t(t-\bar
 t)(z_{i,t-1}-\bar z_{i,t-1})\right)^2}}} \\
 &=& \frac{1}{\hat \sigma_{\varepsilon,i}}\Bigg((\delta_i-1)\sqrt{\left(\sum_t
 (z_{i,t-1}-\bar z_{i,t-1})^2\right)-\frac{\left(\sum_t(t-\bar
 t)(z_{i,t-1}-\bar z_{i,t-1})\right)^2}{\sum_t (t-\bar t)^2}} \\
 &&+\frac{\left(\sum_t (t-\bar t)^2\right)\left(\sum_t
 (z_{i,t-1}-\bar z_{i,t-1})((1-\delta_i)\beta_{1,i}(t-\bar t)+(\varepsilon_{it}-\bar \varepsilon_{it}))\right)}
 {\sqrt{\sum_t (t-\bar t)^2}\sqrt{\left(\sum_t (t-\bar t)^2\right)\left(\sum_t
 (z_{i,t-1}-\bar z_{i,t-1})^2\right)-\left(\sum_t(t-\bar
 t)(z_{i,t-1}-\bar z_{i,t-1})\right)^2}}
 \end{eqnarray*}
 \begin{eqnarray*}
 &&-\frac{\left(\sum_t(t-\bar
 t)(z_{i,t-1}-\bar z_{i,t-1})\right)\left(\sum_t(t-\bar
 t)((1-\delta_i)\beta_{1,i}(t-\bar t)+(\varepsilon_{it}-\bar \varepsilon_{it}))\right)}{\sqrt{\sum_t (t-\bar t)^2}\sqrt{\left(\sum_t (t-\bar t)^2\right)\left(\sum_t
 (z_{i,t-1}-\bar z_{i,t-1})^2\right)-\left(\sum_t(t-\bar
 t)(z_{i,t-1}-\bar z_{i,t-1})\right)^2}}\Bigg) \\
 &=& \frac{1}{\hat \sigma_{\varepsilon,i}}\Bigg(-\frac{c_i}{T N^{1/4}}\sqrt{\left(\sum_t
 (z_{i,t-1}-\bar z_{i,t-1})^2\right)-\frac{\left(\sum_t(t-\bar
 t)(z_{i,t-1}-\bar z_{i,t-1})\right)^2}{\sum_t (t-\bar t)^2}} \\
 &&+\frac{\left(\sum_t (t-\bar t)^2\right)\left(\sum_t
 (z_{i,t-1}-\bar z_{i,t-1})(\varepsilon_{it}-\bar \varepsilon_{it})\right)-\left(\sum_t(t-\bar
 t)(z_{i,t-1}-\bar z_{i,t-1})\right)\left(\sum_t(t-\bar
 t)(\varepsilon_{it}-\bar \varepsilon_{it})\right)}{\sqrt{\sum_t (t-\bar t)^2}\sqrt{\left(\sum_t (t-\bar t)^2\right)\left(\sum_t
 (z_{i,t-1}-\bar z_{i,t-1})^2\right)-\left(\sum_t(t-\bar
 t)(z_{i,t-1}-\bar z_{i,t-1})\right)^2}}\Bigg) \\
 &=& \frac{1}{\hat \sigma_{\varepsilon,i}}\Bigg(-\frac{c_i}{T N^{1/4}}\sqrt{\left(\sum_t
 (y_{i,t-1}-\bar y_{i,t-1})^2\right)-\frac{\left(\sum_t(t-\bar
 t)(y_{i,t-1}-\bar y_{i,t-1})\right)^2}{\sum_t (t-\bar t)^2}} \\
 &&+\frac{\left(\sum_t (t-\bar t)^2\right)\left(\sum_t
 (y_{i,t-1}-\bar y_{i,t-1})(\varepsilon_{it}-\bar \varepsilon_{it})\right)-\left(\sum_t(t-\bar
 t)(y_{i,t-1}-\bar y_{i,t-1})\right)\left(\sum_t(t-\bar
 t)(\varepsilon_{it}-\bar \varepsilon_{it})\right)}{\sqrt{\sum_t (t-\bar t)^2}\sqrt{\left(\sum_t (t-\bar t)^2\right)\left(\sum_t
 (y_{i,t-1}-\bar y_{i,t-1})^2\right)-\left(\sum_t(t-\bar
 t)(y_{i,t-1}-\bar y_{i,t-1})\right)^2}}\Bigg) \\
 &\Rightarrow& -\frac{c_i}{N^{1/4}}\sqrt{\int_0^1
 K_{i,c_i}^\mu(r)^2dr-12\left(\int_0^1
 (r-\frac{1}{2})K_{i,c_i}^\mu(r)dr\right)^2} \\
 &&+\frac{\int_0^1
 K_{i,c_i}^\mu(r)dW_i(r)-12\int_0^1
 (r-\frac{1}{2})K_{i,c_i}^\mu(r)dr\int_0^1 (r-\frac{1}{2})dW_i(r)}{\sqrt{\int_0^1
 K_{i,c_i}^\mu(r)^2dr-12\left(\int_0^1
 (r-\frac{1}{2})K_{i,c_i}^\mu(r)dr\right)^2}}=\frac{\tilde U_{3i}}{\sqrt{\tilde V_{3i}}},
 \end{eqnarray*}}
 where $K_{i,c_i}^\mu(r)=K_{i,c_i}(r)-\int_0^1 K_{i,c_i}(s)ds$, and
 $K_{i,c_i}(r) = \int_0^r e^{-c_iN^{-1/4}(r-s)}dW_i(s)$.

 Now, we apply our approach. From (\ref{A.14}), we have
 {\footnotesize
 \begin{eqnarray*}
 && \left.\frac{\partial }{\partial u}\psi_{3,i}(u,-v)\right|_{u=0} \\
 &=& -\frac{1}{2}\Bigg[e^{-c_iN^{-\frac{1}{4}}}\Big[\frac{(c_iN^{-1/4})^5
 +8v((c_iN^{-1/4})^2-3c_iN^{-1/4}-3)}{(-2v-c_i^2N^{-1/2})^2}\frac{\sin\sqrt{-2v-c_i^2N^{-1/2}}}{\sqrt{-2v-c_i^2N^{-1/2}}} \\
 &&+\frac{24(-4(c_iN^{-1/4}+1)
 v^2)}{(-2v-c_i^2N^{-1/2})^3}\Bigg(\frac{\sin\sqrt{-2v-c_i^2N^{-1/2}}}{\sqrt{-2v-c_i^2N^{-1/2}}}
 +\frac{\cos\sqrt{-2v-c_i^2N^{-1/2}}}{-2v-c_i^2N^{-1/2}}-\frac{1}{-2v-c_i^2N^{-1/2}}\Bigg) \\
 &&+\Bigg(\frac{c_i^4N^{-1}}{(-2v-c_i^2N^{-1/2})^2}-
 \frac{8(c_i^3N^{-3/4}2v)}{(-2v-c_i^2N^{-1/2})^3}\Bigg)\cos\sqrt{-2v-c_i^2N^{-1/2}}
 -\frac{4(2c_i^2N^{-1/2}v(c_iN^{-1/4}+3))}{(-2v-c_i^2N^{-1/2})^3}\Big]\Bigg]^{-1/2} \\
 &&-\frac{1}{2}\Bigg[e^{-c_iN^{-\frac{1}{4}}}\Big[\frac{(c_iN^{-1/4})^5
 +8v((c_iN^{-1/4})^2-3c_iN^{-1/4}-3)}{(-2v-c_i^2N^{-1/2})^2}\frac{\sin\sqrt{-2v-c_i^2N^{-1/2}}}{\sqrt{-2v-c_i^2N^{-1/2}}} \\
 &&+\frac{24(-4(c_iN^{-1/4}+1)
 v^2)}{(-2v-c_i^2N^{-1/2})^3}\Bigg(\frac{\sin\sqrt{-2v-c_i^2N^{-1/2}}}{\sqrt{-2v-c_i^2N^{-1/2}}}
 +\frac{\cos\sqrt{-2v-c_i^2N^{-1/2}}}{-2v-c_i^2N^{-1/2}}-\frac{1}{-2v-c_i^2N^{-1/2}}\Bigg) \\
 &&+\Bigg(\frac{c_i^4N^{-1}}{(-2v-c_i^2N^{-1/2})^2}-
 \frac{8(c_i^3N^{-3/4}2v)}{(-2v-c_i^2N^{-1/2})^3}\Bigg)\cos\sqrt{-2v-c_i^2N^{-1/2}}
 -\frac{4(2c_i^2N^{-1/2}v(c_iN^{-1/4}+3))}{(-2v-c_i^2N^{-1/2})^3}\Big]\Bigg]^{-3/2} \\
 &&\times e^{-c_iN^{-\frac{1}{4}}}\Big[\frac{-c_i^4N^{-1}}{(-2v-c_i^2N^{-1/2})^2}\frac{\sin\sqrt{-2v-c_i^2N^{-1/2}}}{\sqrt{-2v-c_i^2N^{-1/2}}}
 +\frac{24c_i^4N^{-1}}{(-2v-c_i^2N^{-1/2})^3}\Bigg(\frac{\sin\sqrt{-2v-c_i^2N^{-1/2}}}{\sqrt{-2v-c_i^2N^{-1/2}}}
 \end{eqnarray*}
 \begin{eqnarray*}
 && +\frac{\cos\sqrt{-2v-c_i^2N^{-1/2}}}{-2v-c_i^2N^{-1/2}}-\frac{1}{-2v-c_i^2N^{-1/2}}\Bigg)
 +\Bigg(-\frac{8c_i^4N^{-1}}{(-2v-c_i^2N^{-1/2})^3}\Bigg)\cos\sqrt{-2v-c_i^2N^{-1/2}} \\
 &&-\frac{4c_i^4N^{-1}}{(-2v-c_i^2N^{-1/2})^3}\Big].
 \end{eqnarray*}}
 Further, from Taylor expansion, we have
 {\footnotesize
 \begin{eqnarray*}
 && \left.\frac{\partial }{\partial u}\psi_{3,i}(u,-v)\right|_{u=0} \nonumber \\
 &=& -\frac{1}{2}\Bigg[(1-c_iN^{-\frac{1}{4}}+\frac{1}{2}c_i^2N^{-\frac{1}{2}}+O(N^{-1}))
 \Big[\frac{8v(c_i^2N^{-1/2}-3c_iN^{-1/4}-3)}{(-2v-c_i^2N^{-1/2})^2}\frac{\sin\sqrt{-2v-c_i^2N^{-1/2}}}{\sqrt{-2v-c_i^2N^{-1/2}}} \nonumber \\
 &&+\frac{24(-4(c_iN^{-1/4}+1)
 v^2)}{(-2v-c_i^2N^{-1/2})^3}\Bigg(\frac{\sin\sqrt{-2v-c_i^2N^{-1/2}}}{\sqrt{-2v-c_i^2N^{-1/2}}}
 +\frac{\cos\sqrt{-2v-c_i^2N^{-1/2}}}{-2v-c_i^2N^{-1/2}}-\frac{1}{-2v-c_i^2N^{-1/2}}\Bigg) \nonumber \\
 &&-\frac{4(6c_i^2N^{-1/2}v)}{(-2v-c_i^2N^{-1/2})^3}\Big]+O(N^{-3/4})\Bigg]^{-1/2}+O(N^{-1}) \nonumber \\
 &=& -\frac{1}{2}\Bigg[\Big[\frac{8v(c_i^2N^{-1/2}-3c_iN^{-1/4}-3)}{(-2v-c_i^2N^{-1/2})^2}
 \frac{\sin\sqrt{-2v-c_i^2N^{-1/2}}}{\sqrt{-2v-c_i^2N^{-1/2}}} \nonumber \\
 &&+\frac{24(-4(c_iN^{-1/4}+1)
 v^2)}{(-2v-c_i^2N^{-1/2})^3}\Bigg(\frac{\sin\sqrt{-2v-c_i^2N^{-1/2}}}{\sqrt{-2v-c_i^2N^{-1/2}}}
 +\frac{\cos\sqrt{-2v-c_i^2N^{-1/2}}}{-2v-c_i^2N^{-1/2}}-\frac{1}{-2v-c_i^2N^{-1/2}}\Bigg) \nonumber \\
 &&-\frac{4(6c_i^2N^{-1/2}v)}{(-2v-c_i^2N^{-1/2})^3}\Big]-\Big[\frac{8v(-3c_i^2N^{-1/2}-3c_iN^{-1/4})}{(-2v-c_i^2N^{-1/2})^2}
 \frac{\sin\sqrt{-2v-c_i^2N^{-1/2}}}{\sqrt{-2v-c_i^2N^{-1/2}}} \nonumber \\
 &&+\frac{24(-4(c_i^2N^{-1/2}+c_iN^{-1/4})
 v^2)}{(-2v-c_i^2N^{-1/2})^3}\Bigg(\frac{\sin\sqrt{-2v-c_i^2N^{-1/2}}}{\sqrt{-2v-c_i^2N^{-1/2}}}
 +\frac{\cos\sqrt{-2v-c_i^2N^{-1/2}}}{-2v-c_i^2N^{-1/2}}-\frac{1}{-2v-c_i^2N^{-1/2}}\Bigg)\Big] \nonumber \\
 &&+\Big[\frac{4v(-3c_i^2N^{-1/2})}{(-2v-c_i^2N^{-1/2})^2}
 \frac{\sin\sqrt{-2v-c_i^2N^{-1/2}}}{\sqrt{-2v-c_i^2N^{-1/2}}}+\frac{24(-2(c_i^2N^{-1/2})v^2)}{(-2v-c_i^2N^{-1/2})^3}
 \Bigg(\frac{\sin\sqrt{-2v-c_i^2N^{-1/2}}}{\sqrt{-2v-c_i^2N^{-1/2}}} \nonumber \\
 &&+\frac{\cos\sqrt{-2v-c_i^2N^{-1/2}}}{-2v-c_i^2N^{-1/2}}-\frac{1}{-2v-c_i^2N^{-1/2}}\Bigg)\Big]+O(N^{-3/4})\Bigg]^{-1/2}+O(N^{-1}) \nonumber \\
 &=& -\frac{1}{2}\Bigg[\Big[\frac{4v(5c_i^2N^{-1/2}-6)}{(-2v-c_i^2N^{-1/2})^2}
 \frac{\sin\sqrt{-2v-c_i^2N^{-1/2}}}{\sqrt{-2v-c_i^2N^{-1/2}}}+\frac{24(-4
 v^2+2c_i^2N^{-1/2}v^2)}{(-2v-c_i^2N^{-1/2})^3}\Bigg(\frac{\sin\sqrt{-2v-c_i^2N^{-1/2}}}{\sqrt{-2v-c_i^2N^{-1/2}}} \nonumber \\
 &&+\frac{\cos\sqrt{-2v-c_i^2N^{-1/2}}}{-2v-c_i^2N^{-1/2}}-\frac{1}{-2v-c_i^2N^{-1/2}}\Bigg)
 -\frac{4(6c_i^2N^{-1/2}v)}{(-2v-c_i^2N^{-1/2})^3}\Big]+O(N^{-3/4})\Bigg]^{-1/2}+O(N^{-1}). \nonumber \\
 \end{eqnarray*}}

 Recall that $\hat t_i^\tau=\frac{\tilde U_{3i}}{\sqrt{\tilde V_{3i}}}$, applying (\ref{A.5}) and
 by the change of variable as $x=\sqrt{2v}$ and a Taylor expansion, we have
 {\footnotesize
 \begin{eqnarray*}
 && E(Z^\tau) \nonumber \\
 &=& (Var(t_0^\tau))^{-1/2} N^{1/2}N^{-1}\sum_{i=1}^N(E(\hat t_i^\tau)-E(t_0^\tau)) \nonumber
 \end{eqnarray*}
 \begin{eqnarray*}
 &=& (Var(t_0^\tau))^{-1/2} N^{1/2} \Bigg(-\frac{1}{\sqrt{2\pi}}\int_0^\infty \Bigg[3f_{22}(x)+(N^{-1}\sum_{i=1}^N c_i^2)N^{-1/2}\Big(\frac{-2\sinh (x)}{x^3}+\frac{18\cosh (x)}{x^4}-\frac{66\sinh (x)}{x^5} \nonumber \\
 &&+\frac{96\cosh (x)-96}{x^6}\Big)+O(N^{-3/4})\Bigg]^{-1/2} dx+O(N^{-1})-E(t_0^\tau)\Bigg) \nonumber \\
 &=& (Var(t_0^\tau))^{-1/2} N^{1/2} \Bigg(-\frac{1}{\sqrt{2\pi}}\int_0^\infty [3f_{22}(x)]^{-1/2} dx+\frac{(N^{-1}\sum_{i=1}^N c_i^2)N^{-1/2}}{2\sqrt{2\pi}}\int_0^\infty [3f_{22}(x)]^{-3/2}\Big(-\frac{2\sinh (x)}{x^3} \nonumber \\
 &&+\frac{18\cosh (x)}{x^4}-\frac{66\sinh (x)}{x^5}+\frac{96\cosh (x)-96}{x^6}\Big)dx+O(N^{-3/4})-E(t_0^\tau)\Bigg) \nonumber \\
 &=& (Var(t_0^\tau))^{-1/2} \Bigg(\frac{c^2}{\sqrt{2\pi}}\int_0^\infty [3f_{22}(x)]^{-3/2}\Big(-\frac{\sinh (x)}{x^3}+\frac{9\cosh (x)}{x^4}-\frac{33\sinh (x)}{x^5}+\frac{48(\cosh (x)-1)}{x^6}\Big)dx \nonumber \\
 &&+O(N^{-1/4})\Bigg),
 \end{eqnarray*}}
 where $f_{22}(x)=4\left(\frac{1}{x^3}\sinh (x)-\frac{2}{x^4}[\cosh (x)-1]\right)$, and $E(t_0^\tau)=-\frac{1}{\sqrt{2\pi}}\int_0^\infty [3f_{22}(x)]^{-1/2} dx$ from equation (7) and page 147 in Nabeya (1999).

\bigskip

\noindent {\bf \large A.2 Edgeworth expansion}

\bigskip

 We give more detailed derivations below for the Edgeworth expansions of LLC and IPS test. For LLC tests, Theorem 2.2 in
 Hall (1992) can be applied. For $\hat t_{\delta,1}$, we have the one term Edgeworth expansion under $H_0$ or $H_1$ as
 \begin{eqnarray}\label{A.21}
 F_{1n}(x) &=& P\left(\hat t_{\delta,1}-\frac{N^{1/2}E(U_{1i})}{\sqrt{E(V_{1i})}}\leq x\right) \nonumber \\
 &=& \Phi(x)-\frac{1}{2}N^{-1/2}\phi(x)[l_{11}E(\bar U_{1i}^2)+2l_{12}E(\bar U_{1i}\bar V_{1i})+l_{22}E(\bar V_{1i}^2)] \nonumber \\
 &&-\frac{1}{6}N^{-1/2}H_2(x)\phi(x)[3l_{11}[E(l_1\bar U_{1i}^2+l_2\bar U_{1i}\bar V_{1i})]^2+3l_{22}[E(l_1\bar U_{1i}\bar V_{1i}+l_2\bar V_{1i}^2)]^2 \nonumber \\
 &&+6l_{12}E(l_1\bar U_{1i}^2+l_2\bar U_{1i}\bar V_{1i})E(l_1\bar U_{1i}\bar V_{1i}+l_2\bar V_{1i}^2)+E(l_1\bar U_{1i}+l_2\bar V_{1i})^3]+O(N^{-1}), \nonumber \\
 \end{eqnarray}
 where $\bar U_{1i}=U_{1i}-E(U_{1i})$, $\bar V_{1i}=V_{1i}-E(V_{1i})$, $l_1=(E(V_{1i}))^{-1/2}$, $l_2=-\frac{1}{2}E(U_{1i})(E(V_{1i}))^{-3/2}$, $l_{11}=0$, $l_{12}=-\frac{1}{2}(E(V_{1i}))^{-3/2}$, $l_{22}=\frac{3}{4}E(U_{1i})(E(V_{1i}))^{-5/2}$, $H_2(x)=x^2-1$, and $\phi(x)$ is the density function of the standard normal distribution.

 Under $H_0$, we can calculate the moments of $U_{1i}$ and $V_{1i}$ from (A.1) with $c_i=0$, which give $E(U_{1i})=0$, $E(V_{1i})=1/2$, $E(U_{1i}^2)=1/2$, and $E(U_{1i}V_{1i})=1/3$. Moreover, these imply $l_1=\sqrt 2$, $l_2=0$, $l_{12}=-\sqrt 2$, and $l_{22}=0$. Hence,
 from (\ref{A.21}), we obtain
 \begin{equation}\label{A.22}
 F_{1n}(x) = \Phi(x)+\frac{\sqrt 2}{3}N^{-1/2}\phi(x)+O(N^{-1}).
 \end{equation}

 Under $H_1$, from (\ref{A.1}) we obtain
 \begin{eqnarray*}
 E(\tilde U_{1i}) &=& -\frac{\bar c}{2}N^{-1/2}+O(N^{-1}), \quad E(\tilde V_{1i})=\frac{1}{2}-\frac{\bar c}{3}N^{-1/2}+O(N^{-1}), \\
 E(\tilde U_{1i}^2) &=& \frac{1}{2}-\bar cN^{-1/2}+O(N^{-1}), \quad E(\tilde V_{1i}^2)=\frac{7}{12}-\frac{13\bar c}{15}N^{-1/2}+O(N^{-1}), \\
 E(\tilde U_{1i}\tilde V_{1i}) &=& \frac{1}{3}-\frac{11\bar c}{12}N^{-1/2}+O(N^{-1}), \quad E(\tilde U_{1i}^3) = 1-\frac{15\bar c}{4}N^{-1/2}+O(N^{-1}), \\
 E(\tilde V_{1i}^3) &=& \frac{139}{120}-\frac{75\bar c}{28}N^{-1/2}+O(N^{-1}), \quad
 E(\tilde U_{1i}^2\tilde V_{1i}) = \frac{11}{12}-3\bar cN^{-1/2}+O(N^{-1}), \\
 E(\tilde U_{1i}\tilde V_{1i}^2) &=& \frac{13}{15}-\frac{323\bar c}{120}N^{-1/2}+O(N^{-1}).
 \end{eqnarray*}
 Further, from (\ref{A.21}), we have
 \begin{eqnarray}\label{A.23}
 F_{1n,c}(x) &=& P\left(\hat t_{\delta,1}+\frac{\bar c}{\sqrt 2}-\frac{\sqrt 2}{6}c^2N^{-1/2}+O(N^{-1})\leq x\right) \nonumber \\
 &=& \Phi(x)+\frac{\sqrt 2}{3}N^{-1/2}\phi(x)-\frac{\sqrt 2\bar c}{12}N^{-1}\phi(x)+O(N^{-1}).
 \end{eqnarray}

 The one term Edgeworth expansion of $\hat t_{\delta,21}$, $\hat t_{\delta,22}$, $\hat t_{\delta,31}$ and $\hat t_{\delta,32}$ could be
 obtained in the same way. The computation could be carried out using the symbolic calculations in MATLAB.

 Next, we consider IPS tests. Theorem 2.1 in Hall (1992) can be applied directly. For $Z$, under $H_0$,
 applying (\ref{A.5}) to (\ref{A.1}) with $c_i=0$, we have
 \begin{eqnarray*}
 E(t_0) &=& -\frac{1}{\sqrt{2\pi}}\int_0^\infty (\cosh(x))^{-1/2}\left(1-\frac{\sinh(x)}{x\cosh(x)}\right) dx, \\
 E(t_0)^2 &=& \frac{1}{4}\int_0^\infty x(\cosh(x))^{-1/2}\left(1-\frac{2\sinh(x)}{x\cosh(x)}+3\left(\frac{\sinh(x)}{x\cosh(x)}\right)^2\right) dx, \\
 E(t_0)^3 &=& -\frac{\sqrt 2}{8\sqrt{\pi}}\int_0^\infty x^2(\cosh(x))^{-1/2}\left(1-\frac{3\sinh(x)}{x\cosh(x)}+9\left(\frac{\sinh(x)}{x\cosh(x)}\right)^2-15\left(\frac{\sinh(x)}{x\cosh(x)}\right)^3\right) dx.
 \end{eqnarray*}
 Then we have the skewness of $t_0$ as $\lambda_1=E(t_0-E(t_0))^3/(Var(t_0))^{3/2}=[E(t_0)^3-3E(t_0)^2E(t_0)+2(E(t_0))^3]/(Var(t_0))^{3/2}$. The standard one term Edgeworth expansion could be obtained as (\ref{4.11}).
 Under $H_1$, applying (\ref{A.5}) to (\ref{A.1}), we have
 \begin{eqnarray*}
 E(\hat t_i) &=& E(t_0)-\frac{\bar c}{2\sqrt{2\pi}}N^{-1/2}\int_0^\infty (\cosh(x))^{-1/2}\left(1-\frac{2\sinh(x)}{x\cosh(x)}+\frac{3(\sinh(x))^2}{x^2(\cosh(x))^2}\right)dx+O(N^{-1}), \\
 E(\hat t_i)^2 &=& E(t_0)^2+\frac{\bar c}{8}N^{-1/2}\int_0^\infty x(\cosh(x))^{-1/2}\Bigg(1-\frac{3\sinh(x)}{x\cosh(x)}+9\left(\frac{\sinh(x)}{x\cosh(x)}\right)^2 \\
 &&-15\left(\frac{\sinh(x)}{x\cosh(x)}\right)^3\Bigg)dx+O(N^{-1}), \\
 E(\hat t_i)^3 &=& E(t_0)^3-\frac{\sqrt 2\bar c}{16\sqrt{\pi}}N^{-1/2}\int_0^\infty x^2(\cosh(x))^{-1/2}\Bigg(1-\frac{4\sinh(x)}{x\cosh(x)}+18\left(\frac{\sinh(x)}{x\cosh(x)}\right)^2 \\
 &&-60\left(\frac{\sinh(x)}{x\cosh(x)}\right)^3+105\left(\frac{\sinh(x)}{x\cosh(x)}\right)^4\Bigg) dx+O(N^{-1}).
 \end{eqnarray*}
 The one term Edgeworth expansion could be obtained as (\ref{4.12}) with
 \[ \lambda_{1,c}=[E(\hat t_i)^3-3E(\hat t_i)^2E(\hat t_i)+2(E(\hat t_i))^3]/(Var(t_0))^{3/2}.\]

 Similarly, for $Z^\mu$, under $H_0$, applying (\ref{A.5}) to (\ref{A.6}) with $c_i=0$, we have
 \begin{eqnarray*}
 E(t_0^\mu) &=& -\frac{1}{\sqrt{2\pi}}\int_0^\infty \left(\frac{\sinh(x)}{x}\right)^{-1/2}dx, \\
 E(t_0^\mu)^2 &=& \frac{1}{4}\int_0^\infty x\left(\frac{\sinh(x)}{x}\right)^{-1/2}\left(1+\frac{4}{x^2}-\frac{8(\cosh(x)-1)}{x^3\sinh(x)}\right) dx, \\
 E(t_0^\mu)^3 &=& -\frac{\sqrt 2}{8\sqrt{\pi}}\int_0^\infty x^2\left(\frac{\sinh(x)}{x}\right)^{-1/2}\left(1+\frac{12}{x^2}-\frac{24(\cosh(x)-1)}{x^3\sinh(x)}\right) dx.
 \end{eqnarray*}
 Then we have the skewness of $t_0^\mu$ as $\lambda_2=E(t_0^\mu-E(t_0^\mu))^3/(Var(t_0^\mu))^{3/2}$.
 The standard one term Edgeworth expansion could be obtained as (\ref{4.13}).
 Under $H_1$, applying (\ref{A.5}) to (\ref{A.6}), we have
 \begin{eqnarray*}
 E(\hat t_i^\mu) &=& E(t_0^\mu)-\frac{\bar c}{2\sqrt{2\pi}}N^{-1/2}\int_0^\infty \left(\frac{\sinh(x)}{x}\right)^{-1/2}\left(1-\frac{2(\cosh(x)-1)}{x\sinh(x)}\right)dx+O(N^{-1}), \\
 E(\hat t_i^\mu)^2 &=& E(t_0^\mu)^2+\frac{\bar c}{8}N^{-1/2}\int_0^\infty x\left(\frac{\sinh(x)}{x}\right)^{-1/2}\Bigg(1-\frac{2(\cosh(x)-1)}{x\sinh(x)}+\frac{4}{x^2}-\frac{32(\cosh(x)-1)}{x^3\sinh(x)} \\
 &&+\frac{48(\cosh(x)-1)^2}{x^4(\sinh(x))^2}\Bigg)dx+O(N^{-1}),
 \end{eqnarray*}
 \begin{eqnarray*}
 E(\hat t_i^\mu)^3 &=& E(t_0^\mu)^3-\frac{\sqrt 2\bar c}{16\sqrt{\pi}}N^{-1/2}\int_0^\infty x^2\left(\frac{\sinh(x)}{x}\right)^{-1/2}\Bigg(1-\frac{2(\cosh(x)-1)}{x\sinh(x)}+\frac{12}{x^2} \\
 &&-\frac{96(\cosh(x)-1)}{x^3\sinh(x)}+\frac{144(\cosh(x)-1)^2}{x^4(\sinh(x))^2}\Bigg) dx+O(N^{-1}).
 \end{eqnarray*}
 The one term Edgeworth expansion could be obtained as (\ref{4.14}) with
 \[ \lambda_{2,c}=[E(\hat t_i^\mu)^3-3E(\hat t_i^\mu)^2E(\hat t_i^\mu)+2(E(\hat t_i^\mu))^3]/(Var(t_0^\mu))^{3/2}.\]

 For $Z^\tau$, under $H_0$, applying (\ref{A.5}) to (\ref{A.14}) with $c_i=0$, we have
 \begin{eqnarray*}
 E(t_0^\tau) &=& -\frac{1}{\sqrt{2\pi}}\int_0^\infty \left[12\left(\frac{1}{x^3}\sinh (x)-\frac{2}{x^4}[\cosh (x)-1]\right)\right]^{-1/2} dx, \\
 E(t_0^\tau)^2 &=& \frac{1}{4}\int_0^\infty x\left[12\left(\frac{1}{x^3}\sinh (x)-\frac{2}{x^4}[\cosh (x)-1]\right)\right]^{-1/2}\Bigg(1+4\bigg(\left(\frac{24}{x^7}+\frac{1}{x^5}\right)\sinh(x) \\
 &&-\left(\frac{24}{x^8}+\frac{8}{x^6}\right)\cosh(x)+\frac{24}{x^8}-\frac{4}{x^6}\bigg)/\left(\frac{1}{x^3}\sinh (x)-\frac{2}{x^4}[\cosh (x)-1]\right)\Bigg) dx, \\
 E(t_0^\tau)^3 &=& -\frac{\sqrt 2}{8\sqrt{\pi}}\int_0^\infty x^2\left[12\left(\frac{1}{x^3}\sinh (x)-\frac{2}{x^4}[\cosh (x)-1]\right)\right]^{-1/2}\Bigg(1+12\bigg(\left(\frac{24}{x^7}+\frac{1}{x^5}\right)\sinh(x) \\
 &&-\left(\frac{24}{x^8}+\frac{8}{x^6}\right)\cosh(x)+\frac{24}{x^8}-\frac{4}{x^6}\bigg)/\left(\frac{1}{x^3}\sinh (x)-\frac{2}{x^4}[\cosh (x)-1]\right)\Bigg) dx.
 \end{eqnarray*}
 Then we have the skewness of $t_0^\tau$ as $\lambda_3=E(t_0^\tau-E(t_0^\tau))^3/(Var(t_0^\tau))^{3/2}$. The standard one term Edgeworth expansion could be obtained as (\ref{4.15}).
 Under $H_1$, applying (\ref{A.5}) to (\ref{A.14}), we have
 {\small
 \begin{eqnarray*}
 E(\hat t_i^\tau) &=& E(t_0^\tau)-\frac{\overline{c^2}}{\sqrt{2\pi}}N^{-1/2}\int_0^\infty \left[12\left(\frac{1}{x^3}\sinh (x)-\frac{2}{x^4}[\cosh (x)-1]\right)\right]^{-3/2}\Bigg(\frac{\sinh(x)}{x^3} \\
 &&-\frac{9\cosh(x)}{x^4}+\frac{33\sinh(x)}{x^5}-\frac{48(\cosh (x)-1)}{x^6}\Bigg) dx+O(N^{-1}), \\
 E(\hat t_i^\tau)^2 &=& E(t_0^\tau)^2+\frac{\overline{c^2}}{4}N^{-1/2}\int_0^\infty x\left[12\left(\frac{1}{x^3}\sinh (x)-\frac{2}{x^4}[\cosh (x)-1]\right)\right]^{-3/2}\Bigg(\frac{\sinh(x)}{x^3} \\
 &&-\frac{9\cosh(x)}{x^4}+\frac{\sinh(x)}{x^5}+\frac{232\cosh(x)}{x^6}+\frac{176}{x^6}-\frac{1080\sinh(x)}{x^7}+\frac{2496\cosh(x)}{x^8} \\
 &&-\frac{192}{x^8}-\frac{4608\sinh(x)}{x^9}+\frac{4608(\cosh (x)-1)}{x^{10}}\Bigg) dx+\overline{c^2}N^{-1/2}\int_0^\infty x\Bigg[12\bigg(\frac{1}{x^3}\sinh(x) \end{eqnarray*}
 \begin{eqnarray*}
 &&-\frac{2}{x^4}[\cosh (x)-1]\bigg)\Bigg]^{-5/2}\Bigg(\frac{9\sinh(x)}{x^3}-\frac{39\cosh(x)}{x^4}+\frac{99\sinh(x)}{x^5}
 +\frac{12}{x^4}-\frac{144(\cosh (x)-1)}{x^6}\Bigg) \\
 &&\times \Bigg(\frac{12\sinh(x)}{x^5}-\frac{96\cosh(x)}{x^6}+\frac{288\sinh(x)}{x^7}
 -\frac{48}{x^6}-\frac{288(\cosh (x)-1)}{x^8}\Bigg) dx+O(N^{-1}), \\
 E(\hat t_i^\tau)^3 &=& E(t_0^\tau)^3-\frac{\sqrt 2\overline{c^2}}{8\sqrt{\pi}}N^{-1/2}\int_0^\infty x^2\left[12\left(\frac{1}{x^3}\sinh (x)-\frac{2}{x^4}[\cosh (x)-1]\right)\right]^{-3/2}\Bigg(\frac{\sinh(x)}{x^3} \\
 &&-\frac{9\cosh(x)}{x^4}-\frac{63\sinh(x)}{x^5}+\frac{792\cosh(x)}{x^6}+\frac{432}{x^6}-\frac{3240\sinh(x)}{x^7}+\frac{7488\cosh(x)}{x^8} \\
 &&-\frac{576}{x^8}-\frac{13824\sinh(x)}{x^9}+\frac{13824(\cosh (x)-1)}{x^{10}}\Bigg) dx-\frac{3\sqrt 2\overline{c^2}}{2\sqrt{\pi}}N^{-1/2}\int_0^\infty x^2\Bigg[12\bigg(\frac{1}{x^3}\sinh(x) \\
 &&-\frac{2}{x^4}[\cosh (x)-1]\bigg)\Bigg]^{-5/2}\Bigg(\frac{9\sinh(x)}{x^3}-\frac{39\cosh(x)}{x^4}+\frac{99\sinh(x)}{x^5}
 +\frac{12}{x^4}-\frac{144(\cosh (x)-1)}{x^6}\Bigg) \\
 &&\times \Bigg(\frac{12\sinh(x)}{x^5}-\frac{96\cosh(x)}{x^6}+\frac{288\sinh(x)}{x^7}
 -\frac{48}{x^6}-\frac{288(\cosh (x)-1)}{x^8}\Bigg) dx+O(N^{-1}).
 \end{eqnarray*}}
 The one term Edgeworth expansion could be obtained as (\ref{4.16}) with
 \[ \lambda_{3,c}=[E(\hat t_i^\tau)^3-3E(\hat t_i^\tau)^2E(\hat t_i^\tau)+2(E(\hat t_i^\tau))^3]/(Var(t_0^\tau))^{3/2}.\]

\section*{References}
\def\bskip{\vskip 0.055in}

\bskip \noindent Anderson, T.W., Darling, D.A., 1952. Asymptotic theory of certain `Goodness of Fit'
criteria based on stochastic processes. Annals of Mathematical Statistics 23, 193-212.

\bskip \noindent Bai, J., Ng, S., 2004. A PANIC attack on unit roots and cointegration. Econometrica 72, 1127-1177.

\bskip \noindent Baltagi, B., Kao, C., 2000. Nonstationary panels, cointegration in panels,
and dynamic panels: A survey. Advances in Econometrics 15, 7-52.

\bskip \noindent Bowman, D., 2002. Efficient tests for autoregressive unit roots in panel data. Mimeo.

\bskip \noindent Breitung, J., 2000. The local power of some unit root tests
for panel data. In: Baltagi, B. ed. Advances in Econometrics: Nonstationary Panels,
Panel Cointegration, and Dynamic Panels 15. Amsterdam: JAI, 161-178.

\bskip \noindent Breitung, J., Das, S., 2005. Panel unit root tests under cross sectional dependence.
Statistica Neerlandica 59, 414-433.

\bskip \noindent Breitung, J., Das, S., 2008. Testing for unit roots in panels with a factor structure.
Econometric Theory 24, 88-108.

\bskip \noindent Breitung, J., Meyer, W., 1994. Testing for unit roots in panel data: Are wages on
different bargaining levels cointegrated? Applied Economics 26, 353-361.

\bskip \noindent Breitung, J., Pesaran, M.H., 2008. Unit roots and
cointegration in panels. In: M\'{a}ty\'{a}s, L., Sevestre, P. eds.
The Econometrics of Panel Data. Springer, 279-322.

\bskip \noindent Chan, N.H., Wei, C.Z., 1987. Asymptotic inference for nearly nonstationary AR(1)
processes. The Annals of Statistics 15, 1050-1063.

\bskip \noindent Choi, I., 2001. Unit root tests in panel data. Journal of
International Money and Finance 20, 249-272.

\bskip \noindent Choi, I., 2006. Nonstationary panels. In: Millis, T.C., Patterson, K. eds.
Palgrave Handbook of Econometrics Vol 1. New York: Palgrave macmillan, 511-539.

\bskip \noindent Dickey, D.A., Fuller, W.A., 1979. Distribution of estimators for autoregressive time
series with a unit root. Journal of the American Statistical Association 74, 427-431.

\bskip \noindent Dickey, D.A., Fuller, W.A., 1981. Likelihood ratio statistics for autoregressive time
series with a unit root. Econometrica 49, 1057-1072.

\bskip \noindent Evans, G.B.A., Savin, N.E., 1981. Testing for unit roots: 1. Econometrica 49, 753-779.


\bskip \noindent Hall, P., 1992. The Bootstrap and Edgeworth Expansion. Springer-Verlag, New York.

\bskip \noindent Hansen, B.E., 2014. Asymptotic moments of autoregressive estimators
with a near unit root and minimax risk. Advances in Econometrics 33, 3-21.




\bskip \noindent Hurlin, C., Mignon, V., 2007. Second generation panel unit root tests.
Mimeo.

\bskip \noindent Im, K.S., Pesaran, M.H., Shin, Y., 2003. Testing for
unit roots in heterogeneous panels. Journal of Econometrics 115, 53-74.

\bskip \noindent Imhof, J.P., 1961. Computing the distribution of quadratic forms in normal
variables. Biometrika 48, 419-426.


\bskip \noindent Levin, A., Lin, C., 1992. Unit root tests in panel data:
Asymptotic and finite-sample properties. University of California working paper.

\bskip \noindent Levin, A., Lin, C., Chu, C.-J., 2002. Unit root tests in panel data:
Asymptotic and finite-sample properties. Journal of Econometrics 108, 1-24.

\bskip \noindent Maddala, G.S., Wu, S., 1999. A comparative study of unit root
tests with panel data and a new simple test. Oxford Bulletin of
Economics and statistics 61, 631-652.

\bskip \noindent Moon, H.R., Perron, B., 2004. Testing for a unit root in panels with dynamic factors.
Journal of Econometrics 122, 81-126.

\bskip \noindent Moon, H.R., Perron, B., 2008. Asymptotic local power of pooled
t-ratio tests for unit roots in panels with fixed effects. Econometrics Journal 11, 80-104.

\bskip \noindent Moon, H.R., Perron, B., Phillips, P.C.B., 2006.
On the Breitung test for panel unit roots and local asymptotic power.
Econometric Theory 22, 1179-1190.

\bskip \noindent Moon, H.R., Perron, B., Phillips, P.C.B., 2007.
Incidental trends and the power of panel unit root tests.
Journal of Econometrics 141, 416-459.

\bskip \noindent Moon, H.R., Phillips, P.C.B., 2004. GMM estimation of autoregressive
roots near unity with panel data. Econometrica 72, 467-522.


\bskip \noindent Nabeya, S., 1999. Asymptotic moments of some unit
root test statistics in the null case. Econometric Theory 15, 139-149.


\bskip \noindent Nabeya, S., Tanaka, K., 1988. Asymptotic theory of a test for the constancy
of regression coefficients against the random walk alternative. The Annals of Statistics 16, 218-235.

\bskip \noindent Nabeya, S., Tanaka, K., 1990a. A general approach to the limiting
distribution for estimators in time series regression with nonstable autoregressive errors.
Econometrica 58, 145-163.

\bskip \noindent Nabeya, S., Tanaka, K., 1990b. Limiting power of unit-root tests in time-series
regression. Journal of Econometrics 46, 247-271.


\bskip \noindent Phillips, P.C.B., 1987a. Time series regression with a unit root. Econometrica 55, 277-301.

\bskip \noindent Phillips, P.C.B., 1987b. Towards a unified asymptotic theory for autoregression.
Biometrika 74, 535-547.

\bskip \noindent Phillips, P.C.B., 2012. Folklore theorems, implicit maps, and
indirect inference. Econometrica 80, 425-454.

\bskip \noindent Phillips, P.C.B., Moon, H.R., 1999. Linear regression limit theory for nonstationary
panel data. Econometrica 67, 1057-1111.

\bskip \noindent Phillips, P.C.B., Moon, H.R., 2000. Nonstationary panel data analysis: An
overview of some recent developments. Econometric Reviews 19, 263-286.

\bskip \noindent Phillips, P.C.B., Perron, P., 1988. Testing for a unit root in time series regression.
Biometrika 75, 335-346.

\bskip \noindent Ploberger, W., Phillips, P.C.B., 2002. Optimal testing for unit roots in panel data. Mimeo.

\bskip \noindent Quah, D., 1994. Exploiting cross-section variations for unit root inference
in dynamic panels. Economics Letters 44, 9-19.

\bskip \noindent Sawa, T., 1972. Finite sample properties of k-class estimators.
Econometrica 40, 653-680.


\bskip \noindent Tanaka, K., 1996. Time series analysis: Nonstationary and noninvertible
distribution theory. Wiley, New York.

\bskip \noindent Westerlund, J., Breitung, J., 2012. Lessons from a
decade of IPS and LLC. Econometric Reviews 32, 547-591.

\bskip \noindent White, J.S., 1958. The limiting distribution of the serial correlation coefficient
in the explosive case. Annals of Mathematical Statistics 29, 1188-1197.

\newpage
\setcounter{section}{0}
\setcounter{equation}{0}
\renewcommand{\theequation}{S.\arabic{equation}}
\renewcommand{\thelemma}{S.\arabic{lemma}}

\begin{center} \Large Supplemental Material to ``A Unified Approach on the Local Power of Panel Unit Root Tests"  \end{center}

\begin{center}
 Zhongwen Liang \\
 Department of Economics \\
 University at Albany, SUNY
\end{center}

\begin{small}

\section{Comparison with the existing method}

To compare the difference between our approach and the existing method, the brief discussion is given in the
following. For $\hat t_{\delta,1}$, by the standard Law of Large Numbers (LLN) and
CLT, we have
 \begin{equation}\label{S.1}
 \hat t_{\delta,1}\Rightarrow -\frac{N^{-1}\sum_{i=1}^N c_iE\left(\int_0^1 W_i(r)^2dr\right)}{\sqrt{E\left(\int_0^1 W_i(r)^2dr\right)}}+\frac{N^{-1/2}\sum_{i=1}^N\int_0^1 K_{i,c_i}(r)dW_i(r)}{\sqrt{N^{-1}\sum_{i=1}^N\int_0^1
 K_{i,c_i}(r)^2dr}}\Rightarrow -\frac{\bar c}{\sqrt 2}+N(0,1),
 \end{equation}
where $\bar c=\lim_{N\to \infty}N^{-1}\sum_{i=1}^N c_i$, and by noticing that $E\left(\int_0^1 W_i(r)^2dr\right)=\int_0^1 rdr=1/2$.
This coincides with the result in Theorem 3.1, and was obtained in Moon et al. (2007) and
Westerlund and Breitung (2012).

Next, we consider $\hat t_{\delta,21}$ and $\hat t_{\delta,22}$. We have as $T\to \infty$
\begin{eqnarray*}
&&\hat t_{\delta,21} \\
&\Rightarrow& -\frac{\sqrt 5}{2}\frac{N^{-1}\sum_{i=1}^N c_i\int_0^1
 K_{i,c_i}^\mu(r)^2dr}{\sqrt{N^{-1}\sum_{i=1}^N\int_0^1
 K_{i,c_i}^\mu(r)^2dr}}+\frac{\sqrt 5}{2}\frac{N^{-1/2}\sum_{i=1}^N \left(\int_0^1
 K_{i,c_i}^\mu(r)dW_i(r)+\frac{1}{2}\right)}{\sqrt{N^{-1}\sum_{i=1}^N\int_0^1
 K_{i,c_i}^\mu(r)^2dr}} \\
 &&-\frac{\sqrt 5}{2}\left(\frac{\sqrt N}{2\sqrt{N^{-1}\sum_{i=1}^N \int_0^1
 K_{i,c_i}^\mu(r)^2dr}}-\sqrt{\frac{3N}{2}}\right),
 \end{eqnarray*}
where $K_{i,c_i}^\mu(r)=K_{i,c_i}(r)-\int_0^1 K_{i,c_i}(s)ds$, and
\begin{eqnarray*}
 K_{i,c_i}(r) &=& \int_0^r e^{-c_iN^{-1/2}(r-s)}dW_i(s)
 =  W_i(r)-c_iN^{-1/2}\int_0^r e^{-c_iN^{-1/2}(r-s)}W_i(s)ds \\
 &=& W_i(r)-c_iN^{-1/2}\int_0^r
 W_i(s)ds+c_i^2N^{-1}\int_0^r (r-s)W_i(s)ds+O_p(N^{-3/2}), \\
 K_{i,c_i}^\mu(r) &=& W_i^\mu(r)-c_iN^{-1/2}\left(\int_0^r
 W_i(s)ds-\int_0^1\!\!\!\int_0^t W_i(s)dsdt\right) \\
 &&+c_i^2N^{-1}\left(\int_0^r (r-s)W_i(s)ds-\int_0^1\!\!\!\int_0^t (r-s)W_i(s)dsdt\right)+O_p(N^{-3/2}),
 \end{eqnarray*}
 where $W_i^\mu(r)=W_i(r)-\int_0^1 W_i(s)ds$.

Hence,
\begin{eqnarray*}
&&\int_0^1 K_{i,c_i}^\mu(r)dW_i(r) \nonumber \\
&=& \int_0^1 W_i^\mu(r)dW_i(r)-c_iN^{-1/2}\int_0^1\left( \int_0^r W_i(s)ds-\int_0^1\!\!\!\int_0^t W_i(s)dsdt\right)dW_i(r) \nonumber \\
&&+c_i^2N^{-1}\int_0^1 \left(\int_0^r (r-s)W_i(s)ds-\int_0^1\!\!\!\int_0^t (t-s)W_i(s)dsdt\right)dW_i(r)+O_p(N^{-3/2}), \nonumber
\end{eqnarray*}
and
\begin{eqnarray*}
&&\int_0^1 K_{i,c_i}^\mu(r)^2 dr \nonumber \\
&=& \int_0^1 W_i^\mu(r)^2dr-2c_iN^{-1/2}\int_0^1 W_i^\mu(r)\left(\int_0^r W_i(s)ds-\int_0^1\!\!\!\int_0^t W_i(s)dsdt\right)dr \nonumber \\
&&+c_i^2N^{-1}\int_0^1 \left(\int_0^r W_i(s)ds-\int_0^1\!\!\!\int_0^t W_i(s)dsdt\right)^2dr \nonumber \\
&&+2c_i^2N^{-1}\int_0^1 W_i^\mu(r)\left(\int_0^r (r-s)W_i(s)ds-\int_0^1\!\!\!\int_0^t (t-s)W_i(s)dsdt\right)dr+O_p(N^{-3/2}).
\end{eqnarray*}
Thus, we have
{\scriptsize
 \begin{eqnarray}
 &&-\frac{\sqrt 5}{2}\frac{N^{-1}\sum_{i=1}^N c_i\int_0^1
 K_{i,c_i}^\mu(r)^2dr}{\sqrt{N^{-1}\sum_{i=1}^N\int_0^1
 K_{i,c_i}^\mu(r)^2dr}}+\frac{\sqrt 5}{2}\frac{N^{-1/2}\sum_{i=1}^N \left(\int_0^1
 K_{i,c_i}^\mu(r)dW_i(r)+\frac{1}{2}\right)}{\sqrt{N^{-1}\sum_{i=1}^N\int_0^1
 K_{i,c_i}^\mu(r)^2dr}} \nonumber\\
 &&-\frac{\sqrt 5}{2}\left(\frac{\sqrt N}{2\sqrt{N^{-1}\sum_{i=1}^N \int_0^1
 K_{i,c_i}^\mu(r)^2dr}}-\sqrt{\frac{3N}{2}}\right) \nonumber\\
 &=& -\frac{\sqrt 5}{2}\frac{N^{-1}\sum_{i=1}^N c_i\int_0^1 W_i^\mu(r)^2dr}{\sqrt{N^{-1}\sum_{i=1}^N\int_0^1 W_i^\mu(r)^2dr}}+O_p(N^{-1/2}) \nonumber \\
 &&+\frac{\sqrt 5}{2}\left(N^{-1}\sum_{i=1}^N \int_0^1 W_i^\mu(r)^2dr-2N^{-3/2}\sum_{i=1}^N c_i\int_0^1 W_i^\mu(r)\left(\int_0^r W_i(s)ds-\int_0^1\!\!\!\int_0^t W_i(s)dsdt\right)dr+O_p(N^{-1})\right)^{-1/2} \nonumber\\
 &&\times\Bigg(N^{-1/2}\sum_{i=1}^N \left(\int_0^1 W_i^\mu(r)dW_i(r)+\frac{1}{2}\right)-N^{-1}\sum_{i=1}^N c_i\int_0^1\left( \int_0^r W_i(s)ds-\int_0^1\!\!\!\int_0^t W_i(s)dsdt\right)dW_i(r)+O_p(N^{-1/2})\Bigg) \nonumber\\
 &&-\frac{\sqrt 5}{2}\Bigg(\sqrt{\frac{3N}{2}}-\sqrt{\frac{27N}{2}}\Bigg(N^{-1}\sum_{i=1}^N \int_0^1 W_i^\mu(r)^2dr-\frac{1}{6}-2N^{-3/2}\sum_{i=1}^N c_i\int_0^1 W_i^\mu(r)\left(\int_0^r W_i(s)ds-\int_0^1\!\!\!\int_0^t W_i(s)dsdt\right)dr \nonumber\\
 &&+O_p(N^{-1})\Bigg)+\frac{\sqrt N}{2}\sum_{j=1}^\infty\frac{\sqrt \pi\left(\frac{1}{6}\right)^{-\frac{3}{2}-j}}{\Gamma(-\frac{1}{2}-j)(j+2)!}\Bigg(N^{-1}\sum_{i=1}^N \int_0^1 W_i^\mu(r)^2dr-\frac{1}{6} \nonumber\\
 &&-2N^{-3/2}\sum_{i=1}^N c_i\int_0^1 W_i^\mu(r)\left(\int_0^r W_i(s)ds-\int_0^1\!\!\!\int_0^t W_i(s)dsdt\right)dr+O_p(N^{-1})\Bigg)^{j+1}-\sqrt{\frac{3N}{2}}+O_p(N^{-1/2})\Bigg) \nonumber\\
 &=& -\frac{\sqrt 5}{2}\frac{N^{-1}\sum_{i=1}^N c_i\int_0^1 W_i^\mu(r)^2dr}{\sqrt{N^{-1}\sum_{i=1}^N\int_0^1 W_i^\mu(r)^2dr}}
 +\frac{\sqrt 5}{2}\left(N^{-1}\sum_{i=1}^N \int_0^1 W_i^\mu(r)^2dr\right)^{-1/2}\Bigg(N^{-1/2}\sum_{i=1}^N \left(\int_0^1 W_i^\mu(r)dW_i(r)+\frac{1}{2}\right) \nonumber\\
 &&-N^{-1}\sum_{i=1}^N c_i\int_0^1\left( \int_0^r W_i(s)ds-\int_0^1\!\!\!\int_0^t W_i(s)dsdt\right)dW_i(r)\Bigg)
 +\frac{\sqrt 5}{4}\left(N^{-1}\sum_{i=1}^N \int_0^1 W_i^\mu(r)^2dr\right)^{-3/2} \nonumber
 \end{eqnarray}
 \begin{eqnarray}\label{S.2}
 &&\times\Bigg(\left[N^{-1}\sum_{i=1}^N \left(\int_0^1 W_i^\mu(r)dW_i(r) +\frac{1}{2}\right)\right]\left[2N^{-1}\sum_{i=1}^N c_i\int_0^1 W_i^\mu(r)\left(\int_0^r W_i(s)ds-\int_0^1\!\!\!\int_0^t W_i(s)dsdt\right)dr\right]\Bigg) \nonumber\\
 &&+\sqrt{\frac{135}{8}}N^{-1/2}\sum_{i=1}^N \left(\int_0^1 W_i^\mu(r)^2dr-\frac{1}{6}\right)-\sqrt{\frac{135}{2}} N^{-1}\sum_{i=1}^N c_i\int_0^1 W_i^\mu(r)\left(\int_0^r W_i(s)ds-\int_0^1\!\!\!\int_0^t W_i(s)dsdt\right)dr \nonumber\\
 &&+O_p(N^{-1/2}) \nonumber\\
 &\Rightarrow& N(0,1)-\sqrt{\frac{5}{24}}\bar c-\frac{\sqrt 5}{2}\left(E( \int_0^1 W^\mu(r)^2dr)\right)^{-1/2}\left(\bar cE(\int_0^1\left(\int_0^r
 W(s)ds-\int_0^1\!\!\!\int_0^t W(s)dsdt\right)dW(r))\right) \nonumber\\
 &&-\sqrt{\frac{135}{2}} \bar cE\left(\int_0^1 W^\mu(r)\left(\int_0^r W(s)ds-\int_0^1\!\!\!\int_0^t W(s)dsdt\right)dr\right)=N(0,1)-\frac{1}{8}\sqrt{\frac{15}{2}}\bar c,
 \end{eqnarray}}
 where
 \begin{eqnarray*}
 E\left( \int_0^1 W^\mu(r)^2dr\right) &=& E\left(\int_0^1 W(r)^2dr\right)-E\left(\int_0^1 W(r)dr\right)^2 \\
 &=& \frac{1}{2}-E\left(\int_0^1\!\!\!\int_0^1 W(s)W(r)dsdr\right) \\
 &=& \frac{1}{2}-\int_0^1\!\!\!\int_0^1 (s\wedge r)dsdr=\frac{1}{2}-\int_0^1\!\!\!\int_0^r sdsdr-\int_0^1\!\!\!\int_r^1 rdsdr=\frac{1}{6}, \\
 \end{eqnarray*}
 \begin{eqnarray*}
 &&E\left(\int_0^1\left(\int_0^r W(s)ds-\int_0^1\!\!\!\int_0^t W(s)dsdt\right)dW(r)\right) \\
 &=& E\left[\int_0^1\!\!\!\int_0^r W(s)dsdW(r)-\int_0^1\!\!\!\int_0^t W(s)dsdtW(1)\right] \\
 &=& E\left[\left.\int_0^r W(s)dsW(r)\right|_0^1-\int_0^1 W(r)^2dr-\int_0^1\!\!\!\int_0^t W(s)dsdtW(1)\right] \\
 &=& E\left[\int_0^1 W(s)dsW(1)-\int_0^1 W(r)^2dr-\int_0^1\!\!\!\int_0^t W(s)dsdtW(1)\right] \\
 &=& \int_0^1 E[W(s)W(1)]ds-\int_0^1 E[W(r)^2]dr-\int_0^1\!\!\!\int_0^t E[W(s)W(1)]dsdt \\
 &=& \int_0^1 sds-\int_0^1 rdr-\int_0^1\!\!\!\int_0^t sdsdt=-\frac{1}{6},
 \end{eqnarray*}
 and
 \begin{eqnarray*}
 &&E\left(\int_0^1 W^\mu(r)\left(\int_0^r W(s)ds-\int_0^1\!\!\!\int_0^t W(s)dsdt\right)dr\right) \\
 &=& E\Bigg[\int_0^1 W(r)\int_0^r W(s)dsdr-\int_0^1 W(r)dr\int_0^1\!\!\!\int_0^t W(s)dsdt-\int_0^1 W(r)dr\int_0^1\!\!\!\int_0^r W(s)dsdr \\
 &&+\int_0^1 W(r)dr\int_0^1\!\!\!\int_0^t W(s)dsdt\Bigg] \\
 &=& E\left[\int_0^1 W(r)\int_0^r W(s)dsdr-\int_0^1 W(r)dr\int_0^1\!\!\!\int_0^t W(s)dsdt\right]
 \end{eqnarray*}
 \begin{eqnarray*}
 &=& \int_0^1\!\!\!\int_0^r E[W(r)W(s)]dsdr-\int_0^1\!\!\!\int_0^1\!\!\!\int_0^t E[W(r)W(s)]dsdtdr \\
 &=& \int_0^1\!\!\!\int_0^r (r\wedge s)dsdr-\int_0^1\!\!\!\int_0^1\!\!\!\int_0^t (r\wedge s)dsdtdr \\
 &=& \int_0^1\!\!\!\int_0^r sdsdr-\left[\int_0^1\!\!\!\int_0^r\!\!\!\int_0^t sdsdtdr+\int_0^1\!\!\!\int_r^1(\int_0^r sds+\int_r^t rds)dtdr\right] \\
 &=& \int_0^1 \frac{1}{2}r^2dr-\left[\int_0^1\frac{1}{6}r^3dr+\int_0^1(-\frac{1}{2}r^2+\frac{1}{2}r)dr\right]
 =\frac{1}{6}-\frac{1}{8}=\frac{1}{24}.
 \end{eqnarray*}

Moreover, from the similar complicated derivations, we have
{\footnotesize
\begin{eqnarray}\label{S.3}
&&\hat t_{\delta,22} \nonumber \\
&\Rightarrow& N(0,1)-\sqrt{\frac{5}{51}}\bar c-\sqrt{\frac{10}{17}}\left(E( \int_0^1 W^\mu(r)^2dr)\right)^{-1/2}\left(\bar cE(\int_0^1\left(\int_0^r
 W(s)ds-\int_0^1\!\!\!\int_0^t W(s)dsdt\right)dW(r))\right) \nonumber \\
 &&-6\sqrt{\frac{10}{17}} \bar c\left(E( \int_0^1 W^\mu(r)^2dr)\right)^{-1/2}E\left(\int_0^1 W^\mu(r)\left(\int_0^r W(s)ds-\int_0^1\!\!\!\int_0^t W(s)dsdt\right)dr\right) \nonumber \\
 &=& N(0,1)-\frac{1}{2}\sqrt{\frac{15}{17}}\bar c.
\end{eqnarray}}
The results (\ref{S.2}) and (\ref{S.3}) were obtained in Moon and Perron (2008),
where we have slightly different derivations here.

For $\hat t_{\delta,31}$ and $\hat t_{\delta,32}$, using the similar derivations, we have
{\small
\begin{eqnarray}\label{S.4}
&&\hat t_{\delta,31} \nonumber \\
&\Rightarrow& N(0,1)-\sqrt{\frac{448}{277}}N^{1/4}\bar c\frac{E[A-12B_5^2]+E[B_1-12B_6B_8]}{\sqrt{E[A-12B_5^2]}}
+\sqrt{\frac{448}{277}}\frac{\overline{c^2}E[B_{10}-12 B_9B_8]}{\sqrt{E[A-12B_5^2]}} \nonumber \\
&&+\sqrt{\frac{448}{277}}\times\frac{15\sqrt{15}}{4} \overline{c^2}\left[E[B_3-12B_6^2]+2E[B_7-12B_5B_9]\right]=N(0,1)-\frac{1}{14}\sqrt{\frac{105}{277}}\overline{c^2},
\end{eqnarray}}
where $\overline{c^2}=\lim_{N\to \infty}N^{-1}\sum_{i=1}^N c_i^2$,
 \begin{eqnarray*}
 A &=& \int_0^1 W^\mu(r)^2dr, \nonumber \\
 B_1 &=& \int_0^1 \left\{\int_0^r W(s)ds-\int_0^1\!\!\!\int_0^t
 W(s)dsdt\right\}dW(r), \nonumber\\
 B_2 &=& \int_0^1 W^\mu(r)dW(r), \nonumber \\
 B_3 &=& \int_0^1 \left\{\int_0^r W(s)ds-\int_0^1\!\!\!\int_0^t W(s)dsdt\right\}^2dr, \nonumber \\
 B_4 &=& \int_0^1 W^\mu(r)\left\{\int_0^r
 W(s)ds-\int_0^1\!\!\!\int_0^t W(s)dsdt\right\}dr, \nonumber \\
 B_5 &=& \int_0^1 (r-\frac{1}{2})W^\mu(r)dr, \nonumber
 \end{eqnarray*}
 \begin{eqnarray*}
 B_6 &=& \int_0^1 (r-\frac{1}{2})\left\{\int_0^r
 W(s)ds-\int_0^1\!\!\!\int_0^t W(s)dsdt\right\}dr, \nonumber \\
 B_7 &=& \int_0^1 W^\mu(r)\left\{\int_0^r (r-s)
 W(s)ds-\int_0^1\!\!\!\int_0^t (t-s) W(s)dsdt\right\}dr, \nonumber \\
 B_8 &=& \int_0^1 (r-\frac{1}{2})dW(r)=\frac{1}{2}W(1)-\int_0^1 W(r)dr, \nonumber \\
 B_9 &=& \int_0^1(r-\frac{1}{2})\left\{\int_0^r (r-s)
 W(s)ds-\int_0^1\!\!\!\int_0^t (t-s) W(s)dsdt\right\}dr, \nonumber \\
 B_{10} &=& \int_0^1\left\{\int_0^r (r-s)
 W(s)ds-\int_0^1\!\!\!\int_0^t (t-s) W(s)dsdt\right\}d W(r),
 \end{eqnarray*}
 and $W^\mu(r)=W(r)-\int_0^1 W(s)ds$. From the complicated calculations, we have
 $E[A-12B_5^2]=\frac{1}{15}$, $E[B_1-12B_6B_8]=-\frac{1}{15}$,
 $E[B_4-12B_5B_6]=0$, $E[B_{10}-12B_9B_8]=0$, $E[B_3-12B_6^2]=\frac{1}{420}$, and $E[B_7-12B_5B_9]=-\frac{1}{420}$.
 Furthermore, we have
{\small
\begin{eqnarray}\label{S.5}
&&\hat t_{\delta,32} \nonumber \\
&\Rightarrow& N(0,1)-\sqrt{\frac{112}{193}}N^{1/4}\bar c\frac{E[A-12B_5^2]+E[B_1-12B_6B_8]}{\sqrt{E[A-12B_5^2]}}
+\sqrt{\frac{112}{193}}\frac{c^2E[B_{10}-12 B_9B_8]}{\sqrt{E[A-12B_5^2]}} \nonumber \\
 &&+\sqrt{\frac{112}{193}}\times\frac{15}{2}\frac{ c^2\left[E[B_3-12B_6^2]+2E[B_7-12B_5B_9]\right]}{\sqrt{E[A-12B_5^2]}}=N(0,1)-\frac{\sqrt{15}}{56}\sqrt{\frac{112}{193}}c^2.
\end{eqnarray}}
The results in (\ref{S.4}) and (\ref{S.5}) were also obtained in Moon and Perron (2004) and Moon et al. (2007).

\section{Bias correction in LLC test}

For LLC tests, in addition to the two bias correction methods mentioned in the paper,
there is another way to correct the bias for Model $3.2'$ and Model $3.3'$.
First, we consider Model $3.2'$. The additional way to correct the bias is to only correct the numerator's bias.
However, the test constructed in this way does not have power in the neighborhood of unity with order $N^{-1/2}T^{-1}$
but with order $N^{-1/4}T^{-1}$ as shown in Moon and Perron (2008).

We need to modify Assumption 3 to

 \noindent {\bf Assumption $3'$} Let $\delta_i=1-c_i/(N^{1/4}T)$ where $c_i\geq 0$,
 $i=1,\dots,N$.

With this type of bias correction, the $t$-statistic is given by
\[
\tilde t_{\delta,23}
= \sqrt 2\frac{\hat \delta_2+\sqrt N(T-1)/(2(\sum_{i=1}^N\sum_{t=2}^T [(y_{i,t-1}-\bar y_{i,t-1})^2/\hat \sigma_{\varepsilon,i}^2]))}{(\sum_{i=1}^N\sum_{t=2}^T [(y_{i,t-1}-\bar y_{i,t-1})^2/\hat \sigma_{\varepsilon,i}^2])^{-1/2}}. \]
Under Assumption $3'$,
\begin{eqnarray*}
\tilde t_{\delta,23}
&=& -\frac{\sqrt 2}{N^{1/4}T}\frac{\sum_{i=1}^N\sum_{t=2}^T c_i[(y_{i,t-1}-\bar y_{i,t-1})^2/\hat \sigma_{\varepsilon,i}^2]}{\sqrt{\sum_{i=1}^N\sum_{t=2}^T [(y_{i,t-1}-\bar y_{i,t-1})^2/\hat \sigma_{\varepsilon,i}^2]}} \\
&&+\sqrt 2\frac{N^{-1/2}T^{-1}\sum_{i=1}^N\sum_{t=2}^T \left[\left((y_{i,t-1}-\bar y_{i,t-1})(\varepsilon_{i,t}-\bar\varepsilon_{i,t})/\hat \sigma_{\varepsilon,i}^2\right)+\frac{1}{2}\right]}{\sqrt{N^{-1}T^{-2}\sum_{i=1}^N\sum_{t=2}^T [(y_{i,t-1}-\bar y_{i,t-1})^2/\hat \sigma_{\varepsilon,i}^2]}}.
\end{eqnarray*}
Hence, as $T\to \infty$
\begin{eqnarray}\label{S.6}
\tilde t_{\delta,23} &\Rightarrow& -\sqrt 2\frac{N^{-3/4}\sum_{i=1}^N c_i\int_0^1 \tilde K_{i,c_i}^\mu(r)^2dr}{\sqrt{N^{-1}\sum_{i=1}^N\int_0^1 \tilde K_{i,c_i}^\mu(r)^2dr}}+\sqrt 2\frac{N^{-1/2}\sum_{i=1}^N \left(\int_0^1 \tilde K_{i,c_i}^\mu(r)dW_i(r)+\frac{1}{2}\right)}{\sqrt{N^{-1}\sum_{i=1}^N\int_0^1 \tilde K_{i,c_i}^\mu(r)^2dr}} \nonumber \\
&\stackrel{def}{=}& \sqrt 2\frac{N^{-1/2}\sum_{i=1}^N (\breve U_{2i}+\frac{1}{2})}{\sqrt{N^{-1}\sum_{i=1}^N \breve V_{2i}}},
\end{eqnarray}
where $\tilde K_{i,c_i}^\mu(r)=\tilde K_{i,c_i}(r)-\int_0^1 \tilde K_{i,c_i}(s)ds$, and
$\tilde K_{i,c_i}(r)= \int_0^r e^{-c_iN^{-1/4}(r-s)}dW_i(s)$.

Our approach can also be applied here. Substituting $\theta=iu/2$, $x=-v/u$ and $c=c_iN^{-1/4}$
into $\varphi_2(\theta;c,1,x)$ in Lemma \ref{lemma2}, we have the joint m.g.f. for $(\breve U_{2i},\breve V_{2i})$ as
{\footnotesize
 \begin{eqnarray}\label{S.7}
 && \tilde \psi_2(u,v) \nonumber \\
 &=& e^{-\frac{u}{2}}\Bigg[e^{-c_iN^{-\frac{1}{4}}}\Bigg[\frac{u^2+2v+c_i^2N^{-1/2}u-c_i^3N^{-3/4}}{2v-c_i^2N^{-1/2}}
 \frac{\sin\sqrt{2v-c_i^2N^{-1/2}}}{\sqrt{2v-c_i^2N^{-1/2}}}-c_i^2N^{-1/2}\frac{\cos\sqrt{2v-c_i^2N^{-1/2}}}{2v-c_i^2N^{-1/2}} \nonumber \\
 &&+(2u^2-4c_iN^{-1/4}v+2c_i^2N^{-1/2}u)\frac{\cos\sqrt{2v-c_i^2N^{-1/2}}-1}{(2v-c_i^2N^{-1/2})^2}\Bigg]\Bigg]^{-1/2}.
 \end{eqnarray}}
 Hence, the joint m.g.f. for $(N^{-1/2}\sum_{i=1}^N \breve U_{2i},N^{-1}\sum_{i=1}^N \breve V_{2i})$ is
 {\small
 \begin{eqnarray*}
 &&\tilde\phi_2(u,v) = \left(\psi_2\left(\frac{u}{\sqrt N},\frac{v}{N}\right)\right)^N \nonumber \\
 &=& e^{-\frac{N\frac{u}{\sqrt N}}{2}}\Bigg[e^{-\sum_{i=1}^N c_iN^{-\frac{1}{4}}}\prod_{i=1}^N\Bigg[\frac{\frac{u^2}{N}+\frac{2v}{N}+c_i^2N^{-1/2}\frac{u}{\sqrt N}-c_i^3N^{-3/4}}{\frac{2v}{N}-c_i^2N^{-1/2}}
 \frac{\sin\sqrt{\frac{2v}{N}-c_i^2N^{-1/2}}}{\sqrt{\frac{2v}{N}-c_i^2N^{-1/2}}} \nonumber \\
 &&-c_i^2N^{-1/2}\frac{\cos\sqrt{\frac{2v}{N}-c_i^2N^{-1/2}}}{\frac{2v}{N}-c_i^2N^{-1/2}}
 +(\frac{2u^2}{N}-4c_iN^{-1/4}\frac{v}{N}+2c_i^2N^{-1/2}\frac{u}{\sqrt N})\frac{\cos\sqrt{\frac{2v}{N}-c_i^2N^{-1/2}}-1}{(\frac{2v}{N}-c_i^2N^{-1/2})^2}\Bigg]\Bigg]^{-1/2}.
 \end{eqnarray*}}

 Further, we have
 \begin{eqnarray}\label{S.8}
 \left.\frac{\partial }{\partial u}\tilde\phi_2(u,-v)\right|_{u=0} &=& -\frac{\sqrt N}{2}e^{-v/6}-\frac{1}{24}ve^{-v/2}N^{-3/4}\sum_{i=1}^N c_i \nonumber\\
 &&+\frac{ve^{-v/6}}{40N}\sum_{i=1}^N c_i^2+\frac{e^{-v/6}}{24N}\sum_{i=1}^N c_i^2+O(N^{-1/4}),
 \end{eqnarray}
 and
 \begin{eqnarray}\label{S.9}
 && \tilde\phi_2(0,-v) \nonumber\\
 &=& \left[e^{-\sum_{i=1}^N c_iN^{-\frac{1}{4}}}e^{\sum_{i=1}^N\log\Big(1+c_iN^{-1/4}+\frac{v}{3N}-\frac{1}{6}c_ivN^{-5/4}
 +\frac{c_i^2v}{10}N^{-3/2}+O(N^{-7/4}))\Big)}\right]^{-1/2} \nonumber\\
 &=& e^{-v/6}+\frac{v e^{-v/6}}{12 N^{5/4}}\sum_{i=1}^N c_i-\frac{ve^{-v/6}}{20N^{3/2}}\sum_{i=1}^N c_i^2+O(N^{-3/4}).
 \end{eqnarray}
 From (\ref{S.6}), applying (\ref{3.1.sawa}) and (\ref{3.1.sawa1}) and plugging in (\ref{S.8}) and (\ref{S.9}), we have
 \begin{eqnarray*}
 E\left(\tilde t_{\delta,23}\right) &=& E\left(\sqrt 2\frac{N^{-1/2}\sum_{i=1}^N \left(\breve U_{2i}+\frac{1}{2}\right)}{\sqrt{N^{-1}\sum_{i=1}^N \breve V_{2i}}}\right) \\
 &=& \frac{\sqrt 2}{\Gamma(\frac{1}{2})}\int_0^\infty \frac{1}{\sqrt v} \left.\frac{\partial }{\partial u}\tilde\phi_2(u,-v)\right|_{u=0}dv+\frac{\sqrt N}{\sqrt 2\Gamma(\frac{1}{2})}\int_0^\infty \frac{1}{\sqrt v} \tilde\phi_2(0,-v)dv \\
 &=& \left(N^{-1}\sum_{i=1}^N c_i^2\right)\frac{\sqrt 2}{24\sqrt \pi}\int_0^\infty \frac{e^{-v/6}}{\sqrt v}dv+O_p(N^{-1/4})=\frac{\sqrt 3}{12}\overline{c^2}+O_p(N^{-1/4}).
 \end{eqnarray*}

 Next, we consider Model $3.3'$. Similarly, we can construct the test statistic by only correcting
 the numerator's bias. However, the test constructed in this way does not have power in the neighborhood of
 unity with order $N^{-1/4}T^{-1}$ but in the neighborhood of unity with order $N^{-1/8}T^{-1}$.

We need to modify Assumption 3 to

 \noindent {\bf Assumption $3''$} Let $\delta_i=1-c_i/(N^{1/8}T)$ where $c_i\geq 0$,
 $i=1,\dots,N$.

 With this type of bias correction, the $t$-statistic is given by
 \[
 \tilde t_{\delta,33}
 = 2\frac{\hat \delta_3+\sqrt N T/(2(\sum_{i=1}^N\sum_{t=2}^T [(y_{i,t-1}-\bar y_{i,t-1})-\frac{\sum_{s=1}^T (s-\bar s)(y_{is}-\bar y_{is})}{\sum_{s=1}^T (s-\bar s)^2}(t-\bar t)]^2/\hat \sigma_{\varepsilon,i}^2))}{(\sum_{i=1}^N\sum_{t=2}^T [(y_{i,t-1}-\bar y_{i,t-1})-\frac{\sum_{s=1}^T (s-\bar s)(y_{is}-\bar y_{is})}{\sum_{s=1}^T (s-\bar s)^2}(t-\bar t)]^2/\hat \sigma_{\varepsilon,i}^2)^{-1/2}}. \]
Under Assumption $3''$, we have
{\small
 \begin{eqnarray*}
 \tilde t_{\delta,33}
 &=& -\frac{2}{N^{1/8}T}\frac{\sum_{i=1}^N c_i\left[\left(\sum_t
 (y_{i,t-1}-\bar y_{i,t-1})^2\right)-\frac{\left(\sum_t(t-\bar
 t)(y_{i,t-1}-\bar y_{i,t-1})\right)^2}{\sum_t (t-\bar t)^2}\right]/\hat \sigma_{\varepsilon,i}^2}{\sqrt{\sum_{i=1}^N\left[\left(\sum_t
 (y_{i,t-1}-\bar y_{i,t-1})^2\right)-\frac{\left(\sum_t(t-\bar
 t)(y_{i,t-1}-\bar y_{i,t-1})\right)^2}{\sum_t (t-\bar t)^2}\right]/\hat \sigma_{\varepsilon,i}^2}} \\
&&+2\left(N^{-1}\sum_{i=1}^N T^{-2}\left[\left(\sum_t(y_{i,t-1}-\bar y_{i,t-1})^2\right)-\frac{\left(\sum_t(t-\bar
 t)(y_{i,t-1}-\bar y_{i,t-1})\right)^2}{\sum_t (t-\bar t)^2}\right]/\hat \sigma_{\varepsilon,i}^2\right)^{-1/2}
 \end{eqnarray*}
 \begin{eqnarray*}
 &&\times\Bigg(N^{-1/2}\sum_{i=1}^N \Bigg[T^{-1}\Bigg(\left(\sum_t
 (y_{i,t-1}-\bar y_{i,t-1})(\varepsilon_{it}-\bar \varepsilon_{it})\right)-\left(\sum_t (t-\bar t)^2\right)^{-1} \\
 &&\times\left(\sum_t(t-\bar
 t)(y_{i,t-1}-\bar y_{i,t-1})\right)\left(\sum_t(t-\bar t)(\varepsilon_{it}-\bar \varepsilon_{it})\right)\Bigg)/\hat \sigma_{\varepsilon,i}^2+\frac{1}{2}\Bigg]\Bigg).
\end{eqnarray*}}
Furthermore, as $T\to \infty$
\begin{eqnarray}\label{S.10}
\tilde t_{\delta,33}
&\Rightarrow& -2\frac{N^{-5/8}\sum_{i=1}^N c_i\left[\int_0^1
 \breve K_{i,c_i}^\mu(r)^2dr-12\left(\int_0^1
 (r-\frac{1}{2})\breve K_{i,c_i}^\mu(r)dr\right)^2\right]}{\sqrt{N^{-1}\sum_{i=1}^N\left[\int_0^1
 \breve K_{i,c_i}^\mu(r)^2dr-12\left(\int_0^1
 (r-\frac{1}{2})\breve K_{i,c_i}^\mu(r)dr\right)^2\right]}} \nonumber\\
 &&+2\frac{N^{-1/2}\sum_{i=1}^N \left(\int_0^1
 \breve K_{i,c_i}^\mu(r)dW_i(r)-12\int_0^1
 (r-\frac{1}{2})\breve K_{i,c_i}^\mu(r)dr\int_0^1 (r-\frac{1}{2})dW_i(r)+\frac{1}{2}\right)}{\sqrt{N^{-1}\sum_{i=1}^N\left[\int_0^1
 \breve K_{i,c_i}^\mu(r)^2dr-12\left(\int_0^1
 (r-\frac{1}{2})\breve K_{i,c_i}^\mu(r)dr\right)^2\right]}} \nonumber\\
 &\stackrel{def}{=}& 2\frac{N^{-1/2}\sum_{i=1}^N \left(\breve U_{3i}+\frac{1}{2}\right)}{\sqrt{N^{-1}\sum_{i=1}^N \breve V_{3i}}},
\end{eqnarray}
where $\breve K_{i,c_i}^\mu(r)=\breve K_{i,c_i}(r)-\int_0^1 \breve K_{i,c_i}(s)ds$, and
$\breve K_{i,c_i}(r) = \int_0^r e^{-c_iN^{-1/8}(r-s)}dW_i(s)$.

Our approach can also be applied here. Substituting $\theta=iu/2$, $x=-v/u$ and $c=c_iN^{-1/8}$ into $\varphi_4(\theta;c,1,x)$ in Lemma \ref{lemma2}, we have the joint m.g.f. for $(\breve U_{3i},\breve V_{3i})$ as
 {\scriptsize
 \begin{eqnarray}\label{S.11}
 && \psi_3(u,v) \nonumber \\
 &=& e^{-\frac{u}{2}}\Bigg[e^{-c_iN^{-\frac{1}{8}}}\Big[\frac{(c_iN^{-1/8})^5-(c_iN^{-1/8})^4u-4((c_iN^{-1/8})^2+3c_iN^{-1/8}+27)u^2
 -8v((c_iN^{-1/8})^2-3c_iN^{-1/8}-3)}{(2v-c_i^2N^{-1/4})^2} \nonumber \\
 &&\times\frac{\sin\sqrt{2v-c_i^2N^{-1/4}}}{\sqrt{2v-c_i^2N^{-1/4}}}+\frac{24((c_iN^{-1/8})^4u+8vu^2-4(c_iN^{-1/8}+1)
 (v^2-3 u^2))}{(2v-c_i^2N^{-1/4})^3}\Bigg(\frac{\sin\sqrt{2v-c_i^2N^{-1/4}}}{\sqrt{2v-c_i^2N^{-1/4}}} \nonumber \\
 &&
 +\frac{\cos\sqrt{2v-c_i^2N^{-1/4}}}{2v-c_i^2N^{-1/4}}-\frac{1}{2v-c_i^2N^{-1/4}}\Bigg)+\Bigg(\frac{c_i^4N^{-1/2}}{(2v-c_i^2N^{-1/4})^2}-
 \frac{8(c_i^4N^{-1/4}u-c_i^3N^{-3/8}2v+4(c_i^2N^{-1/4}+3c_iN^{-1/8}+6)u^2)}{(2v-c_i^2N^{-1/4})^3}\Bigg) \nonumber \\
 &&\times\cos\sqrt{2v-c_i^2N^{-1/4}}-\frac{4(c_i^4N^{-1/2}u+4(c_i^2N^{-1/4}+3c_iN^{-1/8}-3)u^2-2c_i^2N^{-1/4}v(c_iN^{-1/8}+3))}
 {(2v-c_i^2N^{-1/4})^3}\Big]\Bigg]^{-1/2}.
 \end{eqnarray}}
 Hence, the joint m.g.f. for $(N^{-1/2}\sum_{i=1}^N \breve U_{3i},N^{-1}\sum_{i=1}^N \breve V_{3i})$ is
 {\tiny
 \begin{eqnarray*}
 &&\tilde \phi_3(u,v) = \left(\psi_3\left(\frac{u}{\sqrt N},\frac{v}{N}\right)\right)^N \nonumber \\
 &=& e^{-\frac{N\frac{u}{\sqrt N}}{2}}\Bigg[e^{-\sum_{i=1}^N c_iN^{-\frac{1}{8}}}\prod_{i=1}^N\Bigg[\frac{(c_iN^{-1/8})^5-(c_iN^{-1/8})^4\frac{u}{\sqrt N}-4((c_iN^{-1/8})^2+3c_iN^{-1/8}+27)\frac{u^2}{N}
 -\frac{8v}{N}((c_iN^{-1/8})^2-3c_iN^{-1/8}-3)}{(\frac{2v}{N}-c_i^2N^{-1/4})^2} \nonumber \\
 &&\times\frac{\sin\sqrt{\frac{2v}{N}-c_i^2N^{-1/4}}}{\sqrt{\frac{2v}{N}-c_i^2N^{-1/4}}}+\frac{24((c_iN^{-1/8})^4\frac{u}{\sqrt N}+8\frac{vu^2}{N^2}-4(c_iN^{-1/8}+1)
 (\frac{v^2}{N^2}-3 \frac{u^2}{N}))}{(\frac{2v}{N}-c_i^2N^{-1/4})^3}\Bigg(\frac{\sin\sqrt{\frac{2v}{N}-c_i^2N^{-1/4}}}{\sqrt{\frac{2v}{N}-c_i^2N^{-1/4}}} \nonumber \\
 &&
 +\frac{\cos\sqrt{\frac{2v}{N}-c_i^2N^{-1/4}}}{\frac{2v}{N}-c_i^2N^{-1/4}}-\frac{1}{\frac{2v}{N}-c_i^2N^{-1/4}}\Bigg)
 +\Bigg(\frac{c_i^4N^{-1/2}}{(\frac{2v}{N}-c_i^2N^{-1/4})^2}-
 \frac{8(c_i^4N^{-1/2}\frac{u}{\sqrt N}-c_i^3N^{-3/8}\frac{2v}{N}+4(c_i^2N^{-1/4}+3c_iN^{-1/8}+6)\frac{u^2}{N})}{(\frac{2v}{N}-c_i^2N^{-1/4})^3}\Bigg) \nonumber \\
 &&\times\cos\sqrt{\frac{2v}{N}-c_i^2N^{-1/4}}-\frac{4(c_i^4N^{-1/2}\frac{u}{\sqrt N}+4(c_i^2N^{-1/4}+3c_iN^{-1/8}-3)\frac{u^2}{N}-2c_i^2N^{-1/4}\frac{v}{N}(c_iN^{-1/8}+3))}
 {(\frac{2v}{N}-c_i^2N^{-1/4})^3}\Bigg]\Bigg]^{-1/2}.
 \end{eqnarray*}}

 Furthermore, we have
 \begin{eqnarray}\label{S.12}
 \left.\frac{\partial }{\partial u}\tilde \phi_3(u,-v)\right|_{u=0} &=& -\frac{\sqrt N}{2}e^{-v/15}-\frac{1}{840}ve^{-v/15}N^{-3/4}\sum_{i=1}^N c_i^2 \nonumber\\
 &&+\frac{17}{60480}ve^{-v/15}N^{-1}\sum_{i=1}^N c_i^4+\frac{1}{1440}e^{-v/15}N^{-1}\sum_{i=1}^N c_i^4+O_p(N^{-1/8}),
 \end{eqnarray}
 and
 \begin{eqnarray}\label{S.13}
 && \tilde \phi_3(0,-v) \nonumber\\
 &=& \Bigg[e^{-\sum_{i=1}^N c_iN^{-\frac{1}{6}}}\exp\Bigg(\sum_{i=1}^N\log\Big(1+c_iN^{-1/8}+\frac{1}{2}c_i^2N^{-1/4}+\frac{1}{6}c_i^3N^{-3/8}+N^{-1/2}\frac{c_i^4}{24}
 +N^{-5/8}\frac{c_i^5}{120} \nonumber\\
 &&+N^{-3/4}\frac{c_i^6}{720}+N^{-7/8}\frac{c_i^7}{7!}+\frac{2v}{15N}+N^{-1}\frac{c_i^8}{8!}+N^{-9/8}\frac{2c_iv}{15}
 +N^{-9/8}\frac{c_i^9}{9!}+N^{-5/4}\frac{13c_i^2v}{210}+N^{-5/4}\frac{c_i^{10}}{10!} \nonumber\\
 &&+N^{-11/8}\frac{11c_i^3v}{630}+N^{-11/8}\frac{c_i^{11}}{11!}+N^{-3/2}\frac{13c_i^4v}{3024}+N^{-3/2}\frac{c_i^{12}}{12!}
 +O(N^{-13/8})\Big)\Bigg)\Bigg]^{-1/2} \nonumber\\
 &=& e^{-v/15}+\frac{v e^{-v/15}}{420 N^{5/4}}\sum_{i=1}^N c_i^2-\frac{17v e^{-v/15}}{30240 N^{3/2}}\sum_{i=1}^N c_i^4+O(N^{-5/8}).
 \end{eqnarray}
 From (\ref{S.10}), applying (\ref{3.1.sawa}) and (\ref{3.1.sawa1}) and plugging in (\ref{S.12}) and (\ref{S.13}), we have
 \begin{eqnarray*}
 E\left(\tilde t_{\delta,33}\right) &=& E\left(2\frac{N^{-1/2}\sum_{i=1}^N \left(\breve U_{3i}+\frac{1}{2}\right)}{\sqrt{N^{-1}\sum_{i=1}^N \breve V_{3i}}}\right) \\
 &=& 2\frac{1}{\Gamma(\frac{1}{2})}\int_0^\infty \frac{1}{\sqrt v} \left.\frac{\partial }{\partial u}\tilde\phi_3(u,-v)\right|_{u=0}dv
 +\frac{\sqrt N}{\Gamma(\frac{1}{2})}\int_0^\infty \frac{1}{\sqrt v} \tilde\phi_3(0,-v)dv \\
 &=& -\sqrt N\frac{1}{\sqrt \pi}\int_0^\infty \frac{e^{-v/15}}{\sqrt v}dv-\left(N^{-5/6}\sum_{i=1}^N c_i^2\right)\frac{1}{420\sqrt \pi}\int_0^\infty \frac{ve^{-v/15}}{\sqrt v}dv \\
 &&+\left(N^{-1}\sum_{i=1}^N c_i^4\right)\frac{17}{30240\sqrt \pi}\int_0^\infty \frac{ve^{-v/15}}{\sqrt v}dv+\left(N^{-1}\sum_{i=1}^N c_i^4\right)\frac{2}{1440\sqrt \pi}\int_0^\infty \frac{e^{-v/15}}{\sqrt v}dv \\
 &&+\sqrt N\frac{1}{\sqrt \pi}\int_0^\infty \frac{e^{-v/15}}{\sqrt v}dv+\left(N^{-5/6}\sum_{i=1}^N c_i^2\right)\frac{1}{420\sqrt \pi}\int_0^\infty \frac{ve^{-v/15}}{\sqrt v}dv \\
 &&-\left(N^{-1}\sum_{i=1}^N c_i^4\right)\frac{17}{30240\sqrt \pi}\int_0^\infty \frac{ve^{-v/15}}{\sqrt v}dv+O_p(N^{-1/8}) \\
 &=& \frac{\sqrt{15}}{720}\overline{c^4}+O_p(N^{-1/8}),
 \end{eqnarray*}
 where $\overline{c^4}=\lim_{N\to \infty}N^{-1}\sum_{i=1}^N c_i^4$.

 The results are summarized in the following proposition.
 \begin{proposition}
 Under Assumptions 1, 2 and 4, when $T\to \infty$ followed by $N\to
 \infty$, we have the following asymptotic results.
\begin{itemize}
\item[(a)] For Model $3.2'$, with the additional Assumption $3'$, $\hat t_{\delta,23}\Rightarrow N(0,1)+\frac{\sqrt 3}{12}\overline{c^2}$;

\item[(b)] For Model $3.3'$, with the additional Assumption $3''$, $\hat t_{\delta,33}\Rightarrow N(0,1)+\frac{\sqrt{15}}{720}\overline{c^4}$.
\end{itemize}
 \end{proposition}

\noindent {\bf Remark S.1} From Assumptions $3'$ and $3''$, we can see that $\hat t_{\delta,23}$
and $\hat t_{\delta,33}$ have local power in the neighborhood of unity with a slower rate, compared with
those of $\hat t_{\delta,21}$, $\hat t_{\delta,22}$, $\hat t_{\delta,31}$, and $\hat t_{\delta,32}$, respectively.
In addition, $\hat t_{\delta,23}$ and $\hat t_{\delta,33}$ have the local power on the right tail rather than
the left tail.

\end{small}

\end{document}